\documentclass[preprint,natbib209]{aastex}
\usepackage{url}
\usepackage{amsmath}
\usepackage{afterpage}
\slugcomment{\apj, accepted}
\shorttitle{Comparing Physical Properties of LAEs and oELGs}
\shortauthors{Hagen et al.}
\newcommand{\eg}{e.g.}
\begin{document}

\title{{\sl HST\/} Emission Line Galaxies at $z \sim 2$: Comparing Physical 
Properties of Lyman Alpha and Optical Emission Line Selected Galaxies}

\author{Alex Hagen\altaffilmark{1}, Gregory R. Zeimann\altaffilmark{1}}
\affil{Department of Astronomy \& Astrophysics, The Pennsylvania State 
University, 525 Davey Lab, University Park, PA 16802}

\email{hagen@psu.edu, grzeimann@psu.edu}

\author{Christoph Behrens}
\affil{Institut f\"ur Astrophysik, Georg-August Universit\"at G\"ottingen, 
Friedrich-Hund-Platz 1, 37077, G\"ottingen, Germany}
\email{cbehren@astro.physik.uni-goettingen.de}

\author{Robin Ciardullo\altaffilmark{1}, 
Henry S. Grasshorn Gebhardt\altaffilmark{1}, Caryl Gronwall\altaffilmark{1}, 
Joanna S. Bridge\altaffilmark{1}, Derek B. Fox\altaffilmark{1}, 
Donald P. Schneider\altaffilmark{1}, Jonathan R. Trump\altaffilmark{1,2}}
\affil{Department of Astronomy \& Astrophysics, The Pennsylvania State 
University, 525 Davey Lab, University Park, PA 16802}

\email{rbc@astro.psu.edu, gebhardt@psu.edu caryl@astro.psu.edu, 
jsbridge@psu.edu dfox@astro.psu.edu, dps@astro.psu.edu}

\author{Guillermo A. Blanc\altaffilmark{3}}
\affil{Departamento de Astronomía, Universidad de Chile, Camino del 
Observatorio 1515, Las Condes, Santiago, Chile}
\email{gblancm@obs.carnegiescience.edu}

\author{Yi-Kuan Chiang, Taylor S. Chonis, Steven L. Finkelstein, Gary J. Hill, 
Shardha Jogee}
\affil{Department of Astronomy, The University of Texas at Austin, 
Austin, TX, 78712}
\email{ykchiang@astro.as.utexas.edu, tschonis@astro.as.utexas.edu, 
stevenf@astro.as.utexas.edu, hill@astro.as.utexas.edu, sj@astro.as.utexas.edu}

\author{Eric Gawiser}
\affil{Department of Physics and Astronomy, Rutgers, 
The State University of New Jersey, Piscataway, NJ 08854}
\email{gawiser@physics.rutgers.edu}

\altaffiltext{1}{Institute for Gravitation and the Cosmos, 
The Pennsylvania State University, University Park, PA 16802}
\altaffiltext{2}{Hubble Fellow}
\altaffiltext{3}{Visiting Astronomer, Observatories of the Carnegie Institution
for Science, 813 Santa Barbara St, Pasadena, CA 91101, USA}

\begin{abstract} 
We compare the physical and morphological properties of $z \sim 2$ Ly$\alpha$ 
emitting galaxies (LAEs) identified in the HETDEX Pilot Survey and narrow band 
studies with those of $z \sim 2$ optical emission line galaxies (oELGs)
identified via {\sl HST\/} WFC3 infrared grism spectroscopy.  
Both sets of galaxies extend over the same range in stellar mass 
($7.5 < \log M/M_{\odot} < 10.5$), size ($0.5 < R < 3.0$~kpc), and 
star-formation rate ($\sim 1 < {\rm SFR} < 100 \, M_{\odot}$~yr$^{-1}$). 
Remarkably, a comparison of  the most commonly used physical and 
morphological parameters --- stellar mass, half-light radius, UV slope, 
star formation rate, ellipticity, nearest neighbor distance, star formation surface 
density, specific star formation rate, [O~III] luminosity, and [O~III] equivalent width --- 
reveals no statistically significant differences between the populations.  
This suggests that the processes and conditions which regulate the escape of 
Ly$\alpha$ from a $z \sim 2$ star-forming galaxy do not depend on these
quantities.  In particular,
the lack of dependence on the UV slope suggests that Ly$\alpha$ emission is not being 
significantly modulated by diffuse dust in the interstellar medium.  
We develop a simple model of Ly$\alpha$ emission that connects LAEs to all 
high-redshift star forming galaxies where the escape of 
Ly$\alpha$ depends on the sightline through the galaxy.
Using this model, we find that mean solid angle for Ly$\alpha$ escape is
$\Omega_{{\rm Ly}\alpha} = 2.4 \pm 0.8$ steradians; this value is consistent 
with those calculated from other studies. 
\end{abstract}



\section{Introduction}
\label{sec:intro}

Ly$\alpha$ emitting galaxies (LAEs) are one of the most important constituents
of the  high-redshift universe. Although rare at $z \lesssim 0.3$ 
\citep{deharveng08, cowie10}, LAEs become more common (and more luminous)
with increasing redshift, and by $z \sim 6$, they comprise the 
bulk of the observed star-forming galaxy population \citep{cassata15, ouchi10, 
ciardullo12, bouwens10}.  The clustering
properties of $z \sim 3$ LAEs suggest a connection with today's $L^*$
galaxies \citep{gawiser07, guaita10}, while LAEs at larger redshifts typically 
have higher biases \citep{ouchi10}.

Ly$\alpha$ emitters are generally considered to be low-mass, dust-poor systems,
as the resonant nature of the Ly$\alpha$ line makes its escape from large, 
dusty systems problematic \citep[e.g.][]{verhamme06, gawiser06, gronwall07, 
finkelstein08, finkelstein09, schaerer11}.  However, the association of LAEs 
with low-mass galaxies is an oversimplification: multiple studies have 
demonstrated that luminous Ly$\alpha$ emitters have a very wide range of 
stellar mass (at least three dex or more), and that not all LAEs
are dust-poor \citep{finkelstein09, nilsson11, hagen14, vargas14}. 
This result calls into question the relationship between LAEs and
other denizens of the high-redshift galaxy zoo.  Hydrodynamic models that 
include Ly$\alpha$ radiative transfer predict that galaxy morphology and 
inclination should play an important role in determining the observability of 
Ly$\alpha$ \citep[e.g.,][]{verhamme12, yajima12, behrens14}, and studies 
have shown that the presence of Ly$\alpha$ is correlated with galaxy size and
ellipticity \citep[e.g.,][]{shibuya14a}.  It is also possible that Ly$\alpha$
emission is facilitated by a low column density of neutral hydrogen 
\citep{shibuya14b, song14}, but merger activity appears unrelated to the 
phenomenon \citep{shibuya14a}.

One difficulty with understanding the systematics of Ly$\alpha$ is that to
date, almost all the comparison samples of non-Ly$\alpha$ emitting star-forming
galaxies have been continuum-selected high-mass systems, such as Lyman-break
galaxies (LBGs), with sizes and star formation rates (SFRs) that are quite different from 
that of the typical LAE \citep[e.g.,][]{malhotra12}.
Only very recently, have LAEs been analyzed alongside of galaxies with a 
similar stellar mass range \citep{song14, hathi15}. 
We address this problem by comparing the physical properties
of photometrically- and spectroscopically-selected $z \sim 2$ LAEs 
\citep{guaita10, adams11} to those of optical emission line galaxies (oELGs) 
at the same redshift \citep{zeimann14}.
Because our oELGs have similar masses, sizes, and star-formation rates as their
LAE counterparts, we can analyze the two populations differentially, and
identify those properties most important for the escape of Ly$\alpha$
photons.

In Section \ref{sec:sample} we discuss our sample selection, 
in Section \ref{sec:analysis} we describe our analysis, and we present our
findings in Section \ref{sec:discussion}.
We adopt the standard concordance cosmology of $h = 0.7$, $\Omega_m = 0.3$, 
$\Omega_\Lambda = 0.7$, and $\Omega_k = 0$ \citep{planck13}.

\section{Sample Selection} 
\label{sec:sample}

Most previous studies of the physical properties of Ly$\alpha$ emitters have 
used continuum-selected galaxies as the experiment's control sample 
\citep[e.g.,][]{malhotra12}.  Such an assignment is clearly 
not ideal: systems such as Lyman-break and $BzK$ galaxies are not only more 
massive than typical LAEs, but their clustering properties indicate a very 
different evolutionary path.  What is needed is a set of galaxies with
roughly the same stellar mass, star-formation rate, and size as the
LAEs under investigation. As our analysis will show, the physical properties
of our sample of oELGs are indeed well-matched to those of our program LAEs.

Our comparison set of $z \sim 2$ oELGs was identified using the
 {\sl Hubble Space Telescope's} WFC3 camera and its
$R=130$ G141 grism.  A full description of these data, procured by
the 3D-HST and AGHAST surveys \citep[GO-11600, 12177, and 
12328;][]{brammer12, weiner14}, is given in \citet{zeimann14}. In brief, 
a total of $\sim 350$~arcmin$^2$ of the COSMOS \citep{COSMOS}, GOODS-N,
and GOODS-S \citep{GOODS} fields were surveyed over the wavelength range 
$1.08~\mu{\rm m} < \lambda < 1.68~\mu$m down to a 50\% monochromatic 
completeness flux limit of $F \sim 10^{-17}$~ergs~cm$^{-2}$~s$^{-1}$.
For each object brighter than F140W = 26, a spectrum was extracted, 
flux-calibrated, checked for contamination, continuum-subtracted,
and visually inspected for evidence of emission lines.   Galaxies
with unambiguous redshifts between $1.90 < z < 2.35$ were then selected
via the identification of at least two of the emission lines of 
[O~II] $\lambda 3727$, [Ne~III] $\lambda 3869$, H$\delta$, H$\gamma$, H$\beta$,
and [O~III] $\lambda\lambda 4959,5007$.   Although $\sim 15\%$ of the region's
galaxies were unmeasurable due to contamination from overlapping spectra, this 
procedure still produced a sample of 245 star-forming galaxies in the targeted redshift range
 with the distinctively-shaped [O III] blend typically being the 
brightest feature. 

The LAEs for our analysis are drawn from two sources.   The first was the 
Hobby Eberly Telescope Dark Energy Experiment (HETDEX) Pilot Survey 
\citep[HPS;][]{adams11, blanc11}, a blind integral field 
spectroscopic study that included 107~arcmin$^2$ of the COSMOS and GOODS-N 
fields.  At $z \sim 2.2$, 50\% of the HPS pointings reached a $5\, \sigma$ 
monochromatic flux limit of $1.3 \times 10^{-16}$~ergs~cm$^{-2}$~s$^{-1}$ 
(or $\log L({\rm Ly}\alpha) = 42.68$~ergs)  and 90\% reached
$2.5 \times 10^{-16}$~ergs~cm$^{-2}$~s$^{-1}$ ($\log L({\rm Ly}\alpha) = 
42.96$~ergs).  Simulations demonstrate that above these flux limits, 
the recovery fraction of emission lines was better than 95\% for equivalent widths 
greater than 5~\AA, and higher than 90\% for equivalent widths as small as 
1~\AA\ \citep{adams11}.  The HPS identified 67 LAEs in the COSMOS and
GOODS-N fields, but for consistency with our comparison sample, we consider
only those 11 LAEs between $1.90 < z < 2.35$.  
Over this redshift range, the luminosity limits of the HPS survey are roughly 
constant with redshift \citep{blanc11}.

Our second source of LAEs is a narrow-band survey for $z \sim 2$
Ly$\alpha$ sources in the Extended Chandra Deep Field South (ECDF-S\null).
By using a 50~\AA\ interference filter and the CTIO 4-m Mosaic camera,
\citet{guaita10} identified LAEs with Ly$\alpha$ fluxes brighter
than $2 \times 10^{-17}$~ergs~cm$^{-2}$~s$^{-1}$ and redshifts
$2.04 \lesssim z \lesssim 2.08$.   (See \citet{ciardullo12} for more details 
on this dataset.) A total of 17 of the LAEs brighter than the
90\% completeness limit of \citet{ciardullo12} fall in the 3D-HST's GOODS-S 
region.  Thus, the combination of the HPS and narrowband datasets yields
a sample of 28 Ly$\alpha$ emitters.

We note that the volume of space covered by both the LAE and oELG surveys is
considerably smaller than that observed by each technique individually.  For 
example, while the total number of $1.90 < z < 2.35$ galaxies found via the 
{\sl HST\/} IR grism is 245, only 63 oELGs fall in regions covered by the HPS 
or the ECDF-S narrow-band survey.  Of these, just 12, or roughly 20\%, are
also classified as LAEs \citep{ciardullo14}.  If we remove these 12 
dual-classification objects from the set of oELGs, we are left with a sample 
of 233 galaxies selected via their optical emission lines, and 28 galaxies 
detected via Ly$\alpha$. For most of the comparisons that follow, we use the entire sample of 233 oELGs
to the 28 LAEs:  only when considering the solid angle of Ly$\alpha$ escape do
we use the subset of 63 oELGs in the regions of survey overlap.

To study the conditions which facilitate the escape of Ly$\alpha$ emission,
one would ideally begin with a large, homogeneously selected sample of 
galaxies and then consider which objects emit in Ly$\alpha$ and which do not.
Unfortunately, while such a procedure works well for the continuum-bright
Lyman break objects \citep{steidel96b, steidel96, kornei10, lefevre15}, it is 
difficult to implement for objects selected via their emission lines.  As will 
be shown below, many of our oELGs and LAEs are quite faint in the continuum,
so comprehensive surveys for both the Ly$\alpha$ and rest-frame optical 
emission lines would require prohibitively deep exposures.  In fact, previous 
surveys have shown that the global escape fraction of Ly$\alpha$ from the 
$z \sim 3$ universe is only $\sim 5\%$, and that there is generally little 
overlap between samples of galaxies selected via their Ly$\alpha$ and Balmer 
emission lines \citep{hayes10, ciardullo14}.  Nevertheless, as we shall show, we can 
still examine the properties which facilitate Ly$\alpha$ escape using our 
Ly$\alpha$ and optical emission line selected galaxies. 

\section{Physical Properties}
\label{sec:analysis}

To determine the photometric properties of both our LAEs and the
emission-line selected comparison sample, we took advantage of the 
multi-wavelength catalog of \citet{skelton14}, which extends over the CANDELS
fields of GOODS-S, GOODS-N, and COSMOS \citep{grogin11}.  This database
begins with the deep, co-added F125W + F140W + F160W images from {\sl HST\/}
and then adds in the results of 30 distinct ground- and space-based imaging 
programs to produce a homogeneous, PSF-matched set of broad- and 
intermediate-band flux densities covering the entire wavelength range from
$0.35~\mu$m to $8.0~\mu$m.  In the COSMOS field, this dataset contains 
photometry in 44 separate bandpasses, with measurements from {\sl HST,}
{\sl Spitzer,} Subaru, and a host of smaller ground-based telescopes. In 
GOODS-N, the data come from five different observatories and include 
22 different bandpasses, while in GOODS-S, six different telescopes
provide flux densities in 40 bandpasses.  For $z \sim 2$ systems, these data
cover the rest-frame far-UV through the rest-frame near-IR and
allow very accurate estimates of star-formation rate, stellar mass, and
stellar reddening under the necessary assumptions about the underlying
stellar population (i.e., stellar metallicity, star formation history, IMF, 
and attenuation law).

\subsection{Star Formation Rate and UV Slope}
\label{sec:beta}
Star formation rates for our LAEs and oELGs 
can be obtained from their UV luminosity densities.  Before doing this, however, 
we must deal with the issue of stellar reddening.  
As detailed by \citet{calzetti01}, rest-frame wavelengths between
$1250~\text{\AA} < \lambda <  2600~\text{\AA} $ sample the Rayleigh-Jeans portion of the hot
stars' spectral energy distributions (SEDs); in the absence of reddening, 
the spectral slope in this region should be well fit by a power law, i.e.,
\begin{equation}
\label{eqn:beta}
L (\lambda) \propto \lambda^{\beta_0}
\end{equation}
where $L_(\lambda)$ is the system's luminosity density.  For steady state
star formation over $\sim 10^8$~yr, $\beta_0 = -2.25$, while for extremely 
young starbursts, $\beta_0$ may be as steep as $-2.70$ \citep{calzetti01, 
reddy10}.  Observed values of the spectral slope larger than $-2.25$ must 
therefore be due either to internal extinction or a rapidly declining SFR\null.)
(The latter possibility is unlikely, given the high luminosities and equivalent
widths of the rest-frame optical emission lines.)
While complications may arise if the reddening curves contain a Milky-Way type
bump at $\sim 2175$~\AA, this feature is usually weak or absent in high-redshift 
star-forming galaxies \citep{kriek13, buat12, zeimann15b}. 

To obtain the galaxies' star formation rates, we therefore fit each object's 
photometric measurements to a power law, by first computing its observed UV
slope ($\beta$) and the luminosity density at 1600~\AA\ ($L_{1600}$) via
simple unweighted least squares, and then estimating the errors on the
parameters using a series of Monte Carlo simulations, with each realization
formed using the quoted errors of the photometry.   After translating the 
observed value of $\beta$ into a total extinction at 1600~\AA\ via
a \citet{calzetti01} attenuation law, $A_{1600} = 2.31 \left( \beta - \beta_0 
\right)$, we applied the extinction correction and inferred the galaxies' star 
formation rates using the local UV-based SFR calibration \citep{hao11, 
murphy11, kennicutt12}, i.e.,
\begin{equation}
\log {\rm SFR}_{\rm UV} = \log L_{1600} - 43.35\text{ (M}_{\odot}~{\rm yr}^{-1})
\label{eq:SFR}
\end{equation}

Although our median measurement error on $\beta$ is only 0.147 and 
that for $\log L_{1600}$ is 0.017, our SFR estimates are subject to an 
additional systematic uncertainty associated with the details of extinction
(i.e., geometry, homogeneity, wavelength dependence).  Most specifically,
our measurements of SFR assume the \citet{calzetti01} obscuration law, which 
is based on observations of a small number of starburst galaxies in the local 
neighborhood.  Although there is substantial evidence to suggest that the law 
has not changed much between $z \sim 0$ and $z \sim 2$
 \citep[e.g.,][]{forsterschreiber09, mannucci09, wuyts13, price14,
zeimann15b}, there are counterexamples \citep[e.g.,][]{erb06, shivaei15}.
Nevertheless, we adopt this relation both in our SFR calculation and for our
measurements of stellar mass.

\subsection{[O~III] Luminosity and Equivalent Width}

[O~III] line fluxes were obtained from the {\sl HST\/} near-IR grism images by
simultaneously fitting a fourth order polynomial continuum and six 
Gaussian-shaped emission line profiles to each extracted G141 grism spectrum 
via the method of maximum likelihood \citep{gebhardt15}.  Included in the list of fitted emission 
lines was the blended [O~III] doublet with the strength of $\lambda 5007$
fixed at 2.98 times that of $\lambda 4959$ \citep{storey2000}.  The [O~III]
$\lambda 5007$ fluxes were then converted to luminosities by assuming
isotropic emission, and corrected for internal extinction using the
stellar differential reddenings obtained from the UV continuum (see 
Section~\ref{sec:beta}) and the \citet{calzetti01} obscuration law.

The calculation of [O~III] equivalent widths was a bit more difficult, 
as it is generally not possible to measure the continuum of a faint $z \sim 2$
galaxy on 3D-HST grism frames.  Instead, continuum estimates were obtained by
converting each galaxy's broadband F140W magnitude \citep{skelton14}
into a flux density, using the grism spectrophotometry to subtract off
the contribution of the emission lines within the bandpass, and 
applying a correction to account for the fact that the extinction which
effects emission lines is generally greater than that which extinguishes
starlight, i.e., 
\begin{equation}
E(B-V)_{\rm stars} \sim 0.44 E(B-V)_{\rm gas}
\end{equation}
\citep{calzetti01}.  Equivalent widths were then deriving by scaling
this photometrically-based continuum measurement to the line fluxes
recorded by the {\sl HST\/} grism. We note that while our [O~III] luminosity is dust-corrected,
our [O~III] equivalent widths are not.

These [O~III] measurements can be used as a check on the UV-derived star formation rates
derived in Section~\ref{sec:beta}.  Both \cite{kennicutt92} and \cite{moustakas06} state that
[O~III] $\lambda 5007$ emission is a poor tracer of a galaxy's star formation rate, as the
line is quite sensitive to both changes in the ionization parameter and the metallicity of the
nebular gas.  Nevertheless, we can compare it to our UV-based SFR estimates, to test for
for the presence of large errors or systematics in our analysis.   This is done in Figure 1.   
From the figure, it is clear that the conversion between
[O~III] luminosity and SFR is roughly consistent with that derived by \citet{ly07} 
(corrected for the difference between \cite{kennicutt98} and \cite{kennicutt12}),
using the photometrically-determined [O~III]/H$\alpha$ line ratios of $z = 0.42$ and
$z = 0.84$ narrow-band selected galaxies.   In addition, according to the  Akritas-Thiel-Sen
estimator \citep{akritas95}, which is a statistic that is insensitive to outliers, the log-log
slope of the relation is $0.43 \pm  0.05$.  Given the issues involved with [O~III]
SFR calibrations, this agreement is excellent.


\begin{figure}
\centering
\includegraphics[scale=0.75]{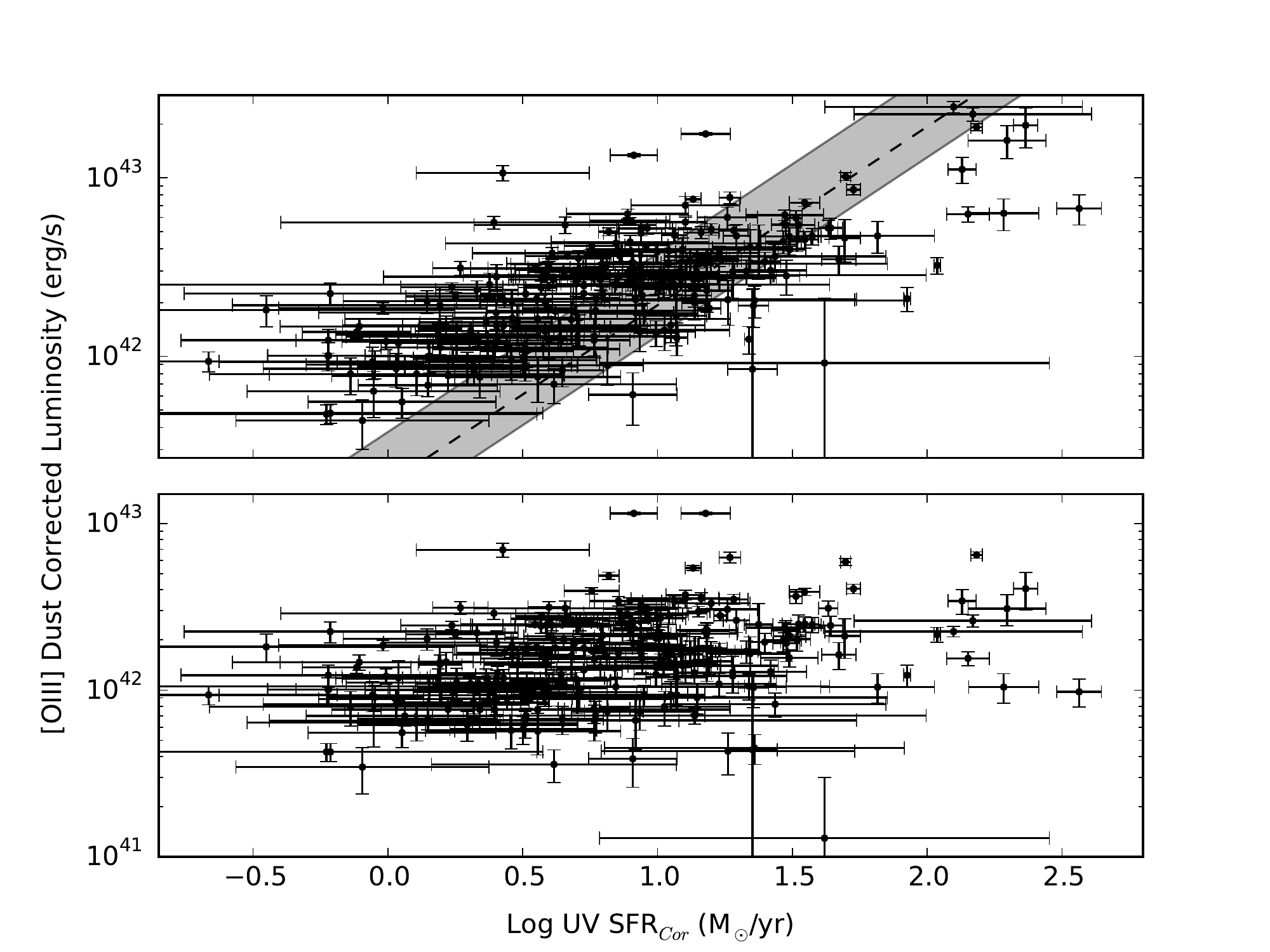}
\caption{Top: Dust corrected UV star formation rate compared to the 
dust corrected [O~III] $\lambda 5007$ line luminosity. The dotted line is the \cite{ly07} calibration and the grey band shows its uncertainty.
 In general, the [O~III] line is not a
good SFR indicator, as it is both metallicity- and ionization-parameter dependent, but we 
use it here as a consistency check on our UV-derived SFRs.  The robust log-log slope of 
the relation is $0.43 \pm 0.0.5$ which, given the systematics of [O~III] emission,
is good agreement.  Bottom:  The same comparison of [O~III] emission to rest-frame
UV flux density, prior to dust correction; this ensures that the quantities are 
statistically independent.  As expected, the scatter in the diagram is significantly
larger.  The shallower slope ($0.2 \pm 0.05$) is consistent with the expected
correlation of extinction with star formation rate \citep[e.g.,][]{bouwens09,bauer11}
}	
\label{fig:sfr_cmp}
\end{figure}

\subsection{Stellar Mass}

The photometric catalog of \citet{skelton14} gives SED-based stellar masses
for our emission-line selected galaxies, but not their associated uncertainties.
Such information is critical for any analysis of continuum-faint targets, as 
in these objects, the photometric errors may propagate into large, asymmetric
uncertainties in the derived parameters.   Consequently, to infer the masses
of our LAEs and oELGs, we performed our own spectral energy distribution
fitting, using the \citet{skelton14} photometric as a base.  We began by using the 
population synthesis models of \cite{bruzual03}, as 
updated in 2007 with an improved treatment of the thermal-pulsating AGB phase. 
(Due to the generally young ages of our systems, the differences between 
the masses derived from the 2003 and 2007 models are minimal.)  
We then adopted a \citet{salpeter55} initial mass function over 
the range $0.1 \, M_{\odot}$ to $100 \, M_{\odot}$, a \citet{calzetti01} dust 
obscuration law, and the \citet{madau95} prescription for absorption by 
intergalactic material.  Since stellar abundances are poorly constrained
by broadband SED measurements, we fixed the metallicity of our models to
$Z = 0.2 \, Z_{\odot}$, which is close to the mean gas-phase oxygen abundance
measured for both LAEs \citep{finkelstein11, nakajima13, song14} and our 
comparison set of oELGs \citep{gebhardt15}.  Also, because the emission lines
and nebular continuum can be an important contributor to the broadband fluxes 
of high-redshift galaxies \citep[e.g.,][]{atek11, schaerer09}, we modeled this 
component using the prescription of \citet{acquaviva11}, with updated 
templates from \citet{acquaviva12}.  Finally, for simplicity, we assumed
that the star formation rates of our galaxies have been constant with
time; at $z \sim 2$, this hypothesis is more appropriate than that 
of a declining SFR \citep[e.g.][]{madau14}. 

In keeping with these assumptions, we did not use any photometric bandpass
redward of rest-frame $3.3~\mu$m, as in this region interstellar medium 
features, such as lines from polycyclic aromatic hydrocarbons (PAHs), which
are not modeled by our code, may dominate.  Similarly, because the 
\citet{madau95} correction for intergalactic absorption is statistical in 
nature and may not alway be modeled properly, all points blueward of Ly$\alpha$ 
were also excluded.  

To perform SED fits, we used the stellar population fitting code {\tt GalMC} 
\citep{acquaviva11}, assigning as the free parameters stellar
mass, reddening, and age.  This Markov-chain 
Monte-Carlo (MCMC) program with a Metropolis-Hastings sampler is not only
much more computationally efficient than traditional grid searches, but it 
returns more realistic errors, as it explores degeneracies between
various parameters \citep{metropolis53, hastings70}.  Upon completion, each chain was analyzed using the
{\tt CosmoMC} program {\tt GetDist} \citep{lewis02}, and, since 
multiple chains were computed for each object, the \citet{gelman92}
$R$ statistic was used to test for convergence via the 
$R -1 < 0.1$ criterion \citep{brooks98}.

As detailed by \citet{conroy13}, changes in the assumed initial mass
function, reddening law, stellar metallicity, and the treatment of the 
thermal pulsing AGB phase can lead to systematic shifts of up to 
$\sim 0.3$~dex in the computed stellar masses of high-redshift galaxies.
However, because our LAE analysis used the exact same set of assumptions 
as that for the oELGs, our comparison between the two
galaxy populations should be valid.

\subsection{Near-UV Morphology: Size, Ellipticity, and Nearest 
Neighbor Distance}
To measure the near-UV size, morphology, and environment of our $z \sim 2$ 
galaxies, we used the deep F814W images of the {\sl HST\/} CANDELS program 
\citep{grogin11, koekemoer11}, and the analysis techniques described in detail 
by \citet{bond09}.   Most of our emission line galaxies are present on these
frames, though, because not all the CANDELS images have yet been released,
5 LAEs and 67 oELGs lack data.  Optimally, these morphological measurements 
should be performed in the rest frame optical (using the F140W WFC3 filter),
as this would probe a larger fraction of the galaxies' stars.  However, at 
this redder wavelength, the instrumental PSF is 2.5 times larger than in the 
rest-frame UV, making morphological measurements of small $z \sim 2$ systems 
that much more difficult.  Moreover, though studies have found that the 
structural parameters of high-redshift galaxies can sometimes change when moving
from the rest-frame UV to the rest-frame optical, such is usually not the case 
for star-forming galaxies with little dust \cite[e.g.][]{conselice14}.  As our 
galaxies are found via recombination lines powered by star formation, they 
fall under this category.  Furthermore, \citet{bond14} showed that the near-UV 
and optical size measurements for LAEs do not show significant differences. 

After creating cutouts around each galaxy, we performed object identifications 
and background subtraction using the routines found in \texttt{SExtractor}
\citep{bertin96}.  We then measured the angular size of each object by using 
the {\tt phot} routine within IRAF to determine their flux-weighted 
centroid and magnitude through a series of circular apertures.  These aperture
magnitudes were used to define the galaxy's half-light radius, with
the uncertainty in the measurement being related to total flux via
\begin{equation}
\frac{\sigma_r}{r} = 0.54 \frac{\sigma_f}{f}
\end{equation}
\citep{bond12}.  
The smallest half-light radius measurable via this analysis
is $\sim 0.75$~kpc, or roughly twice the resolution of {\sl HST\/} in
the F814W filter. 

Next, we examined object morphology to investigate the orientation of our 
galaxies.  In disk systems, the probability of escape for Ly$\alpha$ photons 
should be a function of viewing angle, with face-on galaxies having lower 
Ly$\alpha$ optical depths than edge-on systems (see Section \ref{toymodel} for 
a more complete discussion).  Though we cannot unambiguously determine the 
inclination of a galaxy at $z \sim 2$, we can derive galaxy morphologies via 
{\tt Galfit} \citep{peng02, peng10, peng11}, and use the resultant 
ellipticities and \citet{sersic63} indices as proxies for inclination and basic 
morphology.  Such analyses have been performed on Ly$\alpha$ emitters by a 
number of authors \citep[e.g.,][]{bond09, gronwall11}, with \citet{shibuya14a} 
hinting at an anti-correlation between Ly$\alpha$ equivalent width and
ellipticity. We note that, while the ellipticities of large galaxies 
can be found via simple ellipse fitting \citep[e.g.][]{weinzirl09}, the 
analysis of our small $z \sim 2$ systems necessitates the more careful 
approach of {\tt Galfit}, which convolves the instrumental PSF with models of 
the original image.  Of course, near the resolution
of the instrument, ellipticity measurements will be highly uncertain, but
the vast majority of our galaxies have half-light radii that more than twice 
this limit.  Thus, our estimates of ellipticity should be reasonably robust.

Finally, to explore the effects of mergers on Ly$\alpha$ emitters, we
used \texttt{SExtractor} to measure the distance of each emission
line galaxy to its nearest projected neighbor.  By examining the frequency
of close pairs and the morphological shapes of Ly$\alpha$ emitters, 
\citet{shibuya14a} concluded that the merger fraction of the class was likely
between 10 and 30\%.  However, without a control sample, \citet{shibuya14a} 
could not place this result in context.  By comparing the distribution of 
separations for our LAEs with that for oELGs with similar masses,
we can measure the relative importance of mergers for the LAE population.  
Moreover, because our measurement is differential in nature, our results
should be relatively insensitive to the exact choice of \texttt{SExtractor} 
parameters.  To select nearest neighbor galaxies, we used a range of detection 
thresholds (from $1.65 \, \sigma$ to $2.5 \, \sigma$ above the background) 
and required that the number of pixels above our threshold be at least five. 
(We report the nearest neighbor distance found using a detection threshold of $1.65\sigma$.)
By modifying these numbers, we could include or exclude faint, low-significance
sources which may (or may not) be real, and thereby change the absolute 
form of the proximity  distributions.  However, the relative difference
between the LAE and oELG distributions remained unaffected.

The galaxies in this study have redshifts between $1.90 < z < 2.35$,
so the exact rest-frame wavelength probed by {\sl HST's\/} F814W filter
varies from galaxy to galaxy.  However, \citet{bond14} demonstrated that 
UV morphological measurements of $z \sim 2$ systems are robust against 
wavelength changes. Similarly, by examining the structure of compact 
massive galaxies and simulating systems as small as those in our sample, 
\citet{davari14} showed that size measurements of high-redshift galaxies
are generally robust.  We do note that one concern with any
high-redshift size measurement is that the results may be affected by
cosmological surface brightness dimming \citep[e.g.,][]{weinzirl11}. However, 
previous experience with measuring the sizes of $z \sim 2$ LAEs has 
demonstrated that even one orbit of {\sl HST\/} broadband data is sufficient 
to avoid this issue \citep{hagen14}.

\subsection{Derived Physical Parameters: Star Formation Rate Surface Density 
and Specific Star Formation Rate}

Our measurements of stellar mass, size, and star formation rate can be
combined to form two other physical parameters which are often used to
describe the physical state of galaxies.  Star formation rate surface density
is simply a galaxy's total star formation rate divided by its area, and,
following \citet{malhotra12} (who used the term star formation intensity),
we define this quantity as $\Sigma_{\rm{SFR}} = {\rm{SFR}} / 2 \pi r^2$, 
where $r$ is the half-light radius.  Similarly, the specific star formation
rate of a galaxy (sSFR) is simply its SFR divided by its stellar mass, and the
units of this quantity, inverse time, give a measure of the system's age.
Both $\Sigma_{\rm{SFR}}$ and sSFR have been heavily used in investigations
of the different types of high-redshift galaxies and their modes of star
formation \citep{malhotra12, nakajima12, rhoads14, song14}.  In particular, 
numerous authors have argued that high-redshift LAEs have some of the highest
specific star formation rates in the universe \citep[e.g.,][]{gawiser07, 
nakajima12}, but such statements arise principally from comparisons between
LAEs and higher-mass systems such as Lyman-break and $BzK$ galaxies.   Our
comparison to oELGs can place this conclusion in context.

\section{Results}
\label{sec:discussion}

\subsection{Distributions of Physical and Morphological Parameters}

\begin{figure}
\centering
\includegraphics[scale=0.8]{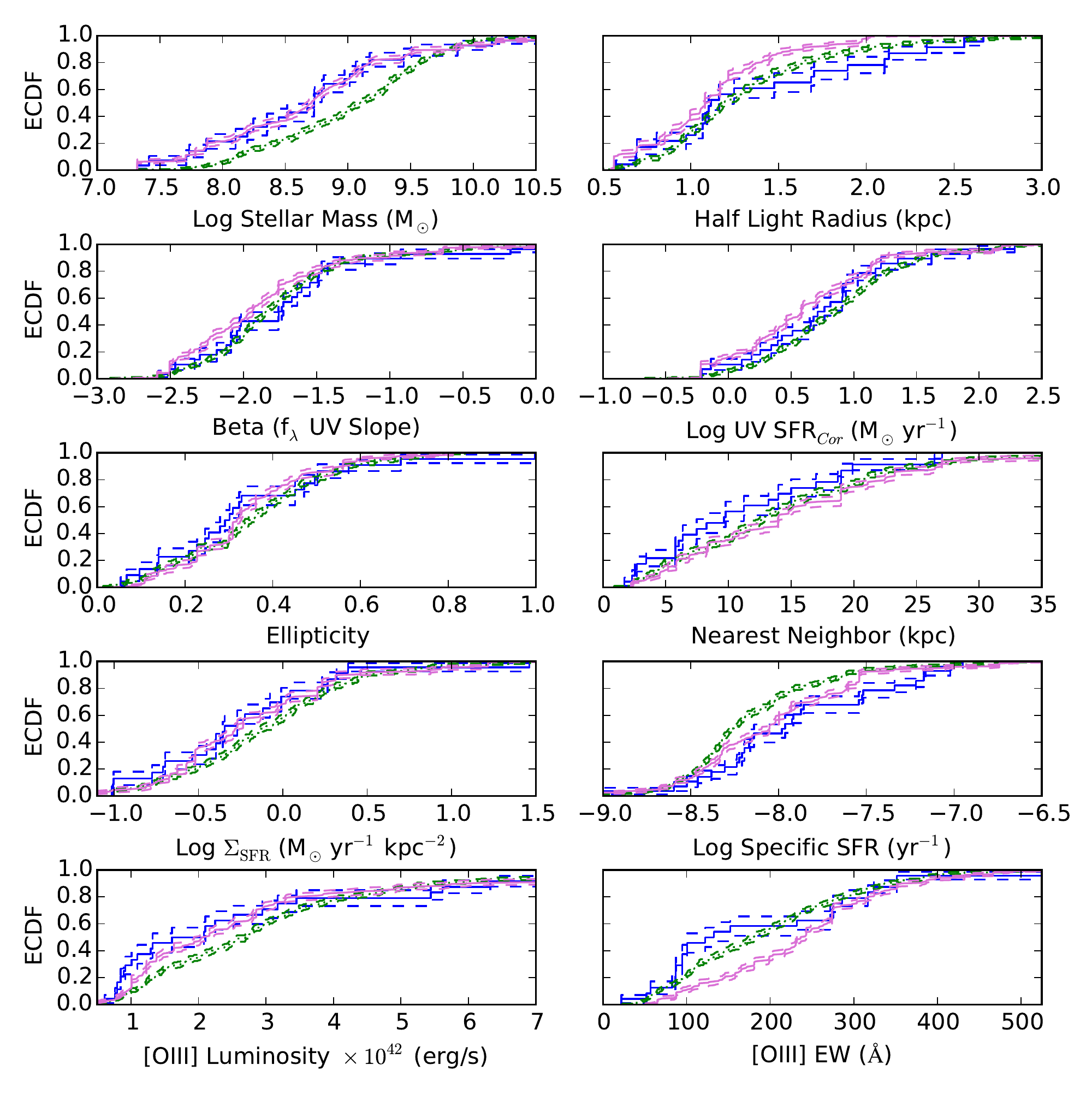}
\caption{Empirical cumulative distribution functions (ECDFs) for 10 
galaxy parameters.  The solid blue line is the distribution for LAEs, 
while the distributions of parameters for oELGs are plotted as a green 
dotted line. The orchid line shows the properties of oELGs selected to 
match the stellar mass distribution of the LAEs.  The dashed lines 
are the 1$\sigma$ asymptotic uncertainties.  The two-sample 
Anderson-Darling test finds the two population are statistically 
identical, meaning that the null hypothesis of a single underlying 
population cannot be rejected for any parameter with more than 99.7\% 
($3 \, \sigma$) confidence. The most discrepant distributions are 
those for stellar mass, half-light radius, and sSFR, but even those do 
not rise to the level of significance. The small LAE sample size 
means a larger difference between the two ECDFs is necessary for 
a significant result.  We also note that the small fraction of 
galaxies with ellipticities greater than 0.4 suggests that at least 
some of the systems under consideration have disk-dominated 
morphologies.}
\label{fig:ecdf}
\end{figure}

Figure \ref{fig:ecdf} presents the empirical cumulative distribution
functions for our LAE and oELG samples, and includes the parameters 
stellar mass, half-light radius, stellar reddening (as measured by the
slope of the UV continuum, $\beta$), star formation rate (as inferred from the
de-reddened flux density at 1600~\AA), ellipticity, and the projected 
distance to the nearest neighbor galaxy.  Also shown are the cumulative
distribution functions for the composite variables of $\Sigma_{\rm{SFR}}$ and 
specific star formation rate. Tables \ref{table:lae} and \ref{table:oelg} 
contain all the measured physical and morphological properties for these 
galaxies.  As illustrated in the figures, our LAEs and oELGs are drawn from 
consistent galactic populations: according to the Kolmogorov-Smirnov and 
Anderson-Darling tests, in no case can the null hypothesis of the two 
distributions being pulled from a single underlying population be rejected with
more than 99.7\% ($3 \sigma$) confidence, far below the $5 \sigma$ standard
common in the physical sciences \citep{kolmogorov33, smirnov48, anderson52}.
Our selection method for LAEs does seem to identify galaxies with slightly 
lower masses than their oELG counterparts, but the offset is not large and 
still below the threshold of statistical significance.
This consistency supports the use of oELGs as an LAE comparison
sample, and suggests a connection between the two galaxy populations.

To ensure that the results describe above are not an artifact of the
differing depths of our LAE and oELG surveys, we created a stellar
mass-matched sample of oELGs for use in our comparison.  To do this, we 
randomly selected a mass from the LAE stellar mass distribution, applied a 
Gaussian uncertainty based on the error on the mass measurement, and then
identified the oELG with the closest stellar mass.   A total of 1,000
mass-matched oELG samples were created in this manner, and their physical and
morphological properties were compared to those of the LAEs via Anderson 
Darling tests.  As shown in Figure \ref{fig:ecdf}, there is no statistically 
significant difference between the LAEs and any of these mass-matched
subsamples of optical emission line galaxies.

\begin{figure}
\centering
\includegraphics[scale=0.8]{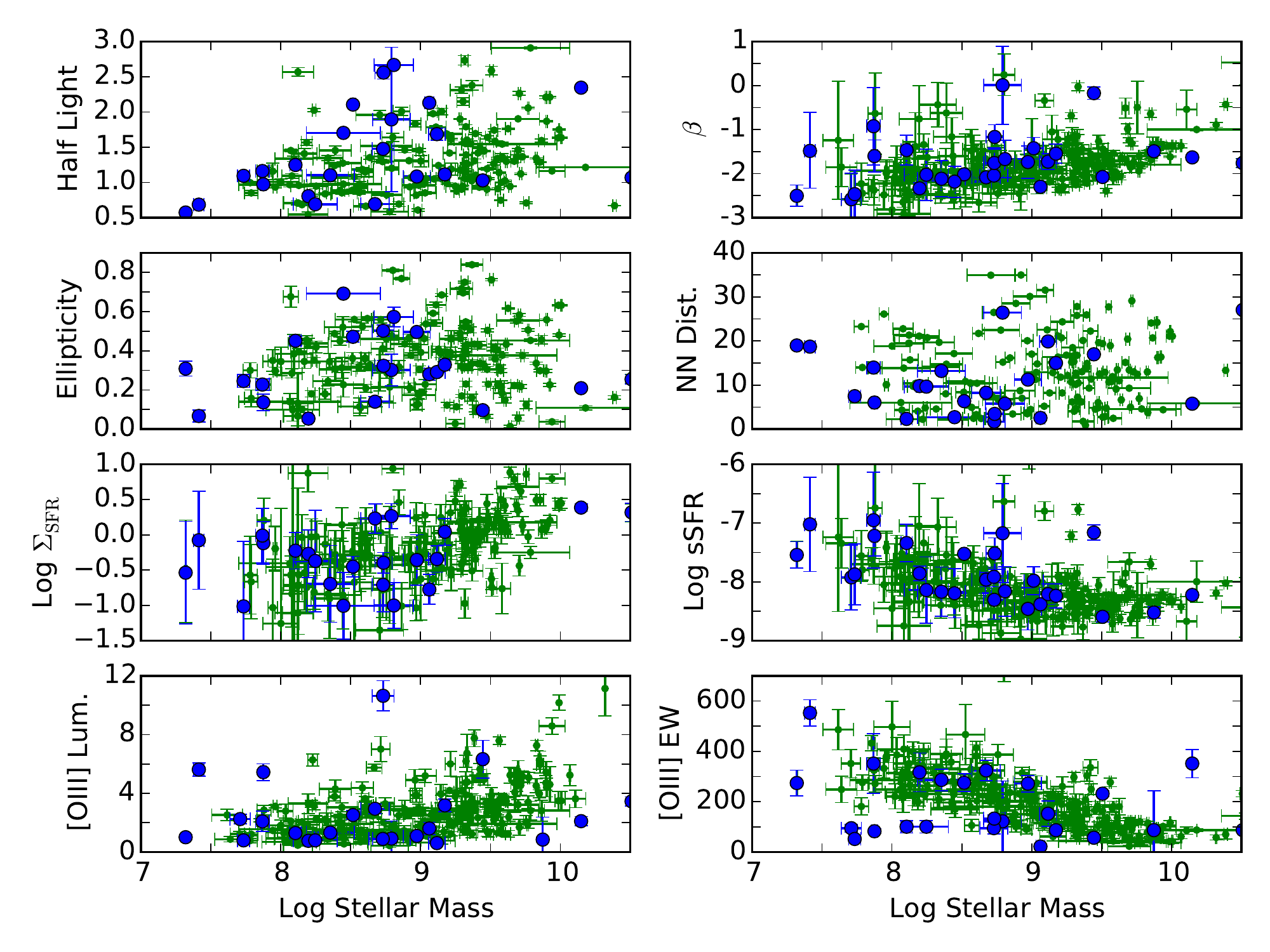}
\caption{The physical and morphological properties of our LAE and oELGs
as a function of stellar mass.   
LAEs (with an without rest-frame optical emission lines) are plotted as blue circles;
the oELGs are plotted as green dots.  The units on each quantity are the same as in 
Figure \ref{fig:ecdf}.  In all cases the distributions of LAE and oELG
physical parameters as a function of mass are indistinguishable; this result suggests that
the two samples are drawn from the same underlying population.  
For stellar mass vs.\  SFR, see Figure~\ref{fig:mainsequence}.}
\label{fig:allvsmass}
\end{figure}

Another way to compare the samples while mitigating the effects of selection 
bias is to analyze the relationships between the various photometric and 
structural parameters.  Since stellar mass has little to no correlation with
observed Ly$\alpha$ line luminosity \citep{hagen14}, we treated this quantity 
as an independent variable, and examined the distribution of physical 
parameters as a function of mass for the two galaxy populations (see 
Figure~\ref{fig:allvsmass}).  For each variable, we found the best-fit line for 
the oELG dataset, subtracted this line from both the oELG and LAE 
distributions, and compared the behavior of the residuals.  In all cases,
the best-fit line through the residuals had a slope consistent 
with zero, and there was no statistically significant difference in the 
distribution of the residuals.  This strongly suggests that the two galaxy 
samples are drawn from the same underlying population.  A similar analysis 
with half-light radius as the independent parameter produces the same result:
if there is a difference between the oELG and Ly$\alpha$  emission line 
populations, it cannot be detected with our present samples.  At least at 
$z \sim  2$, LAEs seem to be randomly drawn from the larger population of 
optical emission line galaxies.

It is somewhat surprising that the distribution of UV slopes 
for galaxies selected via their optical emission lines is so similar to
that found for the LAE population.  Many papers have suggested that Ly$\alpha$
emission is regulated by dust \cite[e.g.,][and references therein]{kornei10, 
atek14}, and the existence of this dust should be revealed through extinction
in the UV\null.  Yet we see no evidence for a deficit of dust obscuration
associated with Ly$\alpha$ galaxies.  Of course, our analysis only 
addresses the global properties of each galaxy; small scale changes in 
the covering fraction of dust due to ISM holes created by supernovae would
be extremely difficult to detect.  Furthermore, the recent work by 
\citet{henry15} argues that Ly$\alpha$ emission in low-redshift Green Pea 
galaxies is actually modulated by H~I column density.  This again would
weaken any supposed connection between UV extinction and Ly$\alpha$ emission.

It is also surprising that we see no significant difference between the [O~III] rest-frame
equivalent width (EW) distributions.   The ratio of an ionization driven emission line,
such as [O~III] $\lambda 5007$, to its underlying starlight-generated optical or near-IR
continuum is a rough proxy for specific star formation rate.  Consequently, a number of
authors have investigated the relationship between rest-frame optical emission-line
equivalent widths and Ly$\alpha$, and have generally found a correlation between the
two parameters.  For example, in the nearby universe, \citet{hayes14} were able to select
their Ly$\alpha$ Reference Sample by limiting their targets to sources with large H$\alpha$
equivalent widths, and \citet{cowie11} found that GALEX-selected LAEs had larger 
optical EWs than a control sample. Additionally, in the $z \sim 2$ universe,
\citet{oteo15} found significant differences in the reddening, stellar mass, UV slope, and 
star formation rate between LAEs and a sample of H$\alpha$ selected galaxies.  
However, this work also found that their H$\alpha$-selected galaxies had properties 
similar to those of  \textit{sBzK}-selected systems.  This suggests a simple explanation
for the discrepancy:  as their Figure~4 shows, their H$\alpha$ systems are systematically 
more massive than their Ly$\alpha$ emitters, and hence likely to be evolving along a 
different evolutionary path.   By creating a stellar mass-matched sample of galaxies
(selected primarily via their [O~III] emission), we have likely avoided this problem.
Alternatively, the difference between our results and those of the previous surveys could 
be result of the brighter Ly$\alpha$ detection limit of the HETDEX Pilot Survey. Finally, we 
note that our analysis is limited by small number statistics; in a few years, when large
LAE samples in the {\sl HST\/} survey fields become available, we will be able to
perform much more stringent tests on the datasets. 

\subsection{A Toy Model of the Solid Angle of Ly$\alpha$ Escape}
\label{toymodel}
By simulating the propagation of Ly$\alpha$ photons through a realistic
ISM embedded within a $\sim 10^{10} \, M_{\odot}$ dark halo, \citet{verhamme12}
found that the likely escape paths for Ly$\alpha$ were anisotropic,
with emission much more likely along directions perpendicular to the
system's disk.  In this scenario, one would expect the presence of Ly$\alpha$
to be correlated with galaxy morphology, as LAEs would be 
preferentially associated with low-inclination systems.  Yet, as 
Figure~\ref{fig:ecdf} illustrates, no such correlation exists:  
the ellipticity distributions of LAEs and oELGs are 
statistically indistinguishable.

Alternatively, \citet{gronke14} and \citet{gronke15} have followed the escape
of Ly$\alpha$ through an ISM consisting of a number of cold, dusty,
star-forming clumps distributed in an hot, ionized plasma.  In a case
such as this, there is no preferred orientation associated with LAEs, as
Ly$\alpha$ photons typically escape through low-optical depth sight lines that
are randomly distributed throughout a galaxy.  Indeed, there is some 
observational support for this idea:  multiple studies have reported 
that when Ly$\alpha$ is observed, its optical depth is not significantly greater
than that for its surrounding far-UV continuum \citep{finkelstein08, blanc11, 
hagen14, song14, vargas14}.  This implies that when Ly$\alpha$ escapes,
it does so without having to undergo a large number of scattering events.

If the escape of Ly$\alpha$ is indeed due to the presence of a series of
randomly distributed holes in the interstellar medium, then we can use our 
data to determine the mean solid angle for Ly$\alpha$ emission.  Of the 63 
oELGs in the HPS footprint, 12 have evidence for Ly$\alpha$ in emission 
\citep{ciardullo14}, implying that $\sim 20\%$ of $z \sim 2$ star-forming 
galaxies have a Ly$\alpha$ escape path in our direction.  If all these systems 
are LAEs when viewed along the appropriate line-of-sight, then the mean solid 
angle for Ly$\alpha$ escape is $\Omega_{Ly\alpha} = 2.4 \pm 0.8$ steradians.
Of course, the lines-of-sight for Ly$\alpha$ escape need not be contiguous, 
but for purposes of visualization, this sky fraction corresponds to an average 
opening angle of $50^\circ \pm 8^\circ$.  For comparison, a narrow-band study 
for $z = 2.2$ star-forming galaxies in the GOODS-S region found 6 out of 55 
H$\alpha$ emitting galaxies were also Ly$\alpha$ sources, for a mean escape 
angle of $1.4 \pm 0.6$ steradians \citep{hayes10}.  Given that this 
double-blind narrow-band survey covers only $\sim 5\%$ of our $1.90 < z < 2.35$
survey volume, and like our study, is limited by small number statistics, 
these results are consistent. 

A further comparison comes from the analysis of \cite{milvangjensen12}, 
who searched for Ly$\alpha$ emission from gamma-ray burst host galaxies between
$1.8 < z < 4.5$.  Recent studies suggest that GRBs at these redshifts trace
UV star formation metrics \citep{greiner15, schulze15}, and as such, they
provide an independent cross-check on our previous calculations.  Out of
a sample of 20 GRB hosts, \cite{milvangjensen12} found Ly$\alpha$ emission
in 7 objects, implying a mean Ly$\alpha$ escape angle of  $\Omega_{Ly\alpha} 
= 4.4 \pm 1.9$ steradians.  Again, this is in rough agreement with our 
previous estimates.

Alternatively, we can compare our estimate of $\Omega_{Ly\alpha}$ to that 
obtained from the adaptive mesh radiative transfer models of \citet{behrens14b}.
These calculations follow Ly$\alpha$ transport in an isolated, turbulent disk
galaxy at time steps of 1.0, 1.5, and 2.0 Gyrs after initialization.   Unlike the
simulations of \citet{verhamme12}, this multiphase, dusty ISM model predicts that 
Ly$\alpha$ escape paths are generated stochastically, and more-or-less independent of 
galaxy morphology.  The result seems consistent with our measurements.

To perform a more quantitative comparison, we re-analyzed the \citet{behrens14b}
models by examining the escape of Ly$\alpha$ along 12,000 separate sight 
lines within the galaxy.   For a galaxy to be considered an LAE, we required 
that the sight line have a rest-frame Ly$\alpha$ equivalent greater than 
20~\AA\  \citep{gronwall07} and an inferred Ly$\alpha$ luminosity larger than 
$10^{41}$~ergs~s$^{-1}$ \citep[i.e., have an intrinsic Case B star formation 
rate greater than $\sim 2 \, M_{\odot}$~yr$^{-1}$;][]{hu98, kennicutt12}.   The 
1.5 and 2.0 Gyr models produce LAEs meeting these criteria with Ly$\alpha$
escape angles of 1.8 and 6.8 steradians, respectively.  Again, this is in
agreement with our observations.  Interestingly, in the 1~Gyr simulation,
no line of sight satisfied the observational criteria, suggesting that a more 
sophisticated model, such as one which includes a changing halo occupation 
fraction, is needed.

Finally, instead of expressing the escape of Ly$\alpha$ in terms of
sight lines and opening angles, it is possible to re-parameterize the analysis
into a duty cycle problem, where the time variable collectively captures all the 
complicated microphysics that enters into the creation of a porous ISM\null.  Such a
calculation has been performed by  \citet{chiang15}, who estimated that an
LAE duty cycle of $\sim 4\%$ best fits the clustering results of the HPS
survey.  Of course, if our detection limits were deeper, this duty cycle estimate
might increase, as more galaxies would be detected via their Ly$\alpha$
emission.  

Data from the upcoming HETDEX \citep{hill08} and Trident
\citep{sandberg15} surveys should greatly improve our understanding of 
Ly$\alpha$ emission from $z \sim 2$ star-forming galaxies,
and allow measurements of $\Omega_{Ly\alpha}$ (or the duty cycle),
as a function of stellar mass, internal reddening, and a host of other parameters.
For example, it is quite possible that the
factors that govern the escape of Ly$\alpha$ depend on the mass of a galaxy.
In low mass systems, feedback can easily outweigh the local gravitational 
potential, causing the distribution of low-column density holes in the ISM 
to be distributed randomly and (somewhat) uniformly; this is the regime
that applies to most our galaxies and to the radiative transfer models of
\citet{behrens14b}.  However, in higher mass systems, the disk structure
may stabilize, making it difficult to blow holes along the plane of the 
system.  In this case, the simulations of \cite{verhamme12} may be
more applicable.  We look forward to the next generation of models detailing
the kinematics of outflows through more complicated geometries which includes
a multi-phase interstellar medium \citep[\eg][]{gronke14}.

\subsection{The place of LAEs and oELGs in the galaxy population}
\begin{figure}
\centering
\includegraphics[scale=0.8]{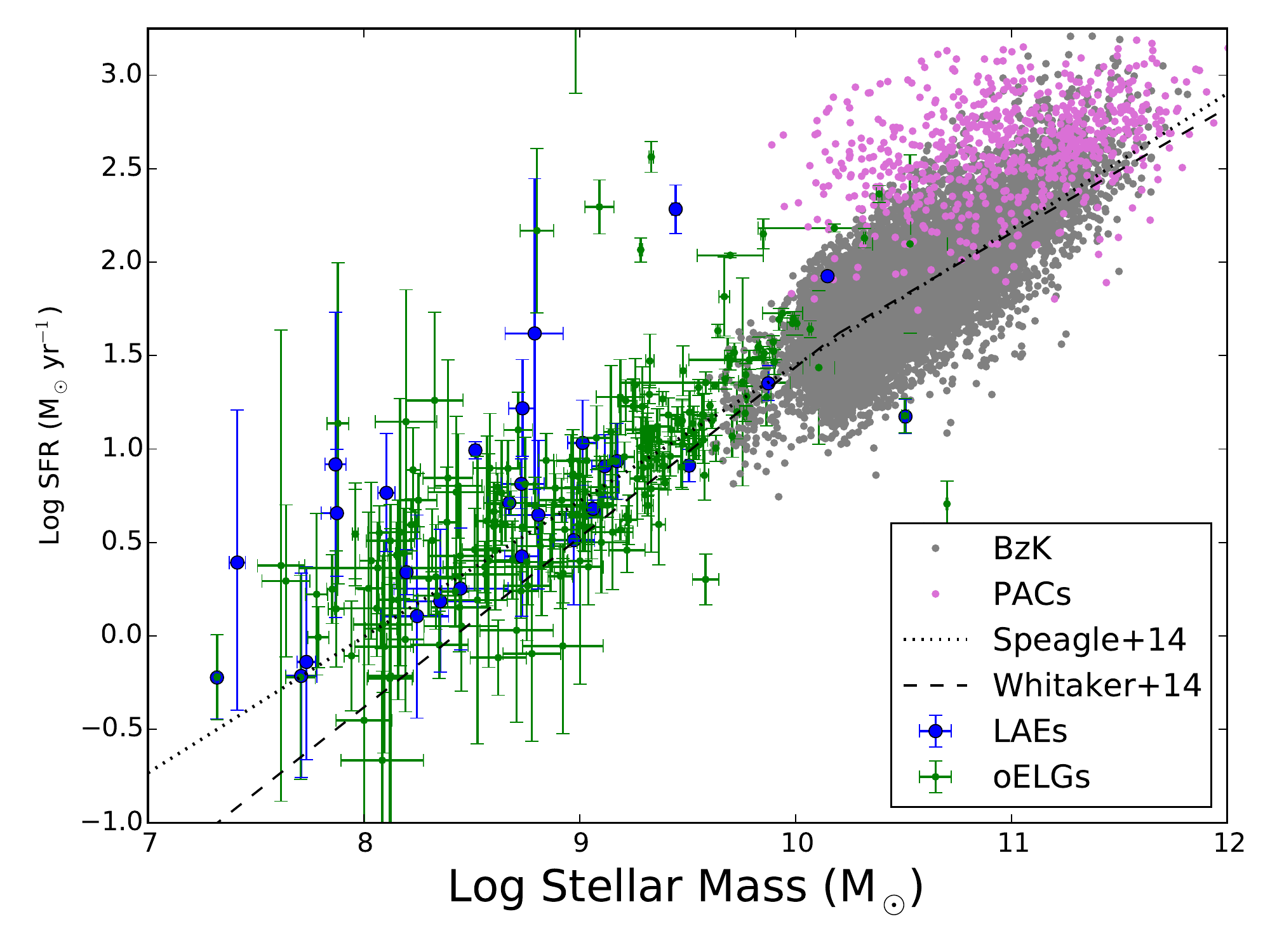} 
\caption{Galaxy star formation rate as a function of stellar mass for our sample of 
LAEs and oELGs.  Also plotted are the continuum-selected samples of 
\citet{rodighiero11}.  Both LAEs and oELG systems lie along, but slightly above the 
low-mass  extrapolation of the $z = 2.1$ ``main sequence'' defined by
\cite{speagle14}; they also lie significantly above the \citet{whitaker14} 
main-sequence line, particularly at the low mass end. The offset could mean
that the galaxies are undergoing starbursts, that there is a change in the slope of the
main sequence at low stellar masses, or that that the sample is biased against
low line-luminosity systems.}
\label{fig:mainsequence}
\end{figure}

Figure \ref{fig:mainsequence} compares the star formation rates and stellar 
masses of our LAEs and oELGs to those of $BzK$ and {\sl Herschel-PACS\/}
galaxies \citep{rodighiero11}.  From the figure, it is clear that the 
emission-line galaxies probe a mass regime that is only examined in the 
deepest continuum-selected surveys.  Both the LAEs and oELGs do lie close to 
the star-forming galaxy ``main sequence'' \citep{speagle14}.   However,
neither population lies {\it on\/} the line: both sets of galaxies lie distinctly 
above the extrapolation of the \citet{speagle14} relation, and above the 
\citet{whitaker14} main sequence line.

There are several possible reasons for this offset.  Perhaps the most 
straightforward explanation is that both galaxy populations are drawn 
from the same subset of high-$z$ star-bursting systems.  However, another
possibility is that the offset between the extrapolated main sequence
and the locus of LAEs and oELGs may simply be due to errors in our estimates
of stellar mass and star formation rate.  \cite{kusakabe15} have argued that 
the flux densities of $z \sim 2$ LAEs are better fit using an SMC-type dust 
law rather than a \citet{calzetti01} obscuration relation.  If were to
adopt such a model, our dust-corrected SFRs would move downward towards 
both the \cite{speagle14} and \citet{whitaker14}  lines.  A third reason for the 
discrepancy may be that a simple extrapolation of the star-forming main sequence 
is incorrect; it is possible that the slope of this relation changes at lower stellar mass.  
Finally, the higher SFRs at a given stellar mass might simply be a selection 
effect, as lower SFR objects are more difficult to detect through their emission
lines.  Larger and deeper samples of both LAEs and oELGs are required to
resolve this question.

\section{Conclusion}
\label{sec:conclusion}

This $z \sim 2$ study finds no significant differences between 
the physical properties of Ly$\alpha$ emitting galaxies and
galaxies selected solely by their rest-frame optical emission lines.
More specifically, the distributions for LAE stellar mass, half-light radius,
star formation rate, ellipticity, nearest neighbor distance, specific
star formation rate, star formation rate surface density, [O~III]
luminosity, and [O~III] equivalent width are not statistically different
from those of their oELG counterparts.  Surprisingly, there is not even
any evidence for a difference between the two populations' UV slopes.  Since
a star-forming galaxy's spectral slope in the UV reflects the presence
of dust and internal extinction, this suggests that the processes which
regulate the escape of Ly$\alpha$ from a $z \sim 2$ galaxy are operating
on a small scale, rather than the global scale probed by our survey.
This small scale escape of Ly$\alpha$ also explains the lack of 
difference we found between the population's [O~III] EW and specific star formation 
rate distributions.  It seems that the presence of Ly$\alpha$ emission 
in a star-forming galaxy is not a strong function of the parameters most
commonly used to describe high-$z$ galaxies.

Since we have found no differences in the physical and morphological properties
studies herein, the question still remains as to what causes Ly$\alpha$ to escape some
galaxies and not others.   As our analysis mainly concerned parameters related to
stellar emission, the data would seem to suggest that at $z \sim 2$, a star-forming
galaxy's interstellar and circumgalactic media are not closely tied to the global distribution
of its stars \citep[e.g.,][]{dijkstra07}.  This matches quite well with recent observational 
results of both nearby LAEs and high-redshift Ly$\alpha$ halos.  For example, 
\citet{herenz15} has argued that in nearby LAEs, it is the turbulence of the ISM that 
makes conditions favorable for Ly$\alpha$ escape.  Similarly, a number of studies of 
Ly$\alpha$ halos surrounding $z \gtrsim 2$ star-forming galaxies have shown the extent 
to which the circumgalactic medium can effect Ly$\alpha$ emission 
\citep[e.g.,][]{matsuda12, momose14, wisotzki15}.  Particularly intriguing are the 
observations of \citet{momose15}, who found an inverse correlation between the scale 
length of a Ly$\alpha$ halo and the total luminosity of the Ly$\alpha$ emission.  This 
suggests that circumgalactic conditions control whether Ly$\alpha$ is scattered out of the 
line of sight into a bright halo, or mostly passes through, creating an LAE\null.  This 
connects well to our toy model in Section \ref{toymodel}.

Though a direct comparison is not possible due sampling issues, we do note
that our results are in general agreement with those of \citet{hathi15}, who 
found physical similarities between LAEs and non-LAEs between $2 < z < 2.5$. 
Our results also agree with the work of \cite{jiang15}, who found
no differences in the distributions of stellar mass, age, and star formation rate
for samples of $z \gtrsim 6$ LAEs and LBGs.
However, our results differ from those of \citet{cooke10}, who claimed
that LBGs with small nearest neighbor separations were more likely to
emit in Ly$\alpha$. One possible reason for this discrepancy is that the
majority of our galaxies are much less massive than Lyman break objects:
while massive galaxies might require an interaction to create an escape 
path for Ly$\alpha$, less organized, lower mass objects might not.
\cite{oteo15} also found differences between the physical properties
(reddening, stellar mass, and star formation rate) of $z \sim 2$ LAEs and 
H$\alpha$ selected systems.  However, once again, this analysis does 
not control for stellar mass: not only is their median galaxy more than a 
dex larger than the systems considered here, but their Balmer-line selected galaxies
are systematically more massive than their LAEs by almost a full dex.
This offset in stellar mass is likely responsible for the difference in results.

While our study is limited by small number statistics, particularly with 
respect to the sample of LAEs, the possibility that $z \sim 2$ Ly$\alpha$ 
emitting galaxies are drawn from the epoch's general star-forming population 
is tantalizing.  Our study suggests that LAEs can be used to probe and 
investigate the population of low mass star forming galaxies, which, sans 
Ly$\alpha$ emission, are very resource intensive to identify.  While this 
reasoning has been used to motivate LAE studies for many years, this work 
provides a justification for this line of attack.  This hypothesis will be 
better tested when the HETDEX survey comes online, as commissioning
data alone will generate more than a thousand LAEs in fields targeted by
the {\sl HST\/} infrared grism. This will allow a rigorous search for 
population differences between the LAEs and other star forming galaxies 
of the epoch, while controlling for stellar mass, size, and star formation 
rate.


\acknowledgments
\section*{Acknowledgments}
We greatly appreciate the very helpful comments and suggestions from our anonymous referee.
This work was supported via NSF through grant AST 09-26641\null.
C. Behrens was supported by the CRC 963 of the German Research Council (DFG).
SJ acknowledges support from NSF grant NSF AST-1413652 and the NASA/JPL SURP 
Program.  We greatly appreciate conversations with the participants in the 
meeting on ``Ly$\alpha$ as an Astrophysical Tool" in September 2013 in 
Stockholm, Sweden.  We acknowledge the Research Computer and 
Cyberinfrastructure Unit of Information Technology Services at 
The Pennsylvania State University for providing computational support and 
resources for this project.  The Institute for Gravitation and the Cosmos is 
supported by the Eberly College of Science and the Office of the Senior Vice
President for Research at the Pennsylvania 
State University. This research has made use of NASA's Astrophysics Data System and the python packages \texttt{IPython}, \texttt{AstroPy}, 
\texttt{NumPy}, \texttt{SciPy}, \texttt{scikit-learn}, and \texttt{matplotlib} \citep{ipython, astropy, numpy, scipy, scikit-learn, matplotlib}.
We also used the GalMC software of \cite{acquaviva11} and the half-light 
radius software from \cite{bond09}.
This work is based on observations taken by the CANDELS Multi-Cycle Treasury 
Program with the NASA/ESA HST, 
which is operated by the Association of Universities for Research in Astronomy,
Inc., under NASA contract NAS5-26555.


\bibliographystyle{apj}                       
\bibliography{lae}

\begin{thebibliography}{}
\expandafter\ifx\csname natexlab\endcsname\relax\def\natexlab#1{#1}\fi

\bibitem[{{Acquaviva}(2012)}]{acquaviva12}
{Acquaviva}, V. 2012, Personal Communication

\bibitem[{{Acquaviva} {et~al.}(2011){Acquaviva}, {Gawiser}, \&
  {Guaita}}]{acquaviva11}
{Acquaviva}, V., {Gawiser}, E., \& {Guaita}, L. 2011, \apj, 737, 47

\bibitem[{{Adams} {et~al.}(2011){Adams}, {Blanc}, {Hill}, {Gebhardt}, {Drory},
  {Hao}, {Bender}, {Byun}, {Ciardullo}, {Cornell}, {Finkelstein}, {Fry},
  {Gawiser}, {Gronwall}, {Hopp}, {Jeong}, {Kelz}, {Kelzenberg}, {Komatsu},
  {MacQueen}, {Murphy}, {Odoms}, {Roth}, {Schneider}, {Tufts}, \&
  {Wilkinson}}]{adams11}
{Adams}, J.~J., {Blanc}, G.~A., {Hill}, G.~J., {et~al.} 2011, \apjs, 192, 5

\bibitem[{{Ade} {et~al.}(2014){Ade}, {Aghanim}, {Armitage-Caplan}, {Arnaud},
  {Ashdown}, {Atrio-Barandela}, {Aumont}, {Baccigalupi}, {Banday}, \&
  et~al.}]{planck13}
{Ade}, P.~A.~R., {Aghanim}, N., {Armitage-Caplan}, C., {et~al.} 2014, \aap,
  571, A16

\bibitem[{Akritas {et~al.}(1995)Akritas, Murphy, \& Lavalley}]{akritas95}
Akritas, M.~G., Murphy, S.~A., \& Lavalley, M.~P. 1995, Journal of the American
  Statistical Association, 90, 170

\bibitem[{Anderson \& Darling(1952)}]{anderson52}
Anderson, T.~W., \& Darling, D.~A. 1952, The Annals of Mathematical Statistics,
  23, 193

\bibitem[{{Astropy Collaboration} {et~al.}(2013){Astropy Collaboration},
  {Robitaille}, {Tollerud}, {Greenfield}, {Droettboom}, {Bray}, {Aldcroft},
  {Davis}, {Ginsburg}, {Price-Whelan}, {Kerzendorf}, {Conley}, {Crighton},
  {Barbary}, {Muna}, {Ferguson}, {Grollier}, {Parikh}, {Nair}, {Unther},
  {Deil}, {Woillez}, {Conseil}, {Kramer}, {Turner}, {Singer}, {Fox}, {Weaver},
  {Zabalza}, {Edwards}, {Azalee Bostroem}, {Burke}, {Casey}, {Crawford},
  {Dencheva}, {Ely}, {Jenness}, {Labrie}, {Lim}, {Pierfederici}, {Pontzen},
  {Ptak}, {Refsdal}, {Servillat}, \& {Streicher}}]{astropy}
{Astropy Collaboration}, {Robitaille}, T.~P., {Tollerud}, E.~J., {et~al.} 2013,
  \aap, 558, A33

\bibitem[{{Atek} {et~al.}(2014){Atek}, {Kunth}, {Schaerer}, {Mas-Hesse},
  {Hayes}, {{\"O}stlin}, \& {Kneib}}]{atek14}
{Atek}, H., {Kunth}, D., {Schaerer}, D., {et~al.} 2014, \aap, 561, A89

\bibitem[{{Atek} {et~al.}(2011){Atek}, {Siana}, {Scarlata}, {Malkan},
  {McCarthy}, {Teplitz}, {Henry}, {Colbert}, {Bridge}, {Bunker}, {Dressler},
  {Fosbury}, {Hathi}, {Martin}, {Ross}, \& {Shim}}]{atek11}
{Atek}, H., {Siana}, B., {Scarlata}, C., {et~al.} 2011, \apj, 743, 121

\bibitem[{{Bauer} {et~al.}(2011){Bauer}, {Conselice}, {P{\'e}rez-Gonz{\'a}lez},
  {Gr{\"u}tzbauch}, {Bluck}, {Buitrago}, \& {Mortlock}}]{bauer11}
{Bauer}, A.~E., {Conselice}, C.~J., {P{\'e}rez-Gonz{\'a}lez}, P.~G., {et~al.}
  2011, \mnras, 417, 289

\bibitem[{{Behrens} \& {Braun}(2014)}]{behrens14b}
{Behrens}, C., \& {Braun}, H. 2014, \aap, 572, A74

\bibitem[{{Behrens} {et~al.}(2014){Behrens}, {Dijkstra}, \&
  {Niemeyer}}]{behrens14}
{Behrens}, C., {Dijkstra}, M., \& {Niemeyer}, J.~C. 2014, \aap, 563, A77

\bibitem[{{Bertin} \& {Arnouts}(1996)}]{bertin96}
{Bertin}, E., \& {Arnouts}, S. 1996, \aaps, 117, 393

\bibitem[{{Blanc} {et~al.}(2011){Blanc}, {Adams}, {Gebhardt}, {Hill}, {Drory},
  {Hao}, {Bender}, {Ciardullo}, {Finkelstein}, {Fry}, {Gawiser}, {Gronwall},
  {Hopp}, {Jeong}, {Kelzenberg}, {Komatsu}, {MacQueen}, {Murphy}, {Roth},
  {Schneider}, \& {Tufts}}]{blanc11}
{Blanc}, G.~A., {Adams}, J.~J., {Gebhardt}, K., {et~al.} 2011, \apj, 736, 31

\bibitem[{{Bond} {et~al.}(2009){Bond}, {Gawiser}, {Gronwall}, {Ciardullo},
  {Altmann}, \& {Schawinski}}]{bond09}
{Bond}, N.~A., {Gawiser}, E., {Gronwall}, C., {et~al.} 2009, \apj, 705, 639

\bibitem[{{Bond} {et~al.}(2012){Bond}, {Gawiser}, {Guaita}, {Padilla},
  {Gronwall}, {Ciardullo}, \& {Lai}}]{bond12}
{Bond}, N.~A., {Gawiser}, E., {Guaita}, L., {et~al.} 2012, \apj, 753, 95

\bibitem[{{Bond} {et~al.}(2014){Bond}, {Gardner}, {de Mello}, {Teplitz},
  {Rafelski}, {Koekemoer}, {Coe}, {Grogin}, {Gawiser}, {Ravindranath}, \&
  {Scarlata}}]{bond14}
{Bond}, N.~A., {Gardner}, J.~P., {de Mello}, D.~F., {et~al.} 2014, \apj, 791,
  18

\bibitem[{{Bouwens} {et~al.}(2009){Bouwens}, {Illingworth}, {Franx}, {Chary},
  {Meurer}, {Conselice}, {Ford}, {Giavalisco}, \& {van Dokkum}}]{bouwens09}
{Bouwens}, R.~J., {Illingworth}, G.~D., {Franx}, M., {et~al.} 2009, \apj, 705,
  936

\bibitem[{{Bouwens} {et~al.}(2010){Bouwens}, {Illingworth}, {Oesch},
  {Stiavelli}, {van Dokkum}, {Trenti}, {Magee}, {Labb{\'e}}, {Franx},
  {Carollo}, \& {Gonzalez}}]{bouwens10}
{Bouwens}, R.~J., {Illingworth}, G.~D., {Oesch}, P.~A., {et~al.} 2010, \apjl,
  709, L133

\bibitem[{{Brammer} {et~al.}(2012){Brammer}, {van Dokkum}, {Franx},
  {Fumagalli}, {Patel}, {Rix}, {Skelton}, {Kriek}, {Nelson}, {Schmidt},
  {Bezanson}, {da Cunha}, {Erb}, {Fan}, {F{\"o}rster Schreiber}, {Illingworth},
  {Labb{\'e}}, {Leja}, {Lundgren}, {Magee}, {Marchesini}, {McCarthy},
  {Momcheva}, {Muzzin}, {Quadri}, {Steidel}, {Tal}, {Wake}, {Whitaker}, \&
  {Williams}}]{brammer12}
{Brammer}, G.~B., {van Dokkum}, P.~G., {Franx}, M., {et~al.} 2012, \apjs, 200,
  13

\bibitem[{Brooks \& Gelman(1998)}]{brooks98}
Brooks, S.~P., \& Gelman, A. 1998, Journal of Computational and Graphical
  Statistics, 7, 434

\bibitem[{{Bruzual} \& {Charlot}(2003)}]{bruzual03}
{Bruzual}, G., \& {Charlot}, S. 2003, \mnras, 344, 1000

\bibitem[{{Buat} {et~al.}(2012){Buat}, {Noll}, {Burgarella}, {Giovannoli},
  {Charmandaris}, {Pannella}, {Hwang}, {Elbaz}, {Dickinson}, {Magdis}, {Reddy},
  \& {Murphy}}]{buat12}
{Buat}, V., {Noll}, S., {Burgarella}, D., {et~al.} 2012, \aap, 545, A141

\bibitem[{{Calzetti}(2001)}]{calzetti01}
{Calzetti}, D. 2001, \pasp, 113, 1449

\bibitem[{{Cassata} {et~al.}(2015){Cassata}, {Tasca}, {Le F{\`e}vre}, {Lemaux},
  {Garilli}, {Le Brun}, {Maccagni}, {Pentericci}, {Thomas}, {Vanzella},
  {Zamorani}, {Zucca}, {Amorin}, {Bardelli}, {Capak}, {Cassar{\`a}},
  {Castellano}, {Cimatti}, {Cuby}, {Cucciati}, {de la Torre}, {Durkalec},
  {Fontana}, {Giavalisco}, {Grazian}, {Hathi}, {Ilbert}, {Moreau}, {Paltani},
  {Ribeiro}, {Salvato}, {Schaerer}, {Scodeggio}, {Sommariva}, {Talia},
  {Taniguchi}, {Tresse}, {Vergani}, {Wang}, {Charlot}, {Contini}, {Fotopoulou},
  {Koekemoer}, {L{\'o}pez-Sanjuan}, {Mellier}, \& {Scoville}}]{cassata15}
{Cassata}, P., {Tasca}, L.~A.~M., {Le F{\`e}vre}, O., {et~al.} 2015, \aap, 573,
  A24

\bibitem[{{Chiang} {et~al.}(2015){Chiang}, {Overzier}, {Gebhardt},
  {Finkelstein}, {Chiang}, {Hill}, {Blanc}, {Drory}, {Chonis}, {Zeimann},
  {Hagen}, {Schneider}, {Jogee}, {Ciardullo}, \& {Gronwall}}]{chiang15}
{Chiang}, Y.-K., {Overzier}, R.~A., {Gebhardt}, K., {et~al.} 2015, \apj, 808,
  37

\bibitem[{{Ciardullo} {et~al.}(2012){Ciardullo}, {Gronwall}, {Wolf},
  {McCathran}, {Bond}, {Gawiser}, {Guaita}, {Feldmeier}, {Treister}, {Padilla},
  {Francke}, {Matkovi{\'c}}, {Altmann}, \& {Herrera}}]{ciardullo12}
{Ciardullo}, R., {Gronwall}, C., {Wolf}, C., {et~al.} 2012, \apj, 744, 110

\bibitem[{{Ciardullo} {et~al.}(2014){Ciardullo}, {Zeimann}, {Gronwall},
  {Gebhardt}, {Schneider}, {Hagen}, {Malz}, {Blanc}, {Hill}, {Drory}, \&
  {Gawiser}}]{ciardullo14}
{Ciardullo}, R., {Zeimann}, G.~R., {Gronwall}, C., {et~al.} 2014, \apj, 796, 64

\bibitem[{{Conroy}(2013)}]{conroy13}
{Conroy}, C. 2013, \araa, 51, 393

\bibitem[{{Conselice}(2014)}]{conselice14}
{Conselice}, C.~J. 2014, \araa, 52, 291

\bibitem[{{Cooke} {et~al.}(2010){Cooke}, {Berrier}, {Barton}, {Bullock}, \&
  {Wolfe}}]{cooke10}
{Cooke}, J., {Berrier}, J.~C., {Barton}, E.~J., {Bullock}, J.~S., \& {Wolfe},
  A.~M. 2010, \mnras, 403, 1020

\bibitem[{{Cowie} {et~al.}(2010){Cowie}, {Barger}, \& {Hu}}]{cowie10}
{Cowie}, L.~L., {Barger}, A.~J., \& {Hu}, E.~M. 2010, \apj, 711, 928

\bibitem[{{Cowie} {et~al.}(2011){Cowie}, {Barger}, \& {Hu}}]{cowie11}
---. 2011, \apj, 738, 136

\bibitem[{{Davari} {et~al.}(2014){Davari}, {Ho}, {Peng}, \& {Huang}}]{davari14}
{Davari}, R., {Ho}, L.~C., {Peng}, C.~Y., \& {Huang}, S. 2014, \apj, 787, 69

\bibitem[{{Deharveng} {et~al.}(2008){Deharveng}, {Small}, {Barlow},
  {P{\'e}roux}, {Milliard}, {Friedman}, {Martin}, {Morrissey}, {Schiminovich},
  {Forster}, {Seibert}, {Wyder}, {Bianchi}, {Donas}, {Heckman}, {Lee},
  {Madore}, {Neff}, {Rich}, {Szalay}, {Welsh}, \& {Yi}}]{deharveng08}
{Deharveng}, J.-M., {Small}, T., {Barlow}, T.~A., {et~al.} 2008, \apj, 680,
  1072

\bibitem[{{Dijkstra} {et~al.}(2007){Dijkstra}, {Lidz}, \&
  {Wyithe}}]{dijkstra07}
{Dijkstra}, M., {Lidz}, A., \& {Wyithe}, J.~S.~B. 2007, \mnras, 377, 1175

\bibitem[{{Erb} {et~al.}(2006){Erb}, {Steidel}, {Shapley}, {Pettini}, {Reddy},
  \& {Adelberger}}]{erb06}
{Erb}, D.~K., {Steidel}, C.~C., {Shapley}, A.~E., {et~al.} 2006, \apj, 647, 128

\bibitem[{{Finkelstein} {et~al.}(2009){Finkelstein}, {Rhoads}, {Malhotra}, \&
  {Grogin}}]{finkelstein09}
{Finkelstein}, S.~L., {Rhoads}, J.~E., {Malhotra}, S., \& {Grogin}, N. 2009,
  \apj, 691, 465

\bibitem[{{Finkelstein} {et~al.}(2008){Finkelstein}, {Rhoads}, {Malhotra},
  {Grogin}, \& {Wang}}]{finkelstein08}
{Finkelstein}, S.~L., {Rhoads}, J.~E., {Malhotra}, S., {Grogin}, N., \& {Wang},
  J. 2008, \apj, 678, 655

\bibitem[{{Finkelstein} {et~al.}(2011){Finkelstein}, {Hill}, {Gebhardt},
  {Adams}, {Blanc}, {Papovich}, {Ciardullo}, {Drory}, {Gawiser}, {Gronwall},
  {Schneider}, \& {Tran}}]{finkelstein11}
{Finkelstein}, S.~L., {Hill}, G.~J., {Gebhardt}, K., {et~al.} 2011, \apj, 729,
  140

\bibitem[{{F{\"o}rster Schreiber} {et~al.}(2009){F{\"o}rster Schreiber},
  {Genzel}, {Bouch{\'e}}, {Cresci}, {Davies}, {Buschkamp}, {Shapiro},
  {Tacconi}, {Hicks}, {Genel}, {Shapley}, {Erb}, {Steidel}, {Lutz},
  {Eisenhauer}, {Gillessen}, {Sternberg}, {Renzini}, {Cimatti}, {Daddi},
  {Kurk}, {Lilly}, {Kong}, {Lehnert}, {Nesvadba}, {Verma}, {McCracken},
  {Arimoto}, {Mignoli}, \& {Onodera}}]{forsterschreiber09}
{F{\"o}rster Schreiber}, N.~M., {Genzel}, R., {Bouch{\'e}}, N., {et~al.} 2009,
  \apj, 706, 1364

\bibitem[{{Gawiser} {et~al.}(2006){Gawiser}, {van Dokkum}, {Gronwall},
  {Ciardullo}, {Blanc}, {Castander}, {Feldmeier}, {Francke}, {Franx},
  {Haberzettl}, {Herrera}, {Hickey}, {Infante}, {Lira}, {Maza}, {Quadri},
  {Richardson}, {Schawinski}, {Schirmer}, {Taylor}, {Treister}, {Urry}, \&
  {Virani}}]{gawiser06}
{Gawiser}, E., {van Dokkum}, P.~G., {Gronwall}, C., {et~al.} 2006, \apjl, 642,
  L13

\bibitem[{{Gawiser} {et~al.}(2007){Gawiser}, {Francke}, {Lai}, {Schawinski},
  {Gronwall}, {Ciardullo}, {Quadri}, {Orsi}, {Barrientos}, {Blanc}, {Fazio},
  {Feldmeier}, {Huang}, {Infante}, {Lira}, {Padilla}, {Taylor}, {Treister},
  {Urry}, {van Dokkum}, \& {Virani}}]{gawiser07}
{Gawiser}, E., {Francke}, H., {Lai}, K., {et~al.} 2007, \apj, 671, 278

\bibitem[{Gelman \& Rubin(1992)}]{gelman92}
Gelman, A., \& Rubin, D.~B. 1992, Statistical Science, 7, 457

\bibitem[{{Giavalisco} {et~al.}(2004){Giavalisco}, {Ferguson}, {Koekemoer},
  {Dickinson}, {Alexander}, {Bauer}, {Bergeron}, {Biagetti}, {Brandt},
  {Casertano}, {Cesarsky}, {Chatzichristou}, {Conselice}, {Cristiani}, {Da
  Costa}, {Dahlen}, {de Mello}, {Eisenhardt}, {Erben}, {Fall}, {Fassnacht},
  {Fosbury}, {Fruchter}, {Gardner}, {Grogin}, {Hook}, {Hornschemeier}, {Idzi},
  {Jogee}, {Kretchmer}, {Laidler}, {Lee}, {Livio}, {Lucas}, {Madau},
  {Mobasher}, {Moustakas}, {Nonino}, {Padovani}, {Papovich}, {Park},
  {Ravindranath}, {Renzini}, {Richardson}, {Riess}, {Rosati}, {Schirmer},
  {Schreier}, {Somerville}, {Spinrad}, {Stern}, {Stiavelli}, {Strolger},
  {Urry}, {Vandame}, {Williams}, \& {Wolf}}]{GOODS}
{Giavalisco}, M., {Ferguson}, H.~C., {Koekemoer}, A.~M., {et~al.} 2004, \apjl,
  600, L93

\bibitem[{{Grasshorn Gebhardt} {et~al.}(2015){Grasshorn Gebhardt}, {Zeimann},
  {Ciardullo}, \& {Hagen}}]{gebhardt15}
{Grasshorn Gebhardt}, H., {Zeimann}, G.~R., {Ciardullo}, R., \& {Hagen}, A.
  2015, \apj, In Press

\bibitem[{{Greiner} {et~al.}(2015){Greiner}, {Fox}, {Schady}, {Kr{\"u}hler},
  {Trenti}, {Cikota}, {Bolmer}, {Elliott}, {Delvaux}, {Perna}, {Afonso},
  {Kann}, {Klose}, {Savaglio}, {Schmidl}, {Schweyer}, {Tanga}, \&
  {Varela}}]{greiner15}
{Greiner}, J., {Fox}, D.~B., {Schady}, P., {et~al.} 2015, \apj, 809, 76

\bibitem[{{Grogin} {et~al.}(2011){Grogin}, {Kocevski}, {Faber}, {Ferguson},
  {Koekemoer}, {Riess}, {Acquaviva}, {Alexander}, {Almaini}, {Ashby}, {Barden},
  {Bell}, {Bournaud}, {Brown}, {Caputi}, {Casertano}, {Cassata}, {Castellano},
  {Challis}, {Chary}, {Cheung}, {Cirasuolo}, {Conselice}, {Roshan Cooray},
  {Croton}, {Daddi}, {Dahlen}, {Dav{\'e}}, {de Mello}, {Dekel}, {Dickinson},
  {Dolch}, {Donley}, {Dunlop}, {Dutton}, {Elbaz}, {Fazio}, {Filippenko},
  {Finkelstein}, {Fontana}, {Gardner}, {Garnavich}, {Gawiser}, {Giavalisco},
  {Grazian}, {Guo}, {Hathi}, {H{\"a}ussler}, {Hopkins}, {Huang}, {Huang},
  {Jha}, {Kartaltepe}, {Kirshner}, {Koo}, {Lai}, {Lee}, {Li}, {Lotz}, {Lucas},
  {Madau}, {McCarthy}, {McGrath}, {McIntosh}, {McLure}, {Mobasher},
  {Moustakas}, {Mozena}, {Nandra}, {Newman}, {Niemi}, {Noeske}, {Papovich},
  {Pentericci}, {Pope}, {Primack}, {Rajan}, {Ravindranath}, {Reddy}, {Renzini},
  {Rix}, {Robaina}, {Rodney}, {Rosario}, {Rosati}, {Salimbeni}, {Scarlata},
  {Siana}, {Simard}, {Smidt}, {Somerville}, {Spinrad}, {Straughn}, {Strolger},
  {Telford}, {Teplitz}, {Trump}, {van der Wel}, {Villforth}, {Wechsler},
  {Weiner}, {Wiklind}, {Wild}, {Wilson}, {Wuyts}, {Yan}, \& {Yun}}]{grogin11}
{Grogin}, N.~A., {Kocevski}, D.~D., {Faber}, S.~M., {et~al.} 2011, \apjs, 197,
  35

\bibitem[{{Gronke} {et~al.}(2015){Gronke}, {Bull}, \& {Dijkstra}}]{gronke15}
{Gronke}, M., {Bull}, P., \& {Dijkstra}, M. 2015, \apj, 812, 123

\bibitem[{{Gronke} \& {Dijkstra}(2014)}]{gronke14}
{Gronke}, M., \& {Dijkstra}, M. 2014, \mnras, 444, 1095

\bibitem[{{Gronwall} {et~al.}(2011){Gronwall}, {Bond}, {Ciardullo}, {Gawiser},
  {Altmann}, {Blanc}, \& {Feldmeier}}]{gronwall11}
{Gronwall}, C., {Bond}, N.~A., {Ciardullo}, R., {et~al.} 2011, \apj, 743, 9

\bibitem[{{Gronwall} {et~al.}(2007){Gronwall}, {Ciardullo}, {Hickey},
  {Gawiser}, {Feldmeier}, {van Dokkum}, {Urry}, {Herrera}, {Lehmer}, {Infante},
  {Orsi}, {Marchesini}, {Blanc}, {Francke}, {Lira}, \& {Treister}}]{gronwall07}
{Gronwall}, C., {Ciardullo}, R., {Hickey}, T., {et~al.} 2007, \apj, 667, 79

\bibitem[{{Guaita} {et~al.}(2010){Guaita}, {Gawiser}, {Padilla}, {Francke},
  {Bond}, {Gronwall}, {Ciardullo}, {Feldmeier}, {Sinawa}, {Blanc}, \&
  {Virani}}]{guaita10}
{Guaita}, L., {Gawiser}, E., {Padilla}, N., {et~al.} 2010, \apj, 714, 255

\bibitem[{{Hagen} {et~al.}(2014){Hagen}, {Ciardullo}, {Gronwall}, {Acquaviva},
  {Bridge}, {Zeimann}, {Blanc}, {Bond}, {Finkelstein}, {Song}, {Gawiser},
  {Fox}, {Gebhardt}, {Malz}, {Schneider}, {Drory}, {Gebhardt}, \&
  {Hill}}]{hagen14}
{Hagen}, A., {Ciardullo}, R., {Gronwall}, C., {et~al.} 2014, \apj, 786, 59

\bibitem[{{Hao} {et~al.}(2011){Hao}, {Kennicutt}, {Johnson}, {Calzetti},
  {Dale}, \& {Moustakas}}]{hao11}
{Hao}, C.-N., {Kennicutt}, R.~C., {Johnson}, B.~D., {et~al.} 2011, \apj, 741,
  124

\bibitem[{Hastings(1970)}]{hastings70}
Hastings, W.~K. 1970, Biometrika, 57, 97

\bibitem[{{Hathi} {et~al.}(2015){Hathi}, {Le F{\`e}vre}, {Ilbert}, {Cassata},
  {Tasca}, {Lemaux}, {Garilli}, {Le Brun}, {Maccagni}, {Pentericci}, {Thomas},
  {Vanzella}, {Zamorani}, {Zucca}, {Amor{\'{\i}}n}, {Bardelli}, {Cassar{\`a}},
  {Castellano}, {Cimatti}, {Cucciati}, {Durkalec}, {Fontana}, {Giavalisco},
  {Grazian}, {Guaita}, {Koekemoer}, {Paltani}, {Pforr}, {Ribeiro}, {Schaerer},
  {Scodeggio}, {Sommariva}, {Talia}, {Tresse}, {Vergani}, {Capak}, {Charlot},
  {Contini}, {Cuby}, {de la Torre}, {Dunlop}, {Fotopoulou},
  {L{\'o}pez-Sanjuan}, {Mellier}, {Salvato}, {Scoville}, {Taniguchi}, \&
  {Wang}}]{hathi15}
{Hathi}, N.~P., {Le F{\`e}vre}, O., {Ilbert}, O., {et~al.} 2015, submitted to
  \aap, ArXiv e-prints, arXiv:1503.01753

\bibitem[{{Hayes} {et~al.}(2010){Hayes}, {{\"O}stlin}, {Schaerer}, {Mas-Hesse},
  {Leitherer}, {Atek}, {Kunth}, {Verhamme}, {de Barros}, \&
  {Melinder}}]{hayes10}
{Hayes}, M., {{\"O}stlin}, G., {Schaerer}, D., {et~al.} 2010, \nat, 464, 562

\bibitem[{{Hayes} {et~al.}(2014){Hayes}, {{\"O}stlin}, {Duval}, {Sandberg},
  {Guaita}, {Melinder}, {Adamo}, {Schaerer}, {Verhamme}, {Orlitov{\'a}},
  {Mas-Hesse}, {Cannon}, {Atek}, {Kunth}, {Laursen}, {Ot{\'{\i}}-Floranes},
  {Pardy}, {Rivera-Thorsen}, \& {Herenz}}]{hayes14}
{Hayes}, M., {{\"O}stlin}, G., {Duval}, F., {et~al.} 2014, \apj, 782, 6

\bibitem[{{Henry} {et~al.}(2015){Henry}, {Scarlata}, {Martin}, \&
  {Erb}}]{henry15}
{Henry}, A., {Scarlata}, C., {Martin}, C.~L., \& {Erb}, D. 2015, \apj, 809, 19

\bibitem[{{Herenz} {et~al.}(2015){Herenz}, {Gruyters}, {Orlitova}, {Hayes},
  {{\"O}stlin}, {Cannon}, {Roth}, {Bik}, {Pardy}, {Ot{\'{\i}}-Floranes},
  {Mas-Hesse}, {Adamo}, {Atek}, {Duval}, {Guaita}, {Kunth}, {Laursen},
  {Melinder}, {Puschnig}, {Rivera-Thorsen}, {Schaerer}, \&
  {Verhamme}}]{herenz15}
{Herenz}, E.~C., {Gruyters}, P., {Orlitova}, I., {et~al.} 2015, \aap, In Press,
  arXiv:1511.05406

\bibitem[{{Hill} {et~al.}(2008){Hill}, {Gebhardt}, {Komatsu}, {Drory},
  {MacQueen}, {Adams}, {Blanc}, {Koehler}, {Rafal}, {Roth}, {Kelz}, {Gronwall},
  {Ciardullo}, \& {Schneider}}]{hill08}
{Hill}, G.~J., {Gebhardt}, K., {Komatsu}, E., {et~al.} 2008, in Astronomical
  Society of the Pacific Conference Series, Vol. 399, Panoramic Views of Galaxy
  Formation and Evolution, ed. T.~{Kodama}, T.~{Yamada}, \& K.~{Aoki}, 115

\bibitem[{{Hu} {et~al.}(1998){Hu}, {Cowie}, \& {McMahon}}]{hu98}
{Hu}, E.~M., {Cowie}, L.~L., \& {McMahon}, R.~G. 1998, \apjl, 502, L99

\bibitem[{Hunter(2007)}]{matplotlib}
Hunter, J.~D. 2007, Computing in Science \& Engineering, 9, 90

\bibitem[{{Jiang} {et~al.}(2015){Jiang}, {Finlator}, {Cohen}, {Egami},
  {Windhorst}, {Fan}, {Dave}, {Kashikawa}, {Mechtley}, {Ouchi}, {Shimasaku}, \&
  {Clement}}]{jiang15}
{Jiang}, L., {Finlator}, K., {Cohen}, S.~H., {et~al.} 2015, ArXiv e-prints,
  arXiv:1511.01519

\bibitem[{Jones {et~al.}(2001)Jones, Oliphant, Peterson, {et~al.}}]{scipy}
Jones, E., Oliphant, T., Peterson, P., {et~al.} 2001, {SciPy}: Open source
  scientific tools for {Python}

\bibitem[{{Kennicutt} \& {Evans}(2012)}]{kennicutt12}
{Kennicutt}, R.~C., \& {Evans}, N.~J. 2012, \araa, 50, 531

\bibitem[{{Kennicutt}(1992)}]{kennicutt92}
{Kennicutt}, Jr., R.~C. 1992, \apj, 388, 310

\bibitem[{{Kennicutt}(1998)}]{kennicutt98}
---. 1998, \araa, 36, 189

\bibitem[{{Koekemoer} {et~al.}(2011){Koekemoer}, {Faber}, {Ferguson}, {Grogin},
  {Kocevski}, {Koo}, {Lai}, {Lotz}, {Lucas}, {McGrath}, {Ogaz}, {Rajan},
  {Riess}, {Rodney}, {Strolger}, {Casertano}, {Castellano}, {Dahlen},
  {Dickinson}, {Dolch}, {Fontana}, {Giavalisco}, {Grazian}, {Guo}, {Hathi},
  {Huang}, {van der Wel}, {Yan}, {Acquaviva}, {Alexander}, {Almaini}, {Ashby},
  {Barden}, {Bell}, {Bournaud}, {Brown}, {Caputi}, {Cassata}, {Challis},
  {Chary}, {Cheung}, {Cirasuolo}, {Conselice}, {Roshan Cooray}, {Croton},
  {Daddi}, {Dav{\'e}}, {de Mello}, {de Ravel}, {Dekel}, {Donley}, {Dunlop},
  {Dutton}, {Elbaz}, {Fazio}, {Filippenko}, {Finkelstein}, {Frazer}, {Gardner},
  {Garnavich}, {Gawiser}, {Gruetzbauch}, {Hartley}, {H{\"a}ussler},
  {Herrington}, {Hopkins}, {Huang}, {Jha}, {Johnson}, {Kartaltepe},
  {Khostovan}, {Kirshner}, {Lani}, {Lee}, {Li}, {Madau}, {McCarthy},
  {McIntosh}, {McLure}, {McPartland}, {Mobasher}, {Moreira}, {Mortlock},
  {Moustakas}, {Mozena}, {Nandra}, {Newman}, {Nielsen}, {Niemi}, {Noeske},
  {Papovich}, {Pentericci}, {Pope}, {Primack}, {Ravindranath}, {Reddy},
  {Renzini}, {Rix}, {Robaina}, {Rosario}, {Rosati}, {Salimbeni}, {Scarlata},
  {Siana}, {Simard}, {Smidt}, {Snyder}, {Somerville}, {Spinrad}, {Straughn},
  {Telford}, {Teplitz}, {Trump}, {Vargas}, {Villforth}, {Wagner}, {Wandro},
  {Wechsler}, {Weiner}, {Wiklind}, {Wild}, {Wilson}, {Wuyts}, \&
  {Yun}}]{koekemoer11}
{Koekemoer}, A.~M., {Faber}, S.~M., {Ferguson}, H.~C., {et~al.} 2011, \apjs,
  197, 36

\bibitem[{Kolmogorov(1933)}]{kolmogorov33}
Kolmogorov, A. 1933, G. Inst. Ital. Attuari, 83

\bibitem[{{Kornei} {et~al.}(2010){Kornei}, {Shapley}, {Erb}, {Steidel},
  {Reddy}, {Pettini}, \& {Bogosavljevi{\'c}}}]{kornei10}
{Kornei}, K.~A., {Shapley}, A.~E., {Erb}, D.~K., {et~al.} 2010, \apj, 711, 693

\bibitem[{{Kriek} \& {Conroy}(2013)}]{kriek13}
{Kriek}, M., \& {Conroy}, C. 2013, \apjl, 775, L16

\bibitem[{{Kusakabe} {et~al.}(2015){Kusakabe}, {Shimasaku}, {Nakajima}, \&
  {Ouchi}}]{kusakabe15}
{Kusakabe}, H., {Shimasaku}, K., {Nakajima}, K., \& {Ouchi}, M. 2015, \apjl,
  800, L29

\bibitem[{{Le F{\`e}vre} {et~al.}(2015){Le F{\`e}vre}, {Tasca}, {Cassata},
  {Garilli}, {Le Brun}, {Maccagni}, {Pentericci}, {Thomas}, {Vanzella},
  {Zamorani}, {Zucca}, {Amorin}, {Bardelli}, {Capak}, {Cassar{\`a}},
  {Castellano}, {Cimatti}, {Cuby}, {Cucciati}, {de la Torre}, {Durkalec},
  {Fontana}, {Giavalisco}, {Grazian}, {Hathi}, {Ilbert}, {Lemaux}, {Moreau},
  {Paltani}, {Ribeiro}, {Salvato}, {Schaerer}, {Scodeggio}, {Sommariva},
  {Talia}, {Taniguchi}, {Tresse}, {Vergani}, {Wang}, {Charlot}, {Contini},
  {Fotopoulou}, {L{\'o}pez-Sanjuan}, {Mellier}, \& {Scoville}}]{lefevre15}
{Le F{\`e}vre}, O., {Tasca}, L.~A.~M., {Cassata}, P., {et~al.} 2015, \aap, 576,
  A79

\bibitem[{Lewis \& Bridle(2002)}]{lewis02}
Lewis, A., \& Bridle, S. 2002, Phys. Rev. D, 66, 103511

\bibitem[{{Ly} {et~al.}(2007){Ly}, {Malkan}, {Kashikawa}, {Shimasaku}, {Doi},
  {Nagao}, {Iye}, {Kodama}, {Morokuma}, \& {Motohara}}]{ly07}
{Ly}, C., {Malkan}, M.~A., {Kashikawa}, N., {et~al.} 2007, \apj, 657, 738

\bibitem[{{Madau}(1995)}]{madau95}
{Madau}, P. 1995, \apj, 441, 18

\bibitem[{{Madau} \& {Dickinson}(2014)}]{madau14}
{Madau}, P., \& {Dickinson}, M. 2014, \araa, 52, 415

\bibitem[{{Malhotra} {et~al.}(2012){Malhotra}, {Rhoads}, {Finkelstein},
  {Hathi}, {Nilsson}, {McLinden}, \& {Pirzkal}}]{malhotra12}
{Malhotra}, S., {Rhoads}, J.~E., {Finkelstein}, S.~L., {et~al.} 2012, \apjl,
  750, L36

\bibitem[{{Mannucci} {et~al.}(2009){Mannucci}, {Cresci}, {Maiolino}, {Marconi},
  {Pastorini}, {Pozzetti}, {Gnerucci}, {Risaliti}, {Schneider}, {Lehnert}, \&
  {Salvati}}]{mannucci09}
{Mannucci}, F., {Cresci}, G., {Maiolino}, R., {et~al.} 2009, \mnras, 398, 1915

\bibitem[{{Matsuda} {et~al.}(2012){Matsuda}, {Yamada}, {Hayashino}, {Yamauchi},
  {Nakamura}, {Morimoto}, {Ouchi}, {Ono}, {Umemura}, \& {Mori}}]{matsuda12}
{Matsuda}, Y., {Yamada}, T., {Hayashino}, T., {et~al.} 2012, \mnras, 425, 878

\bibitem[{{Metropolis} {et~al.}(1953){Metropolis}, {Rosenbluth}, {Rosenbluth},
  {Teller}, \& {Teller}}]{metropolis53}
{Metropolis}, N., {Rosenbluth}, A.~W., {Rosenbluth}, M.~N., {Teller}, A.~H., \&
  {Teller}, E. 1953, \jcp, 21, 1087

\bibitem[{{Milvang-Jensen} {et~al.}(2012){Milvang-Jensen}, {Fynbo}, {Malesani},
  {Hjorth}, {Jakobsson}, \& {M{\o}ller}}]{milvangjensen12}
{Milvang-Jensen}, B., {Fynbo}, J.~P.~U., {Malesani}, D., {et~al.} 2012, \apj,
  756, 25

\bibitem[{{Momose} {et~al.}(2014){Momose}, {Ouchi}, {Nakajima}, {Ono},
  {Shibuya}, {Shimasaku}, {Yuma}, {Mori}, \& {Umemura}}]{momose14}
{Momose}, R., {Ouchi}, M., {Nakajima}, K., {et~al.} 2014, \mnras, 442, 110

\bibitem[{{Momose} {et~al.}(2015){Momose}, {Ouchi}, {Nakajima}, {Ono},
  {Shibuya}, {Shimasaku}, {Yuma}, {Mori}, \& {Umemura}}]{momose15}
---. 2015, ArXiv e-prints, arXiv:1509.09001

\bibitem[{{Moustakas} {et~al.}(2006){Moustakas}, {Kennicutt}, \&
  {Tremonti}}]{moustakas06}
{Moustakas}, J., {Kennicutt}, Jr., R.~C., \& {Tremonti}, C.~A. 2006, \apj, 642,
  775

\bibitem[{{Murphy} {et~al.}(2011){Murphy}, {Condon}, {Schinnerer}, {Kennicutt},
  {Calzetti}, {Armus}, {Helou}, {Turner}, {Aniano}, {Beir{\~a}o}, {Bolatto},
  {Brandl}, {Croxall}, {Dale}, {Donovan Meyer}, {Draine}, {Engelbracht},
  {Hunt}, {Hao}, {Koda}, {Roussel}, {Skibba}, \& {Smith}}]{murphy11}
{Murphy}, E.~J., {Condon}, J.~J., {Schinnerer}, E., {et~al.} 2011, \apj, 737,
  67

\bibitem[{{Nakajima} {et~al.}(2013){Nakajima}, {Ouchi}, {Shimasaku},
  {Hashimoto}, {Ono}, \& {Lee}}]{nakajima13}
{Nakajima}, K., {Ouchi}, M., {Shimasaku}, K., {et~al.} 2013, \apj, 769, 3

\bibitem[{{Nakajima} {et~al.}(2012){Nakajima}, {Ouchi}, {Shimasaku}, {Ono},
  {Lee}, {Foucaud}, {Ly}, {Dale}, {Salim}, {Finn}, {Almaini}, \&
  {Okamura}}]{nakajima12}
---. 2012, \apj, 745, 12

\bibitem[{{Nilsson} {et~al.}(2011){Nilsson}, {{\"O}stlin}, {M{\o}ller},
  {M{\"o}ller-Nilsson}, {Tapken}, {Freudling}, \& {Fynbo}}]{nilsson11}
{Nilsson}, K.~K., {{\"O}stlin}, G., {M{\o}ller}, P., {et~al.} 2011, \aap, 529,
  A9

\bibitem[{Oliphant(2007)}]{numpy}
Oliphant, T. 2007, Computing in Science \& Engineering, 9, 10

\bibitem[{{Oteo} {et~al.}(2015){Oteo}, {Sobral}, {Ivison}, {Smail}, {Best},
  {Cepa}, \& {P{\'e}rez-Garc{\'{\i}}a}}]{oteo15}
{Oteo}, I., {Sobral}, D., {Ivison}, R.~J., {et~al.} 2015, \mnras, 452, 2018

\bibitem[{{Ouchi} {et~al.}(2010){Ouchi}, {Shimasaku}, {Furusawa}, {Saito},
  {Yoshida}, {Akiyama}, {Ono}, {Yamada}, {Ota}, {Kashikawa}, {Iye}, {Kodama},
  {Okamura}, {Simpson}, \& {Yoshida}}]{ouchi10}
{Ouchi}, M., {Shimasaku}, K., {Furusawa}, H., {et~al.} 2010, \apj, 723, 869

\bibitem[{Pedregosa {et~al.}(2011)Pedregosa, Varoquaux, Gramfort, Michel,
  Thirion, Grisel, Blondel, Prettenhofer, Weiss, Dubourg, Vanderplas, Passos,
  Cournapeau, Brucher, Perrot, \& Duchesnay}]{scikit-learn}
Pedregosa, F., Varoquaux, G., Gramfort, A., {et~al.} 2011, Journal of Machine
  Learning Research, 12, 2825

\bibitem[{{Peng} {et~al.}(2002){Peng}, {Ho}, {Impey}, \& {Rix}}]{peng02}
{Peng}, C.~Y., {Ho}, L.~C., {Impey}, C.~D., \& {Rix}, H.-W. 2002, \aj, 124, 266

\bibitem[{{Peng} {et~al.}(2010){Peng}, {Ho}, {Impey}, \& {Rix}}]{peng10}
---. 2010, \aj, 139, 2097

\bibitem[{{Peng} {et~al.}(2011){Peng}, {Ho}, {Impey}, \& {Rix}}]{peng11}
---. 2011, {GALFIT: Detailed Structural Decomposition of Galaxy Images},
  astrophysics Source Code Library, ascl:1104.010

\bibitem[{P\'erez \& Granger(2007)}]{ipython}
P\'erez, F., \& Granger, B.~E. 2007, Computing in Science and Engineering, 9,
  21

\bibitem[{{Price} {et~al.}(2014){Price}, {Kriek}, {Brammer}, {Conroy},
  {F{\"o}rster Schreiber}, {Franx}, {Fumagalli}, {Lundgren}, {Momcheva},
  {Nelson}, {Skelton}, {van Dokkum}, {Whitaker}, \& {Wuyts}}]{price14}
{Price}, S.~H., {Kriek}, M., {Brammer}, G.~B., {et~al.} 2014, \apj, 788, 86

\bibitem[{{Reddy} {et~al.}(2010){Reddy}, {Erb}, {Pettini}, {Steidel}, \&
  {Shapley}}]{reddy10}
{Reddy}, N.~A., {Erb}, D.~K., {Pettini}, M., {Steidel}, C.~C., \& {Shapley},
  A.~E. 2010, \apj, 712, 1070

\bibitem[{{Rhoads} {et~al.}(2014){Rhoads}, {Malhotra}, {Richardson},
  {Finkelstein}, {Fynbo}, {McLinden}, \& {Tilvi}}]{rhoads14}
{Rhoads}, J.~E., {Malhotra}, S., {Richardson}, M.~L.~A., {et~al.} 2014, \apj,
  780, 20

\bibitem[{{Rodighiero} {et~al.}(2011){Rodighiero}, {Daddi}, {Baronchelli},
  {Cimatti}, {Renzini}, {Aussel}, {Popesso}, {Lutz}, {Andreani}, {Berta},
  {Cava}, {Elbaz}, {Feltre}, {Fontana}, {F{\"o}rster Schreiber},
  {Franceschini}, {Genzel}, {Grazian}, {Gruppioni}, {Ilbert}, {Le Floch},
  {Magdis}, {Magliocchetti}, {Magnelli}, {Maiolino}, {McCracken}, {Nordon},
  {Poglitsch}, {Santini}, {Pozzi}, {Riguccini}, {Tacconi}, {Wuyts}, \&
  {Zamorani}}]{rodighiero11}
{Rodighiero}, G., {Daddi}, E., {Baronchelli}, I., {et~al.} 2011, \apjl, 739,
  L40

\bibitem[{{Salpeter}(1955)}]{salpeter55}
{Salpeter}, E.~E. 1955, \apj, 121, 161

\bibitem[{{Sandberg} {et~al.}(2015){Sandberg}, {Guaita}, {{\"O}stlin}, {Hayes},
  \& {Kiaeerad}}]{sandberg15}
{Sandberg}, A., {Guaita}, L., {{\"O}stlin}, G., {Hayes}, M., \& {Kiaeerad}, F.
  2015, \aap, 580, A91

\bibitem[{{Schaerer} \& {de Barros}(2009)}]{schaerer09}
{Schaerer}, D., \& {de Barros}, S. 2009, \aap, 502, 423

\bibitem[{{Schaerer} {et~al.}(2011){Schaerer}, {Hayes}, {Verhamme}, \&
  {Teyssier}}]{schaerer11}
{Schaerer}, D., {Hayes}, M., {Verhamme}, A., \& {Teyssier}, R. 2011, \aap, 531,
  A12

\bibitem[{{Schulze} {et~al.}(2015){Schulze}, {Chapman}, {Hjorth}, {Levan},
  {Jakobsson}, {Bj{\"o}rnsson}, {Perley}, {Kr{\"u}hler}, {Gorosabel}, {Tanvir},
  {de Ugarte Postigo}, {Fynbo}, {Milvang-Jensen}, {M{\o}ller}, \&
  {Watson}}]{schulze15}
{Schulze}, S., {Chapman}, R., {Hjorth}, J., {et~al.} 2015, \apj, 808, 73

\bibitem[{{Scoville} {et~al.}(2007){Scoville}, {Aussel}, {Brusa}, {Capak},
  {Carollo}, {Elvis}, {Giavalisco}, {Guzzo}, {Hasinger}, {Impey}, {Kneib},
  {LeFevre}, {Lilly}, {Mobasher}, {Renzini}, {Rich}, {Sanders}, {Schinnerer},
  {Schminovich}, {Shopbell}, {Taniguchi}, \& {Tyson}}]{COSMOS}
{Scoville}, N., {Aussel}, H., {Brusa}, M., {et~al.} 2007, \apjs, 172, 1

\bibitem[{{S{\'e}rsic}(1963)}]{sersic63}
{S{\'e}rsic}, J.~L. 1963, Boletin de la Asociacion Argentina de Astronomia La
  Plata Argentina, 6, 41

\bibitem[{{Shibuya} {et~al.}(2014{\natexlab{a}}){Shibuya}, {Ouchi}, {Nakajima},
  {Yuma}, {Hashimoto}, {Shimasaku}, {Mori}, \& {Umemura}}]{shibuya14a}
{Shibuya}, T., {Ouchi}, M., {Nakajima}, K., {et~al.} 2014{\natexlab{a}}, \apj,
  785, 64

\bibitem[{{Shibuya} {et~al.}(2014{\natexlab{b}}){Shibuya}, {Ouchi}, {Nakajima},
  {Hashimoto}, {Ono}, {Rauch}, {Gauthier}, {Shimasaku}, {Goto}, {Mori}, \&
  {Umemura.}}]{shibuya14b}
---. 2014{\natexlab{b}}, \apj, 788, 74

\bibitem[{{Shivaei} {et~al.}(2015){Shivaei}, {Reddy}, {Steidel}, \&
  {Shapley}}]{shivaei15}
{Shivaei}, I., {Reddy}, N.~A., {Steidel}, C.~C., \& {Shapley}, A.~E. 2015,
  \apj, 804, 149

\bibitem[{{Skelton} {et~al.}(2014){Skelton}, {Whitaker}, {Momcheva}, {Brammer},
  {van Dokkum}, {Labb{\'e}}, {Franx}, {van der Wel}, {Bezanson}, {Da Cunha},
  {Fumagalli}, {F{\"o}rster Schreiber}, {Kriek}, {Leja}, {Lundgren}, {Magee},
  {Marchesini}, {Maseda}, {Nelson}, {Oesch}, {Pacifici}, {Patel}, {Price},
  {Rix}, {Tal}, {Wake}, \& {Wuyts}}]{skelton14}
{Skelton}, R.~E., {Whitaker}, K.~E., {Momcheva}, I.~G., {et~al.} 2014, \apjs,
  214, 24

\bibitem[{Smirnov(1948)}]{smirnov48}
Smirnov, N. 1948, The Annals of Mathematical Statistics, 19, 279

\bibitem[{{Song} {et~al.}(2014){Song}, {Finkelstein}, {Gebhardt}, {Hill},
  {Drory}, {Ashby}, {Blanc}, {Bridge}, {Chonis}, {Ciardullo}, {Fabricius},
  {Fazio}, {Gawiser}, {Gronwall}, {Hagen}, {Huang}, {Jogee}, {Livermore},
  {Salmon}, {Schneider}, {Willner}, \& {Zeimann}}]{song14}
{Song}, M., {Finkelstein}, S.~L., {Gebhardt}, K., {et~al.} 2014, \apj, 791, 3

\bibitem[{{Speagle} {et~al.}(2014){Speagle}, {Steinhardt}, {Capak}, \&
  {Silverman}}]{speagle14}
{Speagle}, J.~S., {Steinhardt}, C.~L., {Capak}, P.~L., \& {Silverman}, J.~D.
  2014, \apjs, 214, 15

\bibitem[{{Steidel} {et~al.}(1996{\natexlab{a}}){Steidel}, {Giavalisco},
  {Dickinson}, \& {Adelberger}}]{steidel96b}
{Steidel}, C.~C., {Giavalisco}, M., {Dickinson}, M., \& {Adelberger}, K.~L.
  1996{\natexlab{a}}, \aj, 112, 352

\bibitem[{{Steidel} {et~al.}(1996{\natexlab{b}}){Steidel}, {Giavalisco},
  {Pettini}, {Dickinson}, \& {Adelberger}}]{steidel96}
{Steidel}, C.~C., {Giavalisco}, M., {Pettini}, M., {Dickinson}, M., \&
  {Adelberger}, K.~L. 1996{\natexlab{b}}, \apjl, 462, L17

\bibitem[{{Storey} \& {Zeippen}(2000)}]{storey2000}
{Storey}, P.~J., \& {Zeippen}, C.~J. 2000, \mnras, 312, 813

\bibitem[{{Vargas} {et~al.}(2014){Vargas}, {Bish}, {Acquaviva}, {Gawiser},
  {Finkelstein}, {Ciardullo}, {Ashby}, {Feldmeier}, {Ferguson}, {Gronwall},
  {Guaita}, {Hagen}, {Koekemoer}, {Kurczynski}, {Newman}, \&
  {Padilla}}]{vargas14}
{Vargas}, C.~J., {Bish}, H., {Acquaviva}, V., {et~al.} 2014, \apj, 783, 26

\bibitem[{{Verhamme} {et~al.}(2012){Verhamme}, {Dubois}, {Blaizot}, {Garel},
  {Bacon}, {Devriendt}, {Guiderdoni}, \& {Slyz}}]{verhamme12}
{Verhamme}, A., {Dubois}, Y., {Blaizot}, J., {et~al.} 2012, \aap, 546, A111

\bibitem[{{Verhamme} {et~al.}(2006){Verhamme}, {Schaerer}, \&
  {Maselli}}]{verhamme06}
{Verhamme}, A., {Schaerer}, D., \& {Maselli}, A. 2006, \aap, 460, 397

\bibitem[{{Weiner} {et~al.}(2014)}]{weiner14}
{Weiner}, B.~J., {et~al.} 2014, in American Astronomical Society Meeting
  Abstracts, Vol. 223, American Astronomical Society Meeting Abstracts \#223,
  \#227.07

\bibitem[{{Weinzirl} {et~al.}(2009){Weinzirl}, {Jogee}, {Khochfar}, {Burkert},
  \& {Kormendy}}]{weinzirl09}
{Weinzirl}, T., {Jogee}, S., {Khochfar}, S., {Burkert}, A., \& {Kormendy}, J.
  2009, \apj, 696, 411

\bibitem[{{Weinzirl} {et~al.}(2011){Weinzirl}, {Jogee}, {Conselice},
  {Papovich}, {Chary}, {Bluck}, {Gr{\"u}tzbauch}, {Buitrago}, {Mobasher},
  {Lucas}, {Dickinson}, \& {Bauer}}]{weinzirl11}
{Weinzirl}, T., {Jogee}, S., {Conselice}, C.~J., {et~al.} 2011, \apj, 743, 87

\bibitem[{{Whitaker} {et~al.}(2014){Whitaker}, {Franx}, {Leja}, {van Dokkum},
  {Henry}, {Skelton}, {Fumagalli}, {Momcheva}, {Brammer}, {Labb{\'e}},
  {Nelson}, \& {Rigby}}]{whitaker14}
{Whitaker}, K.~E., {Franx}, M., {Leja}, J., {et~al.} 2014, \apj, 795, 104

\bibitem[{{Wisotzki} {et~al.}(2015){Wisotzki}, {Bacon}, {Blaizot},
  {Brinchmann}, {Herenz}, {Schaye}, {Bouch{\'e}}, {Cantalupo}, {Contini},
  {Carollo}, {Caruana}, {Courbot}, {Emsellem}, {Kamann}, {Kerutt}, {Leclercq},
  {Lilly}, {Patr{\'{\i}}cio}, {Sandin}, {Steinmetz}, {Straka}, {Urrutia},
  {Verhamme}, {Weilbacher}, \& {Wendt}}]{wisotzki15}
{Wisotzki}, L., {Bacon}, R., {Blaizot}, J., {et~al.} 2015, ArXiv e-prints,
  arXiv:1509.05143

\bibitem[{{Wuyts} {et~al.}(2013){Wuyts}, {F{\"o}rster Schreiber}, {Nelson},
  {van Dokkum}, {Brammer}, {Chang}, {Faber}, {Ferguson}, {Franx}, {Fumagalli},
  {Genzel}, {Grogin}, {Kocevski}, {Koekemoer}, {Lundgren}, {Lutz}, {McGrath},
  {Momcheva}, {Rosario}, {Skelton}, {Tacconi}, {van der Wel}, \&
  {Whitaker}}]{wuyts13}
{Wuyts}, S., {F{\"o}rster Schreiber}, N.~M., {Nelson}, E.~J., {et~al.} 2013,
  \apj, 779, 135

\bibitem[{{Yajima} {et~al.}(2012){Yajima}, {Li}, {Zhu}, {Abel}, {Gronwall}, \&
  {Ciardullo}}]{yajima12}
{Yajima}, H., {Li}, Y., {Zhu}, Q., {et~al.} 2012, \apj, 754, 118

\bibitem[{{Zeimann} {et~al.}(2014){Zeimann}, {Ciardullo}, {Gebhardt},
  {Gronwall}, {Schneider}, {Hagen}, {Bridge}, {Feldmeier}, \&
  {Trump}}]{zeimann14}
{Zeimann}, G.~R., {Ciardullo}, R., {Gebhardt}, H., {et~al.} 2014, \apj, 790,
  113

\bibitem[{{Zeimann} {et~al.}(2015){Zeimann}, {Ciardullo}, {Gronwall}, {Bridge},
  {Brooks}, {Fox}, {Gawiser}, {Gebhardt}, {Hagen}, {Schneider}, \&
  {Trump}}]{zeimann15b}
{Zeimann}, G.~R., {Ciardullo}, R., {Gronwall}, C., {et~al.} 2015, \apj, in
  press, arXiv:1511.00651

\end{thebibliography}

\begin{deluxetable}{cccccccccccc}
\tablewidth{0pt}
\tablecaption{LAE Physical and Morphological Properties \label{table:lae}}
\rotate
\tabletypesize{\tiny}
\tablehead{\colhead{RA} &\colhead{Dec} &\colhead{Log $M$} & \colhead{Half Light} &\colhead{$\beta$} & \colhead{Log} &\colhead{Ellipticity} &\colhead{Nearest} &\colhead{$\Sigma_{\rm SFR}$} &\colhead{Log sSFR} &\colhead{[O~III] Lum.} & \colhead{[O~III] EW} \\
           \colhead{(2000)} &\colhead{(2000)} &\colhead{($M_{\odot}$)} &\colhead{(kpc)} &&                \colhead{SFR$_{\rm Cor}$} &  &\colhead{Neighbor (kpc)}&\colhead{($M_{\odot}$~yr$^{-1}$~kpc$^{-2}$)} &\colhead{(yr$^{-1}$)} & \colhead{$10^{41}$erg/s} &\colhead{\AA} }
\startdata
150.05908 & 2.24059 &$ 8.52\pm0.02 $&$ 2.10\pm0.02 $&$ -2.02\pm0.05 $&$ 0.99\pm0.05 $&$ 0.47\pm0.03 $& 6.35 &$ 0.35\pm0.04 $&$ -7.52\pm0.05 $&$ 25.19\pm2.96 $&$ 275\pm35 $ \\
150.07777 & 2.25000 &$ 8.79\pm0.13 $&$ 1.89\pm1.02 $&$ 0.01\pm0.89 $&$ 1.62\pm0.83 $&$ 0.30\pm0.08 $& 26.45 &$ 1.84\pm10.75 $&$ -7.17\pm0.84 $&$ 9.18\pm12.02 $&$ 122\pm160 $ \\
150.09915 & 2.21952 &$ 7.32\pm0.01 $&$ 0.57\pm0.00 $&$ -2.50\pm0.24 $&$ -0.22\pm0.23 $&$ 0.31\pm0.04 $& 18.99 &$ 0.29\pm0.20 $&$ -7.54\pm0.23 $&$ 10.09\pm1.81 $&$ 274\pm50 $ \\
150.11349 & 2.29202 &$ 8.67\pm0.10 $&$ 0.69\pm0.00 $&$ -2.08\pm0.07 $&$ 0.71\pm0.06 $&$ 0.14\pm0.02 $& 8.26 &$ 1.71\pm0.28 $&$ -7.96\pm0.12 $&$ 29.46\pm3.28 $&$ 324\pm39 $ \\
150.14150 & 2.22104 &$ 9.06\pm0.03 $&$ 2.13\pm0.01 $&$ -2.30\pm0.05 $&$ 0.68\pm0.05 $&$ 0.28\pm0.03 $& 2.54 &$ 0.17\pm0.02 $&$ -8.38\pm0.06 $&$ 16.07\pm4.26 $&$ 22\pm5 $ \\
150.17009 & 2.30639 &$ 10.15\pm0.00 $&$ 2.34\pm0.03 $&$ -1.63\pm0.01 $&$ 1.93\pm0.01 $&$ 0.21\pm0.01 $& 5.77 &$ 2.44\pm0.08 $&$ -8.22\pm0.01 $&$ 21.02\pm3.20 $&$ 351\pm56 $ \\
189.18378 & 62.23608 &$ 9.51\pm0.02 $& -- &$ -2.08\pm0.09 $&$ 0.91\pm0.09 $& -- & -- & -- &$ -8.59\pm0.09 $&$ 133.73\pm2.68 $&$ 231\pm12 $ \\
189.20890 & 62.23363 &$ 9.87\pm0.02 $& -- &$ -1.49\pm0.10 $&$ 1.35\pm0.09 $& -- & -- & -- &$ -8.52\pm0.09 $&$ 8.47\pm15.28 $&$ 86\pm156 $ \\
189.22592 & 62.22660 &$ 7.71\pm0.07 $& -- &$ -2.58\pm0.59 $&$ -0.21\pm0.55 $& -- & -- & -- &$ -7.92\pm0.55 $&$ 22.46\pm3.23 $&$ 94\pm14 $ \\
189.26814 & 62.24616 &$ 10.51\pm0.01 $&$ 1.07\pm0.01 $&$ -1.76\pm0.10 $&$ 1.17\pm0.09 $&$ 0.25\pm0.02 $& 27.05 &$ 2.08\pm0.49 $&$ -9.33\pm0.09 $&$ 34.51\pm3.11 $&$ 87\pm8 $ \\
189.29592 & 62.19447 &$ 8.73\pm0.10 $& -- &$ -2.04\pm0.14 $&$ 0.81\pm0.13 $& -- & -- & -- &$ -7.92\pm0.17 $&$ 8.93\pm2.04 $&$ 94\pm22 $ \\
53.09089 & -27.94874 &$ 9.12\pm0.06 $&$ 1.69\pm0.09 $&$ -1.73\pm0.18 $&$ 0.91\pm0.17 $&$ 0.29\pm0.02 $& 19.91 &$ 0.45\pm0.21 $&$ -8.21\pm0.18 $&$ 6.10\pm1.99 $&$ 152\pm50 $ \\
53.12923 & -27.91744 &$ 7.73\pm0.04 $&$ 1.09\pm0.01 $&$ -2.47\pm0.56 $&$ -0.14\pm0.52 $&$ 0.25\pm0.03 $& 7.48 &$ 0.10\pm0.22 $&$ -7.87\pm0.52 $&$ 7.96\pm1.86 $&$ 51\pm12 $ \\
53.18964 & -27.89532 &$ 9.17\pm0.04 $&$ 1.11\pm0.01 $&$ -1.55\pm0.22 $&$ 0.94\pm0.20 $&$ 0.33\pm0.01 $& 15.00 &$ 1.10\pm0.66 $&$ -8.24\pm0.21 $&$ 31.75\pm4.43 $&$ 86\pm12 $ \\
53.06526 & -27.87589 &$ 7.41\pm0.04 $&$ 0.68\pm0.00 $&$ -1.48\pm0.86 $&$ 0.39\pm0.80 $&$ 0.07\pm0.03 $& 18.72 &$ 0.84\pm4.49 $&$ -7.02\pm0.80 $&$ 56.17\pm4.55 $&$ 552\pm52 $ \\
53.25288 & -27.85100 &$ 8.35\pm0.17 $&$ 1.10\pm0.03 $&$ -2.12\pm0.41 $&$ 0.19\pm0.38 $& -- & 13.20 &$ 0.20\pm0.28 $&$ -8.17\pm0.42 $&$ 13.24\pm1.84 $&$ 286\pm42 $ \\
53.08466 & -27.85090 &$ 8.97\pm0.10 $&$ 1.08\pm0.02 $&$ -1.74\pm0.37 $&$ 0.51\pm0.35 $&$ 0.50\pm0.01 $& 11.24 &$ 0.44\pm0.54 $&$ -8.46\pm0.36 $&$ 10.97\pm1.21 $&$ 271\pm32 $ \\
53.04775 & -27.84069 &$ 8.20\pm0.02 $&$ 0.81\pm0.02 $&$ -2.33\pm0.13 $&$ 0.34\pm0.12 $&$ 0.05\pm0.01 $& 9.78 &$ 0.54\pm0.17 $&$ -7.86\pm0.12 $&$ 7.65\pm1.82 $&$ 316\pm76 $ \\
53.17574 & -27.81654 &$ 7.88\pm0.02 $&$ 0.97\pm0.01 $&$ -1.60\pm0.36 $&$ 0.66\pm0.34 $&$ 0.14\pm0.04 $& 6.03 &$ 0.76\pm0.91 $&$ -7.22\pm0.34 $&$ 54.42\pm5.70 $&$ 82\pm9 $ \\
53.16744 & -27.78179 &$ 7.87\pm0.05 $&$ 1.16\pm0.02 $&$ -0.92\pm0.88 $&$ 0.92\pm0.82 $&$ 0.23\pm0.03 $& 13.97 &$ 0.98\pm5.42 $&$ -6.95\pm0.82 $&$ 20.83\pm6.94 $&$ 351\pm118 $ \\
53.22010 & -27.76466 &$ 9.44\pm0.02 $&$ 1.03\pm0.02 $&$ -0.17\pm0.14 $&$ 2.28\pm0.13 $&$ 0.10\pm0.01 $& 16.98 &$ 28.91\pm10.06 $&$ -7.16\pm0.13 $&$ 63.25\pm12.76 $&$ 56\pm11 $ \\
53.01713 & -27.75076 &$ 8.10\pm0.04 $&$ 1.25\pm0.02 $&$ -1.47\pm0.34 $&$ 0.77\pm0.32 $&$ 0.45\pm0.02 $& 2.32 &$ 0.60\pm0.64 $&$ -7.34\pm0.32 $&$ 12.93\pm3.14 $&$ 100\pm24 $ \\
53.01662 & -27.73105 &$ 8.25\pm0.16 $&$ 0.69\pm0.00 $&$ -2.03\pm0.58 $&$ 0.11\pm0.54 $&$ 1.00\pm 0.1 $& 9.64 &$ 0.43\pm1.06 $&$ -8.14\pm0.57 $&$ 7.96\pm1.93 $&$ 100\pm24 $ \\
53.18055 & -27.72649 &$ 8.73\pm0.08 $&$ 1.48\pm0.02 $&$ -1.76\pm0.34 $&$ 0.43\pm0.32 $&$ 0.50\pm0.03 $& 1.70 &$ 0.20\pm0.21 $&$ -8.31\pm0.33 $&$ 106.39\pm10.33 $&$ 132\pm14 $ \\
53.05734 & -27.71681 &$ 8.45\pm0.26 $&$ 1.70\pm0.02 $&$ -2.18\pm0.35 $&$ 0.25\pm0.33 $&$ 0.69\pm0.01 $& 2.71 &$ 0.10\pm0.11 $&$ -8.19\pm0.42 $& -- & --  \\
53.27985 & -27.68197 &$ 8.81\pm0.14 $&$ 2.67\pm0.03 $&$ -1.67\pm0.43 $&$ 0.65\pm0.40 $&$ 0.57\pm0.05 $& 5.77 &$ 0.10\pm0.15 $&$ -8.16\pm0.42 $& -- & --  \\
53.25242 & -27.67421 &$ 9.01\pm0.07 $& -- &$ -1.42\pm0.24 $&$ 1.03\pm0.23 $& -- & -- & -- &$ -7.98\pm0.24 $& -- & --  \\
53.14147 & -27.67202 &$ 8.73\pm0.06 $&$ 2.56\pm0.09 $&$ -1.17\pm0.28 $&$ 1.22\pm0.26 $&$ 0.32\pm0.03 $& 3.45 &$ 0.40\pm0.33 $&$ -7.52\pm0.27 $& -- & --  \\
\enddata
\tablecomments{Units on all the quantities are shown in the table headers. Log $M$ denotes the $\log_{10}$ stellar mass in $M_\odot$ assuming a Salpeter IMF, constant star formation rate, a \citet{calzetti01} extinction law, and CB07 stellar population models. The half-light radius, in units of kpc, is measured in the rest-frame NUV using HST F814W images. $\beta$ is the UV spectral slope as parametrized in Equation \ref{eqn:beta}. Log SFR$_{\rm Cor}$ is the $\log_{10}$  star formation rate in units of $M_\odot$/yr corrected using the \citet{calzetti01} extinction law with reddening measured from $\beta$. Projected nearest neighbor distance is measured in kpc, using the redshift of the galaxy in question. $\Sigma_{\rm SFR}$ is dust corrected star formation rate surface density measured in units of $M_{\odot}$~yr$^{-1}$~kpc$^{-2}$. Log sSFR shows the $\log_{10}$ specific (dust-corrected) star formation rate. [O~III] luminosity has also been corrected for extinction using the
\citet{calzetti01} law; the [O~III] EW is in the rest frame, and is not dust corrected.}
\end{deluxetable}

\begin{deluxetable}{cccccccccccc}
\tablewidth{0pt}
\tablecaption{oELG Physical and Morphological Properties \label{table:oelg}}
\rotate
\tabletypesize{\tiny}
\tablehead{\colhead{RA} &\colhead{Dec} &\colhead{Log $M$} & \colhead{Half Light} &\colhead{$\beta$} & \colhead{Log} &\colhead{Ellipticity} &\colhead{Nearest} &\colhead{$\Sigma_{\rm SFR}$} &\colhead{Log sSFR} &\colhead{[O~III] Lum.} & \colhead{[O~III] EW} \\
           \colhead{(2000)} &\colhead{(2000)} &\colhead{($M_{\odot}$)} &\colhead{(kpc)} &&                \colhead{SFR$_{\rm Cor}$} &  &\colhead{Neighbor (kpc)}&\colhead{($M_{\odot}$~yr$^{-1}$~kpc$^{-2}$)} &\colhead{(yr$^{-1}$)} & \colhead{$10^{41}$erg/s} &\colhead{\AA} }
\startdata 
150.06848 & 2.38355 &$ 10.18\pm0.35 $&$ 1.21\pm0.01 $&$ -1.00\pm0.02 $&$ 2.18\pm0.02 $&$ 0.11\pm0.01 $& 5.85 &$ 16.45\pm0.84 $&$ -8.00\pm0.35 $&$ 191.70\pm8.38 $&$ 87\pm5 $ \\
150.08272 & 2.38670 &$ 9.32\pm0.03 $&$ 2.73\pm0.07 $&$ -2.25\pm0.06 $&$ 0.70\pm0.06 $&$ 0.75\pm0.01 $& 4.72 &$ 0.11\pm0.02 $&$ -8.62\pm0.06 $&$ 25.43\pm3.79 $&$ 199\pm31 $ \\
150.14810 & 2.21751 &$ 9.70\pm0.01 $&$ 1.46\pm0.02 $&$ -1.25\pm0.06 $&$ 1.48\pm0.05 $&$ 0.06\pm0.02 $& 13.41 &$ 2.23\pm0.30 $&$ -8.22\pm0.05 $&$ 54.65\pm7.64 $&$ 120\pm17 $ \\
150.07210 & 2.32402 &$ 9.58\pm0.39 $&$ 1.54\pm0.03 $&$ -1.54\pm0.06 $&$ 1.36\pm0.06 $&$ 0.38\pm0.01 $& 11.69 &$ 1.51\pm0.21 $&$ -8.23\pm0.40 $&$ 19.24\pm4.71 $&$ 43\pm10 $ \\
150.08967 & 2.33157 &$ 9.52\pm0.01 $&$ 1.27\pm0.02 $&$ -1.62\pm0.07 $&$ 1.14\pm0.06 $&$ 0.41\pm0.01 $& 12.89 &$ 1.36\pm0.22 $&$ -8.39\pm0.06 $&$ 21.90\pm3.09 $&$ 86\pm12 $ \\
150.06555 & 2.34114 &$ 8.25\pm0.09 $&$ 0.95\pm0.04 $&$ -1.64\pm0.15 $&$ 0.73\pm0.14 $&$ 0.42\pm0.03 $& 20.75 &$ 0.94\pm0.37 $&$ -7.53\pm0.17 $&$ 22.52\pm2.40 $&$ 263\pm31 $ \\
150.19652 & 2.29569 &$ 9.99\pm0.00 $&$ 1.75\pm0.01 $&$ -1.62\pm0.02 $&$ 1.70\pm0.02 $&$ 0.48\pm0.01 $& 22.14 &$ 2.59\pm0.11 $&$ -8.29\pm0.02 $&$ 101.67\pm5.19 $&$ 116\pm8 $ \\
150.17356 & 2.30004 &$ 9.94\pm0.09 $&$ 1.16\pm0.01 $&$ -1.39\pm0.03 $&$ 1.73\pm0.03 $&$ 0.04\pm0.01 $& 4.40 &$ 6.27\pm0.38 $&$ -8.21\pm0.10 $&$ 85.82\pm5.69 $&$ 101\pm8 $ \\
150.17881 & 2.30712 &$ 9.48\pm0.02 $&$ 1.05\pm0.01 $&$ -1.90\pm0.04 $&$ 1.07\pm0.04 $&$ 0.23\pm0.01 $& 12.53 &$ 1.69\pm0.17 $&$ -8.41\pm0.04 $&$ 12.71\pm2.63 $&$ 50\pm10 $ \\
150.16102 & 2.31007 &$ 8.62\pm0.04 $&$ 1.17\pm0.03 $&$ -1.79\pm0.08 $&$ 0.80\pm0.07 $&$ 0.56\pm0.01 $& 21.72 &$ 0.73\pm0.14 $&$ -7.82\pm0.09 $&$ 31.70\pm5.60 $&$ 277\pm50 $ \\
150.17587 & 2.31348 &$ 9.39\pm0.02 $&$ 1.31\pm0.03 $&$ -1.85\pm0.08 $&$ 0.91\pm0.07 $&$ 0.43\pm0.01 $& 13.89 &$ 0.76\pm0.14 $&$ -8.48\pm0.08 $&$ 33.27\pm5.15 $&$ 172\pm28 $ \\
150.09134 & 2.26111 &$ 8.34\pm0.11 $&$ 1.35\pm0.03 $&$ -2.46\pm0.11 $&$ 0.22\pm0.10 $&$ 0.24\pm0.02 $& 19.70 &$ 0.14\pm0.04 $&$ -8.12\pm0.15 $&$ 14.77\pm2.04 $&$ 209\pm30 $ \\
150.08623 & 2.21483 &$ 9.42\pm0.05 $&$ 0.93\pm0.03 $&$ -1.55\pm0.11 $&$ 0.96\pm0.11 $&$ 0.43\pm0.02 $& 22.25 &$ 1.68\pm0.47 $&$ -8.46\pm0.12 $&$ 52.18\pm4.04 $&$ 335\pm30 $ \\
150.09914 & 2.21955 &$ 7.32\pm0.01 $&$ 0.56\pm0.03 $&$ -2.50\pm0.24 $&$ -0.22\pm0.23 $&$ 0.31\pm0.04 $& 18.98 &$ 0.30\pm0.21 $&$ -7.54\pm0.23 $&$ 10.09\pm1.81 $&$ 274\pm50 $ \\
150.07175 & 2.22722 &$ 9.30\pm0.04 $&$ 2.15\pm0.05 $&$ -1.67\pm0.09 $&$ 1.09\pm0.08 $&$ 0.70\pm0.01 $& 4.07 &$ 0.42\pm0.09 $&$ -8.22\pm0.09 $&$ 27.06\pm3.75 $&$ 111\pm16 $ \\
150.07997 & 2.23517 &$ 9.33\pm0.03 $&$ 1.26\pm0.02 $&$ -1.86\pm0.05 $&$ 1.06\pm0.05 $&$ 0.32\pm0.02 $& 10.48 &$ 1.16\pm0.14 $&$ -8.27\pm0.06 $&$ 47.72\pm3.84 $&$ 193\pm18 $ \\
150.10841 & 2.43971 &$ 9.85\pm0.02 $&$ 1.29\pm0.01 $&$ -1.69\pm0.02 $&$ 1.51\pm0.02 $&$ 0.30\pm0.01 $& 24.08 &$ 3.12\pm0.17 $&$ -8.34\pm0.03 $&$ 59.61\pm5.77 $&$ 93\pm10 $ \\
150.10192 & 2.44912 &$ 9.37\pm0.08 $&$ 2.38\pm0.07 $&$ -1.73\pm0.09 $&$ 1.04\pm0.08 $&$ 0.84\pm0.01 $& 1.76 &$ 0.31\pm0.06 $&$ -8.33\pm0.11 $&$ 33.10\pm6.65 $&$ 101\pm20 $ \\
150.09980 & 2.44953 &$ 8.89\pm0.10 $&$ 1.46\pm0.04 $&$ -1.87\pm0.08 $&$ 0.79\pm0.08 $&$ 0.44\pm0.01 $& 28.53 &$ 0.46\pm0.09 $&$ -8.10\pm0.13 $&$ 29.20\pm5.87 $&$ 101\pm20 $ \\
150.17698 & 2.33446 &$ 9.55\pm0.02 $&$ 1.34\pm0.02 $&$ -1.60\pm0.05 $&$ 1.33\pm0.04 $&$ 0.40\pm0.01 $& 27.75 &$ 1.89\pm0.20 $&$ -8.22\pm0.05 $&$ 29.95\pm4.63 $&$ 79\pm12 $ \\
150.17731 & 2.33779 &$ 9.36\pm0.07 $&$ 1.55\pm0.04 $&$ -1.45\pm0.10 $&$ 1.12\pm0.09 $&$ 0.33\pm0.03 $& 13.03 &$ 0.88\pm0.21 $&$ -8.24\pm0.11 $&$ 34.33\pm5.30 $&$ 79\pm12 $ \\
150.15133 & 2.47150 &$ 9.22\pm0.08 $&$ 1.19\pm0.09 $&$ -2.12\pm0.13 $&$ 0.46\pm0.12 $&$ 0.46\pm0.04 $& 24.35 &$ 0.32\pm0.11 $&$ -8.76\pm0.15 $&$ 20.43\pm2.87 $&$ 260\pm38 $ \\
150.09839 & 2.26593 &$ 9.32\pm0.04 $&$ 1.16\pm0.01 $&$ -1.57\pm0.05 $&$ 1.29\pm0.05 $&$ 0.55\pm0.01 $& 25.81 &$ 2.33\pm0.28 $&$ -8.03\pm0.07 $&$ 47.29\pm7.30 $&$ 149\pm24 $ \\
150.12323 & 2.28411 &$ 9.56\pm0.01 $&$ 1.14\pm0.01 $&$ -1.86\pm0.03 $&$ 1.13\pm0.03 $&$ 0.32\pm0.01 $& 18.94 &$ 1.66\pm0.12 $&$ -8.43\pm0.03 $&$ 75.82\pm2.73 $&$ 276\pm17 $ \\
150.11352 & 2.29210 &$ 8.68\pm0.12 $&$ 0.69\pm0.02 $&$ -2.08\pm0.07 $&$ 0.71\pm0.06 $&$ 0.14\pm0.02 $& 8.26 &$ 1.71\pm0.28 $&$ -7.97\pm0.14 $&$ 29.46\pm3.28 $&$ 324\pm39 $ \\
150.09535 & 2.28722 &$ 9.03\pm0.08 $&$ 0.85\pm0.02 $&$ -1.71\pm0.10 $&$ 0.94\pm0.09 $&$ 0.23\pm0.02 $& 12.88 &$ 1.90\pm0.44 $&$ -8.09\pm0.12 $&$ 51.90\pm4.59 $&$ 274\pm27 $ \\
150.11090 & 2.28737 &$ 8.08\pm0.07 $&$ 1.09\pm0.03 $&$ -2.65\pm0.11 $&$ 0.04\pm0.11 $&$ 0.29\pm0.02 $& 22.78 &$ 0.15\pm0.04 $&$ -8.04\pm0.13 $&$ 11.85\pm3.20 $&$ 215\pm59 $ \\
150.12145 & 2.23102 &$ 9.32\pm0.02 $&$ 1.56\pm0.02 $&$ -1.96\pm0.05 $&$ 1.01\pm0.04 $&$ 0.32\pm0.01 $& 12.52 &$ 0.67\pm0.07 $&$ -8.31\pm0.05 $&$ 28.16\pm3.65 $&$ 131\pm18 $ \\
150.10091 & 2.23657 &$ 9.72\pm0.01 $&$ 1.12\pm0.01 $&$ -1.32\pm0.05 $&$ 1.52\pm0.05 $&$ 0.22\pm0.01 $& 29.13 &$ 4.15\pm0.51 $&$ -8.20\pm0.05 $&$ 45.40\pm6.82 $&$ 89\pm14 $ \\
150.10604 & 2.24106 &$ 8.56\pm0.16 $&$ 0.98\pm0.03 $&$ -2.17\pm0.16 $&$ 0.33\pm0.15 $&$ 0.11\pm0.05 $& 10.38 &$ 0.36\pm0.15 $&$ -8.23\pm0.22 $&$ 23.64\pm2.86 $&$ 329\pm43 $ \\
150.09754 & 2.24231 &$ 8.99\pm0.12 $&$ 0.96\pm0.03 $&$ -1.79\pm0.14 $&$ 0.65\pm0.13 $&$ 0.17\pm0.02 $& 30.10 &$ 0.77\pm0.28 $&$ -8.34\pm0.18 $&$ 29.85\pm2.73 $&$ 278\pm29 $ \\
150.09679 & 2.20854 &$ 8.76\pm0.11 $&$ 0.82\pm0.02 $&$ -2.45\pm0.11 $&$ 0.27\pm0.10 $&$ 0.29\pm0.03 $& 6.76 &$ 0.44\pm0.12 $&$ -8.49\pm0.15 $&$ 31.23\pm2.80 $&$ 386\pm39 $ \\
150.13862 & 2.37917 &$ 8.13\pm0.06 $&$ 0.96\pm0.02 $&$ -2.47\pm0.09 $&$ 0.19\pm0.08 $&$ 0.40\pm0.02 $& 15.72 &$ 0.27\pm0.06 $&$ -7.94\pm0.10 $&$ 14.25\pm2.03 $&$ 266\pm40 $ \\
150.17932 & 2.21977 &$ 9.40\pm0.02 $&$ 1.10\pm0.01 $&$ -2.22\pm0.04 $&$ 0.82\pm0.04 $&$ 0.09\pm0.02 $& 14.91 &$ 0.86\pm0.08 $&$ -8.58\pm0.04 $&$ 49.86\pm2.54 $&$ 303\pm21 $ \\
150.18666 & 2.23196 &$ 9.63\pm0.02 $&$ 1.76\pm0.01 $&$ -2.06\pm0.02 $&$ 1.34\pm0.02 $&$ 0.62\pm0.00 $& 10.99 &$ 1.12\pm0.04 $&$ -8.29\pm0.02 $&$ 12.46\pm2.21 $&$ 180\pm33 $ \\
150.18768 & 2.23139 &$ 8.56\pm0.19 $&$ 0.86\pm0.03 $&$ -2.13\pm0.15 $&$ 0.37\pm0.14 $&$ 0.32\pm0.02 $& 10.48 &$ 0.51\pm0.19 $&$ -8.19\pm0.24 $&$ 11.73\pm2.08 $&$ 180\pm33 $ \\
150.16302 & 2.24749 &$ 9.90\pm0.01 $&$ 2.21\pm0.02 $&$ -1.51\pm0.04 $&$ 1.57\pm0.03 $&$ 0.30\pm0.01 $& 24.23 &$ 1.22\pm0.10 $&$ -8.32\pm0.03 $&$ 47.05\pm5.01 $&$ 64\pm7 $ \\
150.18888 & 2.23938 &$ 7.79\pm0.05 $&$ 0.85\pm0.04 $&$ -2.45\pm0.18 $&$ -0.01\pm0.16 $&$ 0.16\pm0.04 $& 14.00 &$ 0.22\pm0.10 $&$ -7.80\pm0.17 $&$ 12.13\pm1.24 $&$ 275\pm31 $ \\
150.15533 & 2.32870 &$ 9.55\pm0.02 $&$ 0.91\pm0.02 $&$ -1.61\pm0.08 $&$ 1.02\pm0.08 $&$ 0.23\pm0.02 $& 10.69 &$ 2.02\pm0.41 $&$ -8.53\pm0.08 $&$ 25.48\pm4.18 $&$ 106\pm18 $ \\
150.13469 & 2.24804 &$ 9.19\pm0.02 $&$ 0.99\pm0.01 $&$ -2.37\pm0.06 $&$ 0.57\pm0.06 $&$ 0.12\pm0.01 $& 21.13 &$ 0.60\pm0.08 $&$ -8.62\pm0.06 $&$ 24.24\pm2.30 $&$ 217\pm23 $ \\
150.13585 & 2.25997 &$ 9.32\pm0.02 $&$ 1.79\pm0.03 $&$ -1.87\pm0.06 $&$ 1.04\pm0.05 $&$ 0.23\pm0.01 $& 27.82 &$ 0.55\pm0.07 $&$ -8.28\pm0.06 $&$ 38.96\pm4.53 $&$ 172\pm21 $ \\
150.14200 & 2.26510 &$ 10.70\pm0.00 $&$ 1.10\pm0.03 $&$ -1.81\pm0.13 $&$ 0.71\pm0.12 $&$ 0.14\pm0.05 $& 8.49 &$ 0.67\pm0.22 $&$ -9.99\pm0.12 $&$ 35.38\pm3.78 $&$ 71\pm8 $ \\
150.15425 & 2.29653 &$ 9.58\pm0.06 $&$ 1.36\pm0.05 $&$ -2.11\pm0.14 $&$ 0.30\pm0.13 $&$ 0.18\pm0.05 $& 2.45 &$ 0.17\pm0.06 $&$ -9.28\pm0.15 $&$ 12.22\pm1.29 $&$ 180\pm21 $ \\
150.11195 & 2.38845 &$ 9.69\pm0.01 $&$ 1.36\pm0.03 $&$ -0.98\pm0.11 $&$ 1.49\pm0.11 $&$ 0.55\pm0.01 $& 20.44 &$ 2.64\pm0.74 $&$ -8.20\pm0.11 $&$ 46.87\pm5.47 $&$ 111\pm14 $ \\
150.08148 & 2.29891 &$ 9.51\pm0.01 $&$ 2.59\pm0.06 $&$ -1.84\pm0.07 $&$ 1.00\pm0.07 $&$ 0.76\pm0.01 $& 19.42 &$ 0.24\pm0.04 $&$ -8.51\pm0.07 $&$ 38.44\pm5.47 $&$ 162\pm24 $ \\
150.08045 & 2.30717 &$ 8.98\pm0.03 $&$ 0.61\pm0.01 $&$ 1.58\pm0.43 $&$ 3.31\pm0.40 $&$ 0.13\pm0.03 $& 11.68 &$ 872.21\pm1331.72 $&$ -5.67\pm0.40 $&$ 619.98\pm147.39 $&$ 262\pm63 $ \\
150.05522 & 2.36939 &$ 9.60\pm0.01 $&$ 1.02\pm0.01 $&$ -1.97\pm0.02 $&$ 1.23\pm0.02 $&$ 0.15\pm0.01 $& 9.09 &$ 2.59\pm0.15 $&$ -8.37\pm0.02 $&$ 35.55\pm2.41 $&$ 114\pm9 $ \\
150.10872 & 2.41218 &$ 9.36\pm0.02 $&$ 1.08\pm0.03 $&$ -1.85\pm0.07 $&$ 0.94\pm0.06 $&$ 0.51\pm0.02 $& 6.20 &$ 1.19\pm0.18 $&$ -8.41\pm0.06 $&$ 21.96\pm3.95 $&$ 126\pm23 $ \\
53.24912 & -27.69392 &$ 7.94\pm0.03 $&$ 1.16\pm0.12 $&$ -2.48\pm0.32 $&$ -0.11\pm0.29 $&$ 0.35\pm0.06 $& 26.12 &$ 0.09\pm0.09 $&$ -8.05\pm0.29 $&$ 14.70\pm1.54 $&$ 332\pm38 $ \\
53.07186 & -27.82070 &$ 9.90\pm0.01 $&$ 1.87\pm0.01 $&$ -1.33\pm0.07 $&$ 1.47\pm0.07 $&$ 0.56\pm0.00 $& 16.25 &$ 1.34\pm0.22 $&$ -8.44\pm0.07 $&$ 44.05\pm6.40 $&$ 85\pm13 $ \\
53.13528 & -27.69762 &$ 9.83\pm0.01 $&$ 1.34\pm0.01 $&$ -1.57\pm0.06 $&$ 1.55\pm0.05 $&$ 0.34\pm0.01 $& 12.98 &$ 3.11\pm0.39 $&$ -8.28\pm0.05 $&$ 45.13\pm4.85 $&$ 78\pm9 $ \\
53.14214 & -27.83269 &$ 8.38\pm0.04 $&$ 1.24\pm0.02 $&$ -1.75\pm0.32 $&$ 0.61\pm0.30 $&$ 0.49\pm0.01 $& 8.02 &$ 0.42\pm0.41 $&$ -7.78\pm0.30 $&$ 37.81\pm2.95 $&$ 381\pm35 $ \\
53.22949 & -27.86480 &$ 9.48\pm0.07 $&$ 1.19\pm0.02 $&$ -1.60\pm0.26 $&$ 1.03\pm0.24 $&$ 0.31\pm0.01 $& 4.83 &$ 1.18\pm0.88 $&$ -8.45\pm0.25 $&$ 14.02\pm3.27 $&$ 51\pm12 $ \\
53.13376 & -27.80791 &$ 8.99\pm0.02 $&$ 1.46\pm0.02 $&$ -1.64\pm0.10 $&$ 0.85\pm0.10 $&$ 0.31\pm0.02 $& 4.46 &$ 0.54\pm0.13 $&$ -8.14\pm0.10 $&$ 28.28\pm2.23 $&$ 196\pm18 $ \\
53.14352 & -27.79522 &$ 8.61\pm0.03 $&$ 1.15\pm0.02 $&$ -2.00\pm0.13 $&$ 0.59\pm0.12 $&$ 0.36\pm0.03 $& 3.25 &$ 0.47\pm0.15 $&$ -8.01\pm0.12 $&$ 30.56\pm3.21 $&$ 381\pm44 $ \\
53.17578 & -27.81656 &$ 8.06\pm0.36 $&$ 0.97\pm0.03 $&$ -1.94\pm0.26 $&$ 0.36\pm0.25 $&$ 0.14\pm0.04 $& 6.02 &$ 0.39\pm0.30 $&$ -7.70\pm0.44 $&$ 15.49\pm2.15 $&$ 286\pm42 $ \\
53.13813 & -27.86549 &$ 9.29\pm0.02 $&$ 0.90\pm0.01 $&$ -2.01\pm0.08 $&$ 0.90\pm0.08 $&$ 0.17\pm0.01 $& 16.59 &$ 1.55\pm0.30 $&$ -8.39\pm0.08 $&$ 28.03\pm1.68 $&$ 192\pm15 $ \\
53.13007 & -27.84289 &$ 8.93\pm0.05 $&$ 1.07\pm0.01 $&$ -2.29\pm0.10 $&$ 0.57\pm0.10 $&$ 0.38\pm0.01 $& 6.99 &$ 0.52\pm0.13 $&$ -8.36\pm0.11 $&$ 27.79\pm1.50 $&$ 314\pm23 $ \\
53.15059 & -27.70789 &$ 8.84\pm0.04 $&$ 0.69\pm0.04 $&$ -1.92\pm0.16 $&$ 0.94\pm0.14 $&$ 0.32\pm0.03 $& 16.17 &$ 2.88\pm1.16 $&$ -7.91\pm0.15 $&$ 24.95\pm2.96 $&$ 236\pm30 $ \\
53.18730 & -27.89793 &$ 8.86\pm0.06 $&$ 2.01\pm0.06 $&$ -2.12\pm0.17 $&$ 0.39\pm0.16 $&$ 0.77\pm0.01 $& 3.44 &$ 0.10\pm0.04 $&$ -8.47\pm0.17 $&$ 10.91\pm2.28 $&$ 138\pm29 $ \\
53.18976 & -27.89537 &$ 9.17\pm0.04 $&$ 1.11\pm0.02 $&$ -1.55\pm0.22 $&$ 0.93\pm0.21 $&$ 0.33\pm0.01 $& 15.01 &$ 1.11\pm0.67 $&$ -8.23\pm0.21 $&$ 13.75\pm3.15 $&$ 94\pm22 $ \\
53.09959 & -27.93800 &$ 9.19\pm0.11 $&$ 1.69\pm0.02 $&$ -1.22\pm0.22 $&$ 1.28\pm0.20 $&$ 0.39\pm0.01 $& 16.36 &$ 1.06\pm0.62 $&$ -7.91\pm0.23 $&$ 29.69\pm4.44 $&$ 105\pm16 $ \\
53.16297 & -27.91688 &$ 9.53\pm0.01 $&$ 1.58\pm0.01 $&$ -2.39\pm0.05 $&$ 1.01\pm0.05 $&$ 0.36\pm0.00 $& 4.75 &$ 0.65\pm0.07 $&$ -8.52\pm0.05 $&$ 26.99\pm3.05 $&$ 99\pm12 $ \\
53.10760 & -27.76926 &$ 8.64\pm0.03 $&$ 1.35\pm0.02 $&$ -1.56\pm0.30 $&$ 0.77\pm0.28 $&$ 0.33\pm0.02 $& 2.42 &$ 0.51\pm0.46 $&$ -7.87\pm0.28 $&$ 27.90\pm2.68 $&$ 284\pm30 $ \\
53.11179 & -27.76718 &$ 9.32\pm0.02 $&$ 1.39\pm0.01 $&$ -1.15\pm0.15 $&$ 1.47\pm0.14 $&$ 0.39\pm0.01 $& 16.76 &$ 2.44\pm0.95 $&$ -7.85\pm0.14 $&$ 61.83\pm4.21 $&$ 151\pm12 $ \\
53.05181 & -27.84883 &$ 9.09\pm0.04 $&$ 1.98\pm0.02 $&$ -1.71\pm0.14 $&$ 0.76\pm0.13 $&$ 0.59\pm0.01 $& 5.19 &$ 0.24\pm0.09 $&$ -8.32\pm0.14 $&$ 12.32\pm2.26 $&$ 107\pm20 $ \\
53.07551 & -27.82831 &$ 9.43\pm0.02 $&$ 1.49\pm0.01 $&$ -1.77\pm0.08 $&$ 1.10\pm0.07 $&$ 0.34\pm0.01 $& 11.36 &$ 0.91\pm0.16 $&$ -8.32\pm0.07 $&$ 56.46\pm3.93 $&$ 249\pm21 $ \\
53.10337 & -27.84163 &$ 8.15\pm0.04 $&$ 0.68\pm0.01 $&$ -2.02\pm0.19 $&$ 0.43\pm0.18 $&$ 0.10\pm0.02 $& 9.74 &$ 0.92\pm0.47 $&$ -7.72\pm0.18 $&$ 20.55\pm1.81 $&$ 397\pm40 $ \\
53.16832 & -27.87671 &$ 9.53\pm0.02 $&$ 1.74\pm0.02 $&$ -1.68\pm0.15 $&$ 1.09\pm0.14 $&$ 0.42\pm0.02 $& 2.87 &$ 0.64\pm0.25 $&$ -8.44\pm0.14 $&$ 29.32\pm5.13 $&$ 104\pm18 $ \\
53.16614 & -27.87466 &$ 9.47\pm0.01 $&$ 0.90\pm0.01 $&$ -1.77\pm0.11 $&$ 1.15\pm0.11 $&$ 0.06\pm0.01 $& 6.01 &$ 2.76\pm0.77 $&$ -8.32\pm0.11 $&$ 27.14\pm3.79 $&$ 103\pm15 $ \\
53.18917 & -27.86309 &$ 9.04\pm0.03 $&$ 1.05\pm0.01 $&$ -2.09\pm0.09 $&$ 0.58\pm0.09 $&$ 0.54\pm0.01 $& 14.81 &$ 0.54\pm0.12 $&$ -8.46\pm0.09 $&$ 11.01\pm1.56 $&$ 139\pm20 $ \\
53.23087 & -27.89474 &$ 8.35\pm0.13 $&$ 1.07\pm0.05 $&$ -2.48\pm0.19 $&$ -0.05\pm0.18 $&$ 0.35\pm0.04 $& 14.13 &$ 0.13\pm0.07 $&$ -8.40\pm0.22 $&$ 9.08\pm1.63 $&$ 331\pm61 $ \\
53.24629 & -27.88908 &$ 9.33\pm0.01 $&$ 1.60\pm0.01 $&$ -0.03\pm0.09 $&$ 2.56\pm0.08 $&$ 0.43\pm0.01 $& 6.29 &$ 22.73\pm4.77 $&$ -6.77\pm0.08 $&$ 67.24\pm12.81 $&$ 53\pm10 $ \\
53.24427 & -27.88427 &$ 8.72\pm0.07 $&$ 1.37\pm0.04 $&$ -1.47\pm0.22 $&$ 1.10\pm0.20 $&$ 0.56\pm0.02 $& 7.55 &$ 1.08\pm0.64 $&$ -7.61\pm0.21 $&$ 70.09\pm8.54 $&$ 303\pm39 $ \\
53.01709 & -27.74002 &$ 9.64\pm0.01 $&$ 0.94\pm0.00 $&$ -1.65\pm0.04 $&$ 1.63\pm0.04 $&$ 0.01\pm0.01 $& 6.64 &$ 7.68\pm0.66 $&$ -8.01\pm0.04 $&$ 52.33\pm5.48 $&$ 82\pm9 $ \\
53.04458 & -27.72971 &$ 9.79\pm0.28 $&$ 2.91\pm0.01 $&$ -1.80\pm0.05 $&$ 1.48\pm0.05 $&$ 0.45\pm0.01 $& 4.60 &$ 0.57\pm0.07 $&$ -8.31\pm0.29 $&$ 28.26\pm6.30 $&$ 37\pm8 $ \\
53.04329 & -27.72713 &$ 9.61\pm0.01 $&$ 1.30\pm0.01 $&$ -2.07\pm0.05 $&$ 1.15\pm0.04 $&$ 0.43\pm0.00 $& 13.47 &$ 1.34\pm0.14 $&$ -8.46\pm0.04 $&$ 34.59\pm2.52 $&$ 135\pm12 $ \\
53.03610 & -27.71394 &$ 9.71\pm0.01 $&$ 2.26\pm0.03 $&$ -1.85\pm0.12 $&$ 1.07\pm0.11 $&$ 0.58\pm0.01 $& 4.97 &$ 0.36\pm0.11 $&$ -8.64\pm0.11 $&$ 16.87\pm5.44 $&$ 53\pm17 $ \\
53.01040 & -27.71408 &$ 9.92\pm0.00 $&$ 2.21\pm0.01 $&$ -1.35\pm0.06 $&$ 1.69\pm0.06 $&$ 0.23\pm0.01 $& 16.37 &$ 1.61\pm0.23 $&$ -8.23\pm0.06 $&$ 46.01\pm12.22 $&$ 58\pm15 $ \\
53.04419 & -27.76833 &$ 8.53\pm0.14 $&$ 1.08\pm0.03 $&$ -1.93\pm0.83 $&$ 0.19\pm0.77 $&$ 0.56\pm0.01 $& 10.82 &$ 0.21\pm1.04 $&$ -8.33\pm0.78 $&$ 19.33\pm4.85 $&$ 466\pm119 $ \\
53.03542 & -27.80838 &$ 8.32\pm0.04 $&$ 0.84\pm0.01 $&$ -1.96\pm0.18 $&$ 0.51\pm0.17 $&$ 0.18\pm0.03 $& 4.57 &$ 0.73\pm0.35 $&$ -7.81\pm0.17 $&$ 22.35\pm1.91 $&$ 336\pm33 $ \\
53.04000 & -27.79442 &$ 9.31\pm0.03 $&$ 1.18\pm0.01 $&$ -1.53\pm0.22 $&$ 0.95\pm0.20 $&$ 0.52\pm0.01 $& 10.14 &$ 1.01\pm0.60 $&$ -8.37\pm0.20 $&$ 17.35\pm2.46 $&$ 100\pm15 $ \\
53.02820 & -27.77938 &$ 8.39\pm0.11 $&$ 1.45\pm0.07 $&$ -0.62\pm0.67 $&$ 0.85\pm0.63 $&$ 0.31\pm0.06 $& 14.74 &$ 0.53\pm1.73 $&$ -7.54\pm0.64 $&$ 42.93\pm6.16 $&$ 389\pm59 $ \\
53.08472 & -27.86133 &$ 9.28\pm0.01 $&$ 1.90\pm0.01 $&$ -0.69\pm0.07 $&$ 2.07\pm0.06 $&$ 0.49\pm0.00 $& 10.72 &$ 5.14\pm0.82 $&$ -7.22\pm0.06 $&$ -7.43\pm6.12 $&$ -13\pm-10 $ \\
53.19357 & -27.84351 &$ 9.33\pm0.02 $&$ 1.27\pm0.01 $&$ -1.91\pm0.09 $&$ 0.91\pm0.09 $&$ 0.53\pm0.00 $& 7.25 &$ 0.80\pm0.18 $&$ -8.42\pm0.09 $&$ 25.68\pm2.44 $&$ 153\pm16 $ \\
53.21752 & -27.82591 &$ 9.37\pm0.03 $&$ 1.02\pm0.02 $&$ -2.04\pm0.23 $&$ 0.60\pm0.22 $&$ 0.16\pm0.02 $& 20.12 &$ 0.60\pm0.39 $&$ -8.77\pm0.22 $&$ 12.27\pm2.65 $&$ 98\pm21 $ \\
53.19595 & -27.82481 &$ 9.29\pm0.02 $&$ 1.17\pm0.02 $&$ -1.18\pm0.23 $&$ 1.23\pm0.21 $&$ 0.32\pm0.01 $& 6.18 &$ 1.95\pm1.23 $&$ -8.06\pm0.21 $&$ 27.48\pm4.34 $&$ 115\pm19 $ \\
53.19229 & -27.82294 &$ 9.00\pm0.04 $&$ 1.05\pm0.01 $&$ -2.08\pm0.10 $&$ 0.59\pm0.10 $&$ 0.18\pm0.01 $& 17.02 &$ 0.57\pm0.14 $&$ -8.41\pm0.10 $&$ 18.52\pm1.78 $&$ 202\pm21 $ \\
53.20340 & -27.81601 &$ 9.77\pm0.01 $&$ 2.06\pm0.01 $&$ -1.61\pm0.06 $&$ 1.40\pm0.06 $&$ 0.51\pm0.01 $& 6.93 &$ 0.94\pm0.14 $&$ -8.37\pm0.06 $&$ 33.92\pm4.38 $&$ 63\pm8 $ \\
53.18804 & -27.74458 &$ 8.17\pm0.09 $&$ 1.41\pm0.07 $&$ -1.38\pm0.77 $&$ 0.56\pm0.71 $&$ 0.35\pm0.04 $& 10.36 &$ 0.29\pm1.21 $&$ -7.61\pm0.72 $&$ 16.23\pm3.86 $&$ 316\pm76 $ \\
53.19244 & -27.73599 &$ 9.29\pm0.08 $&$ 2.31\pm0.05 $&$ -1.49\pm0.13 $&$ 1.10\pm0.12 $&$ 0.72\pm0.01 $& 18.50 &$ 0.38\pm0.12 $&$ -8.19\pm0.14 $&$ 27.95\pm6.58 $&$ 113\pm27 $ \\
53.18128 & -27.73417 &$ 9.03\pm0.03 $&$ 0.98\pm0.01 $&$ -2.06\pm0.09 $&$ 0.77\pm0.09 $&$ 0.46\pm0.01 $& 11.26 &$ 0.98\pm0.22 $&$ -8.25\pm0.09 $&$ 21.14\pm1.48 $&$ 211\pm18 $ \\
53.19071 & -27.72948 &$ 8.20\pm0.14 $&$ 0.55\pm0.04 $&$ -0.76\pm0.76 $&$ 1.15\pm0.71 $&$ 0.44\pm0.03 $& 21.11 &$ 7.51\pm30.62 $&$ -7.05\pm0.72 $&$ 32.97\pm6.68 $&$ 230\pm48 $ \\
53.18109 & -27.72681 &$ 9.67\pm0.02 $&$ 1.30\pm0.02 $&$ -0.51\pm0.23 $&$ 1.82\pm0.21 $&$ 0.22\pm0.01 $& 15.14 &$ 6.15\pm3.85 $&$ -7.85\pm0.21 $&$ 47.24\pm9.53 $&$ 56\pm11 $ \\
53.17942 & -27.72628 &$ 9.15\pm0.03 $&$ 2.01\pm0.05 $&$ -1.78\pm0.18 $&$ 0.94\pm0.17 $&$ 0.69\pm0.01 $& 12.81 &$ 0.34\pm0.17 $&$ -8.21\pm0.17 $&$ 31.75\pm6.84 $&$ 204\pm45 $ \\
53.18195 & -27.71891 &$ 9.30\pm0.01 $&$ 0.95\pm0.01 $&$ -2.19\pm0.11 $&$ 0.76\pm0.10 $&$ 0.36\pm0.01 $& 15.34 &$ 1.01\pm0.27 $&$ -8.55\pm0.10 $&$ 41.53\pm1.76 $&$ 297\pm19 $ \\
53.11838 & -27.71293 &$ 10.01\pm0.00 $&$ 1.63\pm0.01 $&$ -1.37\pm0.07 $&$ 1.67\pm0.06 $&$ 0.63\pm0.00 $& 21.09 &$ 2.80\pm0.44 $&$ -8.33\pm0.06 $&$ 34.89\pm6.44 $&$ 42\pm8 $ \\
53.10605 & -27.71073 &$ 9.14\pm0.03 $&$ 1.06\pm0.03 $&$ -1.05\pm0.38 $&$ 1.09\pm0.35 $&$ 0.39\pm0.02 $& 15.16 &$ 1.76\pm2.23 $&$ -8.05\pm0.36 $&$ 39.68\pm6.29 $&$ 189\pm31 $ \\
53.10193 & -27.70666 &$ 9.00\pm0.06 $&$ 1.30\pm0.08 $&$ -1.46\pm0.71 $&$ 0.40\pm0.66 $&$ 0.39\pm0.04 $& 12.37 &$ 0.24\pm0.84 $&$ -8.60\pm0.66 $&$ 17.57\pm3.38 $&$ 165\pm32 $ \\
53.09685 & -27.70879 &$ 9.31\pm0.02 $&$ 1.06\pm0.01 $&$ -1.66\pm0.16 $&$ 1.05\pm0.15 $&$ 0.40\pm0.01 $& 12.71 &$ 1.58\pm0.64 $&$ -8.27\pm0.15 $&$ 24.50\pm2.52 $&$ 127\pm14 $ \\
53.09352 & -27.80971 &$ 9.26\pm0.02 $&$ 0.88\pm0.01 $&$ -1.21\pm0.15 $&$ 1.34\pm0.14 $&$ 0.11\pm0.01 $& 12.17 &$ 4.58\pm1.74 $&$ -7.91\pm0.14 $&$ 40.65\pm10.62 $&$ 112\pm29 $ \\
53.09040 & -27.80183 &$ 8.61\pm0.09 $&$ 0.66\pm0.02 $&$ -1.67\pm0.34 $&$ 0.46\pm0.32 $&$ 0.21\pm0.02 $& 10.42 &$ 1.04\pm1.12 $&$ -8.15\pm0.33 $&$ 9.51\pm2.17 $&$ 203\pm47 $ \\
53.08913 & -27.79275 &$ 8.22\pm0.03 $&$ 0.82\pm0.01 $&$ -2.02\pm0.15 $&$ 0.60\pm0.14 $&$ 0.18\pm0.02 $& 2.23 &$ 0.93\pm0.35 $&$ -7.62\pm0.14 $&$ 32.80\pm2.43 $&$ 343\pm30 $ \\
53.08884 & -27.78168 &$ 8.92\pm0.11 $&$ 1.31\pm0.01 $&$ -1.87\pm0.13 $&$ 0.73\pm0.12 $&$ 0.47\pm0.02 $& 8.71 &$ 0.50\pm0.15 $&$ -8.19\pm0.16 $&$ 24.85\pm2.31 $&$ 229\pm24 $ \\
53.11297 & -27.77869 &$ 8.24\pm0.01 $&$ 2.02\pm0.02 $&$ -2.42\pm0.08 $&$ 0.60\pm0.08 $&$ 0.35\pm0.01 $& 4.82 &$ 0.15\pm0.03 $&$ -7.64\pm0.08 $&$ 31.41\pm2.62 $&$ 229\pm22 $ \\
53.14352 & -27.79522 &$ 8.60\pm0.03 $&$ 1.15\pm0.02 $&$ -2.00\pm0.12 $&$ 0.59\pm0.12 $&$ 0.36\pm0.03 $& 3.25 &$ 0.46\pm0.14 $&$ -8.02\pm0.12 $&$ 30.51\pm3.21 $&$ 381\pm44 $ \\
53.13915 & -27.78615 &$ 8.12\pm0.10 $&$ 0.71\pm0.03 $&$ -2.13\pm0.84 $&$ -0.23\pm0.78 $&$ 0.07\pm0.05 $& 19.49 &$ 0.19\pm0.94 $&$ -8.35\pm0.79 $&$ 4.74\pm0.59 $&$ 297\pm39 $ \\
53.12986 & -27.78225 &$ 9.29\pm0.02 $&$ 1.18\pm0.01 $&$ -1.63\pm0.17 $&$ 1.01\pm0.16 $&$ 0.46\pm0.01 $& 15.20 &$ 1.18\pm0.53 $&$ -8.27\pm0.16 $&$ 25.33\pm2.79 $&$ 144\pm17 $ \\
53.14781 & -27.77136 &$ 7.96\pm0.01 $&$ 0.94\pm0.01 $&$ -1.99\pm0.26 $&$ 0.55\pm0.24 $&$ 0.30\pm0.02 $& 10.08 &$ 0.64\pm0.47 $&$ -7.41\pm0.24 $&$ 31.10\pm1.17 $&$ 378\pm23 $ \\
53.14778 & -27.76562 &$ 9.41\pm0.04 $&$ 1.42\pm0.01 $&$ -1.73\pm0.10 $&$ 1.18\pm0.09 $&$ 0.33\pm0.01 $& 9.31 &$ 1.20\pm0.28 $&$ -8.23\pm0.10 $&$ 23.34\pm5.64 $&$ 71\pm17 $ \\
53.10372 & -27.71409 &$ 9.48\pm0.01 $&$ 1.64\pm0.02 $&$ -1.27\pm0.14 $&$ 1.42\pm0.13 $&$ 0.50\pm0.01 $& 19.63 &$ 1.55\pm0.56 $&$ -8.06\pm0.13 $&$ 30.27\pm5.26 $&$ 58\pm10 $ \\
53.11838 & -27.71293 &$ 9.99\pm0.03 $&$ 1.63\pm0.01 $&$ -1.37\pm0.07 $&$ 1.67\pm0.06 $&$ 0.63\pm0.00 $& 21.09 &$ 2.80\pm0.44 $&$ -8.31\pm0.07 $&$ 34.89\pm6.44 $&$ 42\pm8 $ \\
53.16881 & -27.79694 &$ 8.19\pm0.04 $&$ 1.57\pm0.04 $&$ -1.94\pm0.26 $&$ 0.44\pm0.24 $&$ 0.45\pm0.02 $& 4.06 &$ 0.18\pm0.13 $&$ -7.74\pm0.24 $&$ 11.71\pm1.68 $&$ 198\pm30 $ \\
53.14832 & -27.79594 &$ 8.57\pm0.05 $&$ 0.94\pm0.03 $&$ -1.48\pm0.49 $&$ 0.62\pm0.45 $&$ 0.11\pm0.03 $& 4.93 &$ 0.74\pm1.36 $&$ -7.95\pm0.46 $&$ 6.97\pm1.56 $&$ 105\pm24 $ \\
53.14352 & -27.79522 &$ 8.60\pm0.03 $&$ 1.15\pm0.02 $&$ -1.89\pm0.17 $&$ 0.67\pm0.16 $&$ 0.36\pm0.03 $& 3.25 &$ 0.56\pm0.24 $&$ -7.94\pm0.16 $&$ 36.40\pm1.45 $&$ 413\pm26 $ \\
53.13915 & -27.78615 &$ 8.12\pm0.10 $&$ 0.71\pm0.03 $&$ -2.11\pm0.85 $&$ -0.21\pm0.79 $&$ 0.07\pm0.05 $& 19.48 &$ 0.19\pm1.00 $&$ -8.34\pm0.79 $&$ 4.80\pm0.60 $&$ 297\pm39 $ \\
53.18222 & -27.78331 &$ 8.96\pm0.04 $&$ 1.83\pm0.05 $&$ -2.00\pm0.16 $&$ 0.65\pm0.15 $&$ 0.42\pm0.02 $& 3.70 &$ 0.21\pm0.09 $&$ -8.32\pm0.16 $&$ 8.36\pm1.60 $&$ 89\pm17 $ \\
53.16746 & -27.78183 &$ 7.88\pm0.05 $&$ 1.17\pm0.04 $&$ -0.64\pm0.92 $&$ 1.14\pm0.86 $&$ 0.23\pm0.03 $& 13.97 &$ 1.60\pm9.98 $&$ -6.74\pm0.86 $&$ 28.54\pm3.16 $&$ 271\pm32 $ \\
53.14889 & -27.77750 &$ 9.76\pm0.02 $&$ 0.71\pm0.01 $&$ -0.50\pm0.60 $&$ 1.36\pm0.56 $&$ 0.12\pm0.02 $& 12.46 &$ 7.24\pm18.81 $&$ -8.40\pm0.56 $&$ 20.56\pm4.17 $&$ 38\pm8 $ \\
53.15381 & -27.76730 &$ 9.45\pm0.01 $&$ 1.45\pm0.01 $&$ -2.10\pm0.05 $&$ 1.14\pm0.05 $&$ 0.33\pm0.01 $& 4.53 &$ 1.04\pm0.12 $&$ -8.32\pm0.05 $&$ 19.94\pm3.95 $&$ 70\pm14 $ \\
53.14781 & -27.77136 &$ 7.96\pm0.01 $&$ 0.94\pm0.01 $&$ -1.99\pm0.30 $&$ 0.54\pm0.27 $&$ 0.30\pm0.02 $& 10.08 &$ 0.64\pm0.56 $&$ -7.42\pm0.27 $&$ 31.02\pm1.17 $&$ 378\pm23 $ \\
188.99576 & 62.20455 &$ 8.61\pm0.05 $& -- &$ -1.93\pm0.16 $&$ 0.61\pm0.15 $& -- & -- & -- &$ -7.99\pm0.16 $&$ 26.93\pm1.79 $&$ 289\pm24 $ \\
189.00248 & 62.21303 &$ 9.15\pm0.09 $& -- &$ -1.90\pm0.33 $&$ 0.56\pm0.31 $& -- & -- & -- &$ -8.60\pm0.32 $&$ 7.67\pm2.15 $&$ 61\pm17 $ \\
189.04899 & 62.21538 &$ 8.02\pm0.06 $& -- &$ -2.08\pm0.44 $&$ 0.25\pm0.41 $& -- & -- & -- &$ -7.77\pm0.41 $&$ 12.90\pm2.82 $&$ 220\pm49 $ \\
189.06248 & 62.22497 &$ 9.02\pm0.06 $& -- &$ -1.97\pm0.20 $&$ 0.64\pm0.19 $& -- & -- & -- &$ -8.38\pm0.20 $&$ 14.55\pm1.80 $&$ 167\pm22 $ \\
189.03772 & 62.23307 &$ 9.85\pm0.01 $& -- &$ -1.52\pm0.06 $&$ 1.49\pm0.06 $& -- & -- & -- &$ -8.36\pm0.06 $&$ 39.70\pm2.70 $&$ 67\pm5 $ \\
189.02381 & 62.22722 &$ 7.87\pm0.04 $& -- &$ -2.37\pm0.33 $&$ 0.15\pm0.31 $& -- & -- & -- &$ -7.73\pm0.31 $&$ 20.34\pm2.99 $&$ 324\pm50 $ \\
189.06606 & 62.22381 &$ 8.33\pm0.12 $& -- &$ -2.20\pm0.30 $&$ 0.32\pm0.28 $& -- & -- & -- &$ -8.02\pm0.30 $&$ 8.81\pm1.25 $&$ 242\pm36 $ \\
189.06248 & 62.22497 &$ 9.02\pm0.06 $& -- &$ -1.97\pm0.20 $&$ 0.64\pm0.19 $& -- & -- & -- &$ -8.39\pm0.20 $&$ 14.54\pm1.80 $&$ 167\pm22 $ \\
189.03772 & 62.23307 &$ 9.85\pm0.01 $& -- &$ -1.52\pm0.06 $&$ 1.49\pm0.06 $& -- & -- & -- &$ -8.36\pm0.06 $&$ 39.67\pm2.70 $&$ 67\pm5 $ \\
189.08526 & 62.24783 &$ 8.45\pm0.17 $& -- &$ -2.26\pm0.38 $&$ 0.05\pm0.35 $& -- & -- & -- &$ -8.40\pm0.39 $&$ 5.56\pm1.05 $&$ 170\pm33 $ \\
189.17219 & 62.29881 &$ 9.90\pm0.01 $& -- &$ -1.35\pm0.07 $&$ 1.52\pm0.06 $& -- & -- & -- &$ -8.38\pm0.06 $&$ 54.39\pm7.18 $&$ 92\pm13 $ \\
189.22353 & 62.29847 &$ 8.03\pm0.05 $& -- &$ -1.83\pm0.45 $&$ 0.40\pm0.42 $& -- & -- & -- &$ -7.63\pm0.42 $&$ 27.88\pm4.80 $&$ 385\pm69 $ \\
189.20491 & 62.30445 &$ 8.96\pm0.07 $& -- &$ -1.71\pm0.15 $&$ 0.94\pm0.14 $& -- & -- & -- &$ -8.02\pm0.15 $&$ 48.99\pm7.23 $&$ 242\pm37 $ \\
189.21937 & 62.30544 &$ 9.45\pm0.03 $& -- &$ -1.87\pm0.07 $&$ 1.16\pm0.06 $& -- & -- & -- &$ -8.29\pm0.07 $&$ 49.37\pm3.89 $&$ 165\pm15 $ \\
189.22689 & 62.30773 &$ 8.75\pm0.07 $& -- &$ -1.41\pm0.27 $&$ 0.81\pm0.25 $& -- & -- & -- &$ -7.94\pm0.26 $&$ 32.84\pm4.27 $&$ 258\pm36 $ \\
189.24072 & 62.31822 &$ 8.23\pm0.03 $& -- &$ -1.34\pm0.24 $&$ 0.89\pm0.23 $& -- & -- & -- &$ -7.34\pm0.23 $&$ 62.74\pm4.06 $&$ 399\pm32 $ \\
189.21432 & 62.32324 &$ 8.87\pm0.07 $& -- &$ -2.25\pm0.19 $&$ 0.52\pm0.17 $& -- & -- & -- &$ -8.36\pm0.19 $&$ 10.73\pm2.09 $&$ 126\pm25 $ \\
189.26024 & 62.32050 &$ 9.77\pm0.01 $& -- &$ -1.94\pm0.05 $&$ 1.19\pm0.04 $& -- & -- & -- &$ -8.57\pm0.04 $&$ 18.56\pm2.28 $&$ 49\pm6 $ \\
189.25123 & 62.32082 &$ 8.43\pm0.13 $& -- &$ -1.17\pm0.43 $&$ 0.77\pm0.41 $& -- & -- & -- &$ -7.66\pm0.43 $&$ 18.05\pm3.87 $&$ 196\pm43 $ \\
189.23959 & 62.32524 &$ 8.33\pm0.13 $& -- &$ -0.44\pm0.50 $&$ 1.26\pm0.47 $& -- & -- & -- &$ -7.07\pm0.49 $&$ 20.77\pm5.83 $&$ 245\pm69 $ \\
189.27839 & 62.32653 &$ 9.77\pm0.01 $& -- &$ -1.83\pm0.06 $&$ 1.28\pm0.05 $& -- & -- & -- &$ -8.49\pm0.05 $&$ 50.69\pm3.57 $&$ 120\pm10 $ \\
189.27377 & 62.32666 &$ 8.82\pm0.17 $& -- &$ -2.10\pm0.28 $&$ 0.37\pm0.26 $& -- & -- & -- &$ -8.45\pm0.31 $&$ 11.84\pm2.06 $&$ 199\pm35 $ \\
189.30246 & 62.32795 &$ 10.07\pm0.01 $& -- &$ -1.37\pm0.05 $&$ 1.64\pm0.04 $& -- & -- & -- &$ -8.43\pm0.04 $&$ 52.36\pm7.06 $&$ 61\pm8 $ \\
189.30526 & 62.32801 &$ 9.10\pm0.11 $& -- &$ -1.64\pm0.29 $&$ 0.50\pm0.27 $& -- & -- & -- &$ -8.60\pm0.29 $&$ 9.87\pm1.87 $&$ 133\pm26 $ \\
189.24342 & 62.32965 &$ 8.92\pm0.19 $& -- &$ -2.32\pm0.50 $&$ -0.05\pm0.47 $& -- & -- & -- &$ -8.98\pm0.50 $&$ 6.38\pm1.84 $&$ 180\pm52 $ \\
189.25594 & 62.33626 &$ 9.33\pm0.08 $& -- &$ -1.47\pm0.36 $&$ 0.79\pm0.34 $& -- & -- & -- &$ -8.54\pm0.35 $&$ 26.93\pm3.47 $&$ 170\pm23 $ \\
189.28220 & 62.33822 &$ 8.19\pm0.09 $& -- &$ -2.49\pm0.42 $&$ -0.02\pm0.39 $& -- & -- & -- &$ -8.21\pm0.40 $&$ 18.55\pm1.36 $&$ 334\pm29 $ \\
189.26735 & 62.34506 &$ 10.11\pm0.07 $& -- &$ -0.54\pm0.44 $&$ 1.44\pm0.41 $& -- & -- & -- &$ -8.67\pm0.42 $&$ 36.27\pm6.00 $&$ 85\pm14 $ \\
189.32224 & 62.34025 &$ 8.62\pm0.13 $& -- &$ -2.66\pm0.22 $&$ -0.12\pm0.20 $& -- & -- & -- &$ -8.74\pm0.24 $&$ 13.68\pm1.09 $&$ 307\pm28 $ \\
189.32373 & 62.34848 &$ 8.82\pm0.11 $& -- &$ -2.15\pm0.13 $&$ 0.48\pm0.12 $& -- & -- & -- &$ -8.34\pm0.17 $&$ 15.59\pm4.10 $&$ 203\pm54 $ \\
189.34988 & 62.35917 &$ 8.58\pm0.16 $& -- &$ -1.97\pm0.61 $&$ 0.40\pm0.57 $& -- & -- & -- &$ -8.18\pm0.59 $&$ 14.86\pm1.89 $&$ 248\pm33 $ \\
189.33272 & 62.36602 &$ 7.62\pm0.11 $& -- &$ -1.24\pm1.34 $&$ 0.38\pm1.26 $& -- & -- & -- &$ -7.24\pm1.27 $&$ 25.23\pm3.96 $&$ 486\pm80 $ \\
189.30218 & 62.37042 &$ 9.04\pm0.07 $& -- &$ -2.24\pm0.22 $&$ 0.37\pm0.20 $& -- & -- & -- &$ -8.67\pm0.21 $&$ 11.07\pm1.60 $&$ 145\pm22 $ \\
189.07819 & 62.17703 &$ 9.57\pm0.05 $& -- &$ -1.73\pm0.12 $&$ 1.18\pm0.11 $& -- & -- & -- &$ -8.39\pm0.12 $&$ 34.24\pm5.66 $&$ 98\pm16 $ \\
189.01411 & 62.18331 &$ 8.73\pm0.10 $& -- &$ -1.94\pm0.26 $&$ 0.58\pm0.24 $& -- & -- & -- &$ -8.15\pm0.26 $&$ 20.22\pm2.06 $&$ 268\pm30 $ \\
189.07820 & 62.17703 &$ 9.57\pm0.06 $& -- &$ -1.73\pm0.12 $&$ 1.18\pm0.11 $& -- & -- & -- &$ -8.38\pm0.13 $&$ 35.94\pm3.57 $&$ 99\pm11 $ \\
189.05803 & 62.19085 &$ 8.06\pm0.19 $& -- &$ -2.22\pm0.28 $&$ 0.15\pm0.26 $& -- & -- & -- &$ -7.91\pm0.32 $&$ 6.89\pm0.94 $&$ 248\pm36 $ \\
189.05556 & 62.19589 &$ 8.88\pm0.13 $& -- &$ -1.74\pm0.25 $&$ 0.70\pm0.24 $& -- & -- & -- &$ -8.18\pm0.27 $&$ 14.58\pm3.32 $&$ 150\pm35 $ \\
189.10401 & 62.20656 &$ 9.08\pm0.07 $& -- &$ -1.56\pm0.18 $&$ 1.06\pm0.17 $& -- & -- & -- &$ -8.02\pm0.18 $&$ 25.36\pm4.31 $&$ 114\pm20 $ \\
189.16919 & 62.21970 &$ 9.02\pm0.14 $& -- &$ -1.81\pm0.25 $&$ 0.69\pm0.24 $& -- & -- & -- &$ -8.32\pm0.28 $&$ 28.01\pm2.66 $&$ 235\pm25 $ \\
189.17537 & 62.22540 &$ 10.53\pm0.17 $& -- &$ 0.52\pm0.51 $&$ 2.10\pm0.48 $& -- & -- & -- &$ -8.43\pm0.51 $&$ 249.27\pm17.80 $&$ 143\pm12 $ \\
189.18707 & 62.22655 &$ 7.64\pm0.11 $& -- &$ -1.85\pm0.44 $&$ 0.29\pm0.41 $& -- & -- & -- &$ -7.34\pm0.42 $&$ 8.76\pm1.79 $&$ 248\pm52 $ \\
189.14760 & 62.24377 &$ 9.67\pm0.01 $& -- &$ -1.68\pm0.06 $&$ 1.37\pm0.05 $& -- & -- & -- &$ -8.30\pm0.05 $&$ 40.71\pm13.69 $&$ 99\pm33 $ \\
189.23168 & 62.24741 &$ 8.12\pm0.11 $&$ 2.57\pm0.06 $&$ -1.85\pm0.21 $&$ 0.51\pm0.20 $&$ 0.13\pm0.16 $& 21.34 &$ 0.08\pm0.04 $&$ -7.61\pm0.22 $&$ 8.89\pm1.63 $&$ 351\pm66 $ \\
189.19774 & 62.25096 &$ 8.42\pm0.06 $&$ 0.89\pm0.01 $&$ -2.46\pm0.20 $&$ 0.24\pm0.19 $&$ 0.45\pm0.02 $& 51.08 &$ 0.35\pm0.19 $&$ -8.19\pm0.20 $&$ 24.43\pm1.41 $&$ 357\pm27 $ \\
189.22412 & 62.25603 &$ 8.44\pm0.13 $&$ 1.27\pm0.02 $&$ -2.54\pm0.26 $&$ 0.15\pm0.24 $&$ 0.52\pm0.05 $& 17.16 &$ 0.14\pm0.10 $&$ -8.29\pm0.27 $&$ 10.05\pm2.02 $&$ 229\pm47 $ \\
189.25813 & 62.26390 &$ 9.58\pm0.02 $&$ 1.00\pm0.01 $&$ -1.99\pm0.14 $&$ 0.86\pm0.13 $&$ 0.21\pm0.03 $& 11.91 &$ 1.15\pm0.41 $&$ -8.72\pm0.13 $&$ 36.51\pm2.47 $&$ 172\pm14 $ \\
189.19239 & 62.26434 &$ 9.38\pm0.02 $& -- &$ -2.00\pm0.04 $&$ 1.27\pm0.04 $& -- & 0.88 & -- &$ -8.12\pm0.04 $&$ 77.50\pm5.79 $&$ 208\pm18 $ \\
189.26161 & 62.26430 &$ 8.73\pm0.08 $&$ 1.30\pm0.02 $&$ -2.29\pm0.16 $&$ 0.24\pm0.15 $&$ 0.41\pm0.03 $& 26.49 &$ 0.17\pm0.07 $&$ -8.49\pm0.17 $&$ 9.73\pm1.16 $&$ 217\pm28 $ \\
189.23716 & 62.27272 &$ 9.11\pm0.07 $&$ 1.78\pm0.02 $&$ -1.99\pm0.12 $&$ 0.80\pm0.11 $&$ 0.63\pm0.01 $& 22.53 &$ 0.31\pm0.09 $&$ -8.31\pm0.13 $&$ 22.31\pm2.16 $&$ 154\pm16 $ \\
189.25422 & 62.27573 &$ 8.78\pm0.13 $&$ 0.59\pm0.04 $&$ -1.98\pm0.50 $&$ -0.09\pm0.47 $&$ 0.21\pm0.08 $& 22.47 &$ 0.37\pm0.73 $&$ -8.87\pm0.49 $&$ 4.37\pm1.35 $&$ 198\pm62 $ \\
189.28491 & 62.27583 &$ 8.44\pm0.11 $&$ 0.85\pm0.02 $&$ -1.60\pm0.30 $&$ 0.80\pm0.28 $&$ 0.31\pm0.05 $& 2.16 &$ 1.39\pm1.25 $&$ -7.63\pm0.30 $&$ 30.44\pm3.09 $&$ 282\pm31 $ \\
189.25994 & 62.27750 &$ 9.21\pm0.04 $&$ 1.18\pm0.01 $&$ -1.47\pm0.12 $&$ 1.26\pm0.11 $&$ 0.39\pm0.03 $& 20.73 &$ 2.06\pm0.60 $&$ -7.95\pm0.12 $&$ 59.99\pm8.37 $&$ 189\pm28 $ \\
189.26627 & 62.27763 &$ 8.09\pm0.19 $&$ 0.31\pm0.01 $&$ -2.91\pm0.51 $&$ -0.67\pm0.48 $&$ 0.10\pm0.06 $& 13.84 &$ 0.35\pm0.70 $&$ -8.75\pm0.51 $&$ 9.37\pm1.21 $&$ 408\pm56 $ \\
189.31006 & 62.28404 &$ 9.51\pm0.04 $&$ 3.73\pm0.06 $&$ -1.44\pm0.12 $&$ 1.20\pm0.11 $& -- & 2.74 &$ 0.18\pm0.05 $&$ -8.31\pm0.12 $&$ 34.42\pm5.06 $&$ 115\pm17 $ \\
189.30022 & 62.28829 &$ 8.00\pm0.13 $&$ 1.01\pm0.04 $&$ -2.82\pm0.75 $&$ -0.45\pm0.70 $&$ 0.35\pm0.07 $& 18.85 &$ 0.06\pm0.22 $&$ -8.45\pm0.71 $&$ 18.19\pm3.59 $&$ 497\pm101 $ \\
189.23436 & 62.28846 &$ 8.30\pm0.17 $& -- &$ -1.97\pm0.22 $&$ 0.31\pm0.20 $& -- & -- & -- &$ -7.99\pm0.27 $&$ 13.89\pm1.44 $&$ 313\pm36 $ \\
189.29888 & 62.28939 &$ 8.09\pm0.14 $&$ 1.34\pm0.04 $&$ -2.36\pm0.61 $&$ -0.06\pm0.57 $&$ 0.38\pm0.10 $& 2.24 &$ 0.08\pm0.21 $&$ -8.15\pm0.58 $&$ 9.33\pm1.92 $&$ 313\pm66 $ \\
189.26949 & 62.29079 &$ 8.80\pm0.08 $&$ 1.64\pm0.06 $&$ 0.24\pm0.47 $&$ 2.17\pm0.44 $&$ 0.81\pm0.01 $& 6.87 &$ 8.72\pm15.32 $&$ -6.63\pm0.45 $&$ 227.08\pm19.43 $&$ 751\pm74 $ \\
189.29635 & 62.29415 &$ 8.07\pm0.05 $&$ 1.45\pm0.02 $&$ -2.08\pm0.16 $&$ 0.55\pm0.15 $&$ 0.68\pm0.05 $& 4.32 &$ 0.27\pm0.11 $&$ -7.52\pm0.16 $&$ 20.98\pm1.49 $&$ 309\pm26 $ \\
189.33446 & 62.29348 &$ 9.39\pm0.02 $&$ 1.69\pm0.01 $&$ -2.09\pm0.07 $&$ 0.91\pm0.07 $&$ 0.11\pm0.02 $& 25.93 &$ 0.45\pm0.07 $&$ -8.48\pm0.07 $&$ 29.16\pm1.98 $&$ 151\pm12 $ \\
189.34123 & 62.30173 &$ 9.70\pm0.15 $&$ 1.90\pm0.00 $&$ -1.78\pm0.01 $&$ 2.04\pm0.01 $&$ 0.56\pm0.00 $& 9.29 &$ 4.79\pm0.13 $&$ -7.66\pm0.15 $&$ 32.30\pm3.37 $&$ 23\pm2 $ \\
189.34254 & 62.30152 &$ 8.58\pm0.08 $&$ 1.18\pm0.03 $&$ -1.45\pm0.31 $&$ 0.90\pm0.29 $&$ 0.52\pm0.02 $& 2.45 &$ 0.90\pm0.86 $&$ -7.69\pm0.30 $&$ 43.70\pm3.95 $&$ 251\pm25 $ \\
189.33936 & 62.30858 &$ 9.13\pm0.04 $&$ 1.24\pm0.01 $&$ -2.19\pm0.07 $&$ 0.70\pm0.06 $&$ 0.41\pm0.02 $& 8.21 &$ 0.52\pm0.08 $&$ -8.43\pm0.08 $&$ 27.49\pm1.35 $&$ 224\pm15 $ \\
189.33991 & 62.30851 &$ 9.22\pm0.02 $&$ 1.61\pm0.01 $&$ -2.35\pm0.06 $&$ 0.64\pm0.05 $&$ 0.47\pm0.01 $& 8.21 &$ 0.27\pm0.04 $&$ -8.58\pm0.06 $&$ 12.42\pm1.43 $&$ 107\pm13 $ \\
189.28801 & 62.31523 &$ 8.91\pm0.06 $& -- &$ -2.45\pm0.19 $&$ 0.32\pm0.17 $& -- & -- & -- &$ -8.59\pm0.18 $&$ 8.21\pm1.35 $&$ 109\pm18 $ \\
189.31805 & 62.32235 &$ 9.83\pm0.01 $& -- &$ -1.53\pm0.06 $&$ 1.55\pm0.06 $& -- & 3.75 & -- &$ -8.29\pm0.06 $&$ 72.55\pm3.55 $&$ 114\pm8 $ \\
189.29659 & 62.31574 &$ 9.05\pm0.06 $& -- &$ -1.76\pm0.18 $&$ 0.58\pm0.17 $& -- & -- & -- &$ -8.47\pm0.18 $&$ 13.80\pm1.64 $&$ 158\pm20 $ \\
189.40167 & 62.32212 &$ 8.79\pm0.05 $&$ 1.20\pm0.02 $&$ -1.78\pm0.17 $&$ 0.70\pm0.15 $&$ 0.48\pm0.02 $& 15.20 &$ 0.55\pm0.23 $&$ -8.10\pm0.16 $&$ 17.60\pm1.68 $&$ 171\pm18 $ \\
189.36074 & 62.32344 &$ 8.76\pm0.19 $&$ 1.52\pm0.04 $&$ -1.81\pm0.45 $&$ 0.40\pm0.42 $&$ 0.46\pm0.07 $& 51.13 &$ 0.17\pm0.28 $&$ -8.36\pm0.46 $&$ 11.22\pm1.80 $&$ 180\pm30 $ \\
189.38253 & 62.32442 &$ 8.45\pm0.15 $&$ 0.98\pm0.02 $&$ -1.99\pm0.40 $&$ 0.43\pm0.37 $&$ 0.23\pm0.06 $& 49.78 &$ 0.45\pm0.61 $&$ -8.02\pm0.40 $&$ 14.84\pm1.83 $&$ 273\pm36 $ \\
189.38390 & 62.32925 &$ 9.10\pm0.06 $&$ 1.05\pm0.02 $&$ -1.81\pm0.27 $&$ 0.71\pm0.25 $&$ 0.54\pm0.02 $& 31.55 &$ 0.74\pm0.58 $&$ -8.39\pm0.26 $&$ 14.27\pm2.81 $&$ 126\pm25 $ \\
189.39256 & 62.33348 &$ 9.73\pm0.01 $&$ 1.36\pm0.01 $&$ -1.75\pm0.15 $&$ 1.20\pm0.14 $&$ 0.38\pm0.01 $& 12.98 &$ 1.35\pm0.53 $&$ -8.53\pm0.14 $&$ 51.47\pm3.77 $&$ 120\pm10 $ \\
189.39343 & 62.33770 &$ 8.97\pm0.04 $&$ 0.82\pm0.01 $&$ -1.84\pm0.26 $&$ 0.87\pm0.24 $&$ 0.12\pm0.03 $& 20.07 &$ 1.75\pm1.27 $&$ -8.10\pm0.24 $&$ 38.64\pm4.69 $&$ 227\pm29 $ \\
189.05848 & 62.14314 &$ 9.85\pm0.01 $& -- &$ -0.64\pm0.08 $&$ 2.15\pm0.08 $& -- & -- & -- &$ -7.70\pm0.08 $&$ 62.60\pm6.10 $&$ 55\pm6 $ \\
189.12640 & 62.16257 &$ 9.63\pm0.01 $& -- &$ -2.07\pm0.07 $&$ 1.00\pm0.07 $& -- & -- & -- &$ -8.63\pm0.07 $&$ 23.86\pm2.97 $&$ 88\pm11 $ \\
189.17192 & 62.18768 &$ 9.47\pm0.02 $& -- &$ -1.96\pm0.12 $&$ 0.90\pm0.11 $& -- & -- & -- &$ -8.57\pm0.11 $&$ 33.36\pm3.10 $&$ 182\pm19 $ \\
189.22943 & 62.22975 &$ 8.97\pm0.02 $&$ 3.52\pm0.02 $&$ -2.10\pm0.10 $&$ 0.85\pm0.09 $& -- & 3.50 &$ 0.09\pm0.02 $&$ -8.11\pm0.09 $&$ 38.93\pm2.56 $&$ 214\pm17 $ \\
189.22578 & 62.22663 &$ 7.71\pm0.07 $& -- &$ -2.59\pm0.59 $&$ -0.22\pm0.55 $& -- & -- & -- &$ -7.93\pm0.55 $&$ 12.29\pm1.87 $&$ 351\pm56 $ \\
189.29907 & 62.22747 &$ 8.92\pm0.04 $&$ 0.94\pm0.01 $&$ -2.08\pm0.17 $&$ 0.33\pm0.16 $&$ 0.21\pm0.03 $& 34.99 &$ 0.39\pm0.17 $&$ -8.59\pm0.16 $&$ 12.65\pm1.29 $&$ 227\pm25 $ \\
189.25837 & 62.23863 &$ 8.71\pm0.17 $&$ 1.96\pm0.06 $&$ -2.21\pm0.53 $&$ 0.03\pm0.49 $&$ 0.30\pm0.11 $& 34.94 &$ 0.04\pm0.09 $&$ -8.68\pm0.52 $&$ 8.54\pm1.84 $&$ 230\pm50 $ \\
189.29665 & 62.24210 &$ 8.51\pm0.06 $&$ 0.89\pm0.01 $&$ -2.24\pm0.24 $&$ 0.46\pm0.22 $&$ 0.36\pm0.03 $& 8.37 &$ 0.58\pm0.38 $&$ -8.05\pm0.23 $&$ 16.38\pm1.54 $&$ 243\pm25 $ \\
189.26806 & 62.24618 &$ 10.51\pm0.02 $&$ 1.08\pm0.01 $&$ -1.76\pm0.10 $&$ 1.18\pm0.09 $&$ 0.26\pm0.02 $& 27.04 &$ 2.05\pm0.48 $&$ -9.33\pm0.09 $&$ 176.06\pm3.53 $&$ 231\pm12 $ \\
189.33323 & 62.24905 &$ 9.21\pm0.05 $&$ 1.42\pm0.01 $&$ -1.92\pm0.09 $&$ 0.96\pm0.08 $&$ 0.37\pm0.02 $& 11.79 &$ 0.72\pm0.14 $&$ -8.25\pm0.09 $&$ 41.88\pm1.81 $&$ 218\pm14 $ \\
189.34656 & 62.26064 &$ 10.39\pm0.01 $&$ 0.67\pm0.00 $&$ -0.43\pm0.05 $&$ 2.36\pm0.04 $&$ 0.16\pm0.01 $& 13.34 &$ 82.01\pm8.97 $&$ -8.02\pm0.05 $&$ 196.77\pm49.67 $&$ 70\pm18 $ \\
189.33931 & 62.27194 &$ 9.25\pm0.07 $&$ 0.95\pm0.01 $&$ -1.36\pm0.16 $&$ 1.23\pm0.15 $&$ 0.03\pm0.02 $& 16.79 &$ 3.01\pm1.21 $&$ -8.02\pm0.16 $&$ 37.92\pm3.54 $&$ 132\pm14 $ \\
189.11442 & 62.11135 &$ 10.32\pm0.01 $& -- &$ -0.89\pm0.06 $&$ 2.13\pm0.05 $& -- & -- & -- &$ -8.19\pm0.05 $&$ 111.35\pm18.62 $&$ 57\pm10 $ \\
189.17800 & 62.11557 &$ 9.09\pm0.07 $& -- &$ -0.34\pm0.15 $&$ 2.30\pm0.14 $& -- & -- & -- &$ -6.80\pm0.16 $&$ 161.82\pm34.02 $&$ 176\pm38 $ \\
189.12000 & 62.11658 &$ 8.09\pm0.14 $& -- &$ -1.96\pm0.39 $&$ 0.06\pm0.37 $& -- & -- & -- &$ -8.03\pm0.39 $&$ 8.98\pm1.14 $&$ 236\pm32 $ \\
189.10132 & 62.11879 &$ 9.23\pm0.04 $& -- &$ -2.30\pm0.14 $&$ 0.62\pm0.13 $& -- & -- & -- &$ -8.61\pm0.14 $&$ 17.87\pm1.46 $&$ 144\pm13 $ \\
189.14746 & 62.13034 &$ 9.32\pm0.04 $& -- &$ -2.15\pm0.09 $&$ 0.99\pm0.08 $& -- & -- & -- &$ -8.32\pm0.09 $&$ 13.58\pm2.22 $&$ 64\pm11 $ \\
189.15377 & 62.13209 &$ 7.85\pm0.03 $& -- &$ -2.34\pm0.20 $&$ 0.25\pm0.18 $& -- & -- & -- &$ -7.60\pm0.19 $&$ 21.64\pm1.09 $&$ 432\pm30 $ \\
189.17982 & 62.16091 &$ 9.27\pm0.03 $& -- &$ -1.53\pm0.23 $&$ 0.84\pm0.22 $& -- & -- & -- &$ -8.42\pm0.22 $&$ 25.31\pm3.37 $&$ 185\pm26 $ \\
189.22796 & 62.16110 &$ 8.66\pm0.04 $& -- &$ -2.07\pm0.10 $&$ 0.60\pm0.09 $& -- & -- & -- &$ -8.06\pm0.10 $&$ 24.62\pm1.26 $&$ 303\pm21 $ \\
189.25155 & 62.16676 &$ 8.69\pm0.08 $& -- &$ -2.24\pm0.23 $&$ 0.41\pm0.21 $& -- & -- & -- &$ -8.28\pm0.22 $&$ 12.66\pm1.34 $&$ 217\pm25 $ \\
189.26131 & 62.16966 &$ 8.67\pm0.06 $& -- &$ -1.66\pm0.16 $&$ 0.90\pm0.15 $& -- & -- & -- &$ -7.77\pm0.16 $&$ 57.39\pm2.37 $&$ 353\pm22 $ \\
189.22217 & 62.17663 &$ 9.54\pm0.01 $& -- &$ -1.69\pm0.09 $&$ 1.02\pm0.08 $& -- & -- & -- &$ -8.52\pm0.08 $&$ 29.78\pm3.43 $&$ 134\pm16 $ \\
189.29343 & 62.17652 &$ 8.16\pm0.09 $& -- &$ -2.02\pm0.57 $&$ 0.19\pm0.53 $& -- & -- & -- &$ -7.97\pm0.54 $&$ 12.27\pm2.02 $&$ 222\pm38 $ \\
189.27959 & 62.19796 &$ 9.14\pm0.05 $& -- &$ -1.89\pm0.11 $&$ 0.93\pm0.10 $& -- & -- & -- &$ -8.21\pm0.11 $&$ 24.45\pm2.86 $&$ 145\pm18 $ \\
189.31272 & 62.20915 &$ 7.78\pm0.05 $&$ 1.00\pm0.02 $&$ -2.23\pm0.46 $&$ 0.22\pm0.43 $&$ 0.30\pm0.04 $& 23.28 &$ 0.27\pm0.46 $&$ -7.56\pm0.43 $&$ 7.80\pm1.41 $&$ 180\pm33 $ \\
189.39515 & 62.21305 &$ 9.87\pm0.02 $&$ 1.59\pm0.02 $&$ -1.34\pm0.17 $&$ 1.28\pm0.15 $&$ 0.46\pm0.02 $& 20.71 &$ 1.19\pm0.51 $&$ -8.59\pm0.16 $&$ 28.14\pm7.02 $&$ 65\pm16 $ \\
189.39009 & 62.23108 &$ 9.57\pm0.02 $&$ 0.75\pm0.01 $&$ -1.71\pm0.14 $&$ 1.05\pm0.13 $&$ 0.30\pm0.01 $& 10.82 &$ 3.19\pm1.08 $&$ -8.52\pm0.13 $&$ 14.93\pm3.01 $&$ 68\pm14 $ \\
\enddata
\tablecomments{For details on specific measurements, please see the note in Table 1.}
\end{deluxetable}

\end{document}